# Portfolios Under Construction

Yin Luo, CFA
yin.luo@db.com

Sheng Wang
sheng.wang@db.com

Javed Jussa
javed.jussa@db.com

Zongye Chen
john.chen@db.com

Miguel-A Alvarez
miguel-a.alvarez@db.com

North America: +1 212 250 8983
Europe: +44 20 754 71684
Asia: +852 2203 6990

# Dynamic allocation: extremes, tail dependence, and regimes

DB Handbook of Portfolio Construction, Part 2

### Introducing a global financial risk regime indicator

By capturing outliers, volatility clustering, and tail dependence in the asset return distribution, we build a sophisticated model to predict the downside risk of the global financial market. We further develop a dynamic regime switching model that can forecast real-time risk regime of the market.

### Dynamic asset allocation with extremes, tail dependence and regime switching

Our GARCH-DCC-Copula risk model can significantly improve both risk- and alpha-based global tactical asset allocation strategies.

### Quantitative equity factor weighting and alternative beta allocation

Our risk regime has strong predictive power of quantitative equity factor performance, which can help equity investors to build better factor models and asset allocation managers to construct more efficient risk premia portfolios.

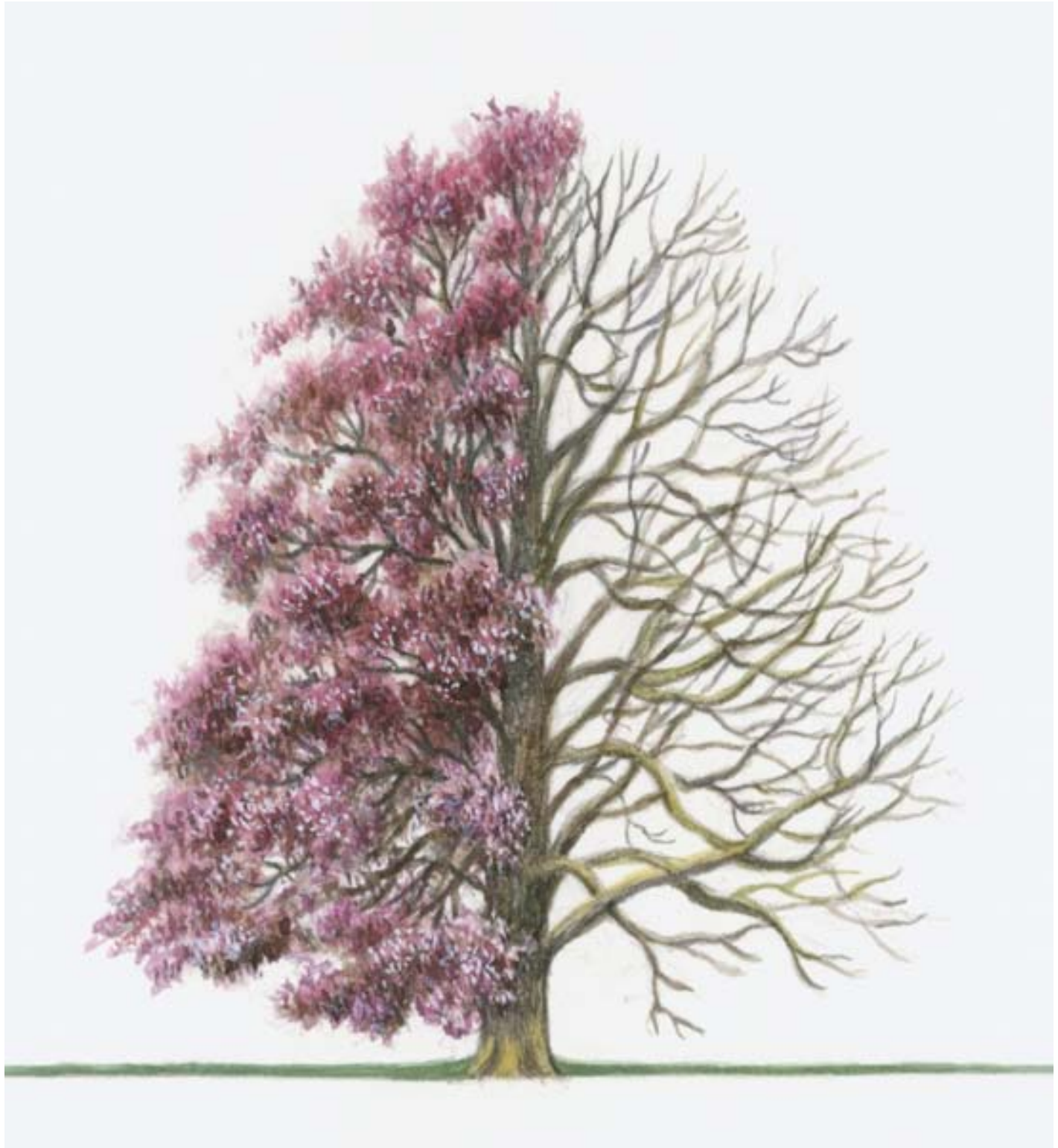

*Source: gettyimages.com*

Table Of Contents

# A letter to our readers

## DB Handbook of Portfolio Construction, Part 2

In Part 1 of our *DB Handbook of Portfolio Construction*[1], we focused on risk-based portfolio construction techniques. We suggested that the heart of portfolio construction was about diversification and risk reduction. We found that asset returns almost never follow the theoretical assumption of multivariate normal distribution. Therefore, traditional risk management and portfolio construction techniques that heavily depend on the normality assumption may not be appropriate. We further developed two alternative portfolio construction techniques – minimum tail dependence portfolio and robust minimum CVaR strategy to better capture the diversification and risk reduction benefit.

In a world where fewer and fewer asset classes provide consistent positive risk premia, and that frequently faces risk-on and risk-off, the traditional static asset allocation[2] may have become obsolete. In Luo, *et al* [2010] "Style rotation", Luo, et *al* [2011] "Quant tactical asset allocation", Luo, *et al* [2013] "DB Handbook of Portfolio Construction, Part 1", for example, we have been publishing extensively in the field of dynamic asset allocation.

In this research, we further expand our previous work on how to incorporate extreme events, tail dependence, and more importantly, shifts in regimes in risk management and portfolio construction.

First of all, we use a representative asset that comprises 50% of equities, 40% of bonds, and 10% of alternatives to proxy the global capital market. We then develop a GARCH-DCC-Copula risk model that accounts for return serial correlation, volatility clustering, dynamic correlation, and tail dependence in asset returns. Our GARCH-DCC-Copula model is next used to predict the future risk of the global capital market, using conditional value-at-risk (CVaR) that capture the downside risk.

Secondly, we develop a dynamic Markov switching model on the global risk indicator (CVaR) to predict whether we are in a high risk or low risk regime in real time. We find our real-time risk regimes can clearly differentiate return and risk for almost all asset classes. We also successfully implement a GTAA model conditional on our risk regime indicator, which significantly improves our predictive power of asset returns. More interestingly, we also find our risk regime indicator can help us to predict the returns of quantitative stock-selection factors and alternative beta strategies; therefore, we can finally bridge top-down asset allocation with bottom-up securities selection.

Lastly, we find our GARCH-DCC-Copula model captures asset return and risk much better than traditional risk models. We use two examples – a real-life 11 asset class allocation and an equity factor weighting – to demonstrate that our

[1] See Luo, *et al* [2013].

[2] When we say "asset allocation", we refer to not only the classic definition of allocating among different asset classes, but also strategies like factor weighting in a quantitatively managed equity portfolio.

GARCH-DCC-Copula model can considerably improve portfolio performance, for both risk-based and alpha-oriented strategies.

The central themes of this research are:

- The global capital market is dynamic with periodic risk-on and risk-off (and high and low correlation)
- Asset return (i.e., risk premium), risk, and comovement are very different in different market regimes
- Asset allocation strategies can benefit from regime shifts
- Quantitative factor weighting decisions can also be refined by incorporating market risk regime and sophisticated risk modeling

Yin, Miguel, Javed, John, and Sheng

**Deutsche Bank Quantitative Strategy Team**

# I. Modeling non-normal asset returns

In this section, we first build a representative static asset mix as a proxy of the global capital market. Then, we discuss various stylized patterns of asset returns and models to properly account for these patterns.

## A simple asset representing the global capital market

First of all, to measure market risk environment, we need to find a simple proxy to represent the global capital market. For that purpose, we use a representative static asset mix as our risky asset. The static asset allocation (SAA) comprises three asset classes: equities, fixed income, and alternatives. We use MSCI World total return index for equities, Deutsche Bank Aggregate Bond index for fixed income, and S&P/GSCI for alternatives. We assign 50% weight to equities, 40% weight to bonds, and 10% to commodities[3].

As shown in Figure 1, equities have the highest cumulative return, followed by bonds, while commodities come last. In terms of Sharpe ratio (see Figure 2), however, bonds clearly dominate – more than double that of stocks and almost six times as high as that of commodities, mostly due to bonds' low volatility. That explains the outperformance and popularity of risk parity type of asset allocation strategies, which tend to overweight fixed income securities. All three asset classes show some levels of extreme downside risks that are beyond what's implied by a normal distribution (see Figure 3). Both univariate and multivariate tests of normality strongly reject the Null hypothesis of a normal distribution for all three asset classes[4]. Figure 4 shows the scatterplot of the three asset classes – again, non-linear comovement pattern is apparent.

[3] This is roughly in line with the representative asset mix in other research, e.g., Wang, Sullivan, and Ge [2012]. We need each asset to have long enough history of daily return data to model our representative global financial market properly and consistently, and that's why we choose only these three simple assets.
[4] To save space, we do not show the results of these statistical tests. Details are available upon request.

Figure 1: Cumulative returns of equities, bonds, commodities, and SAA portfolio

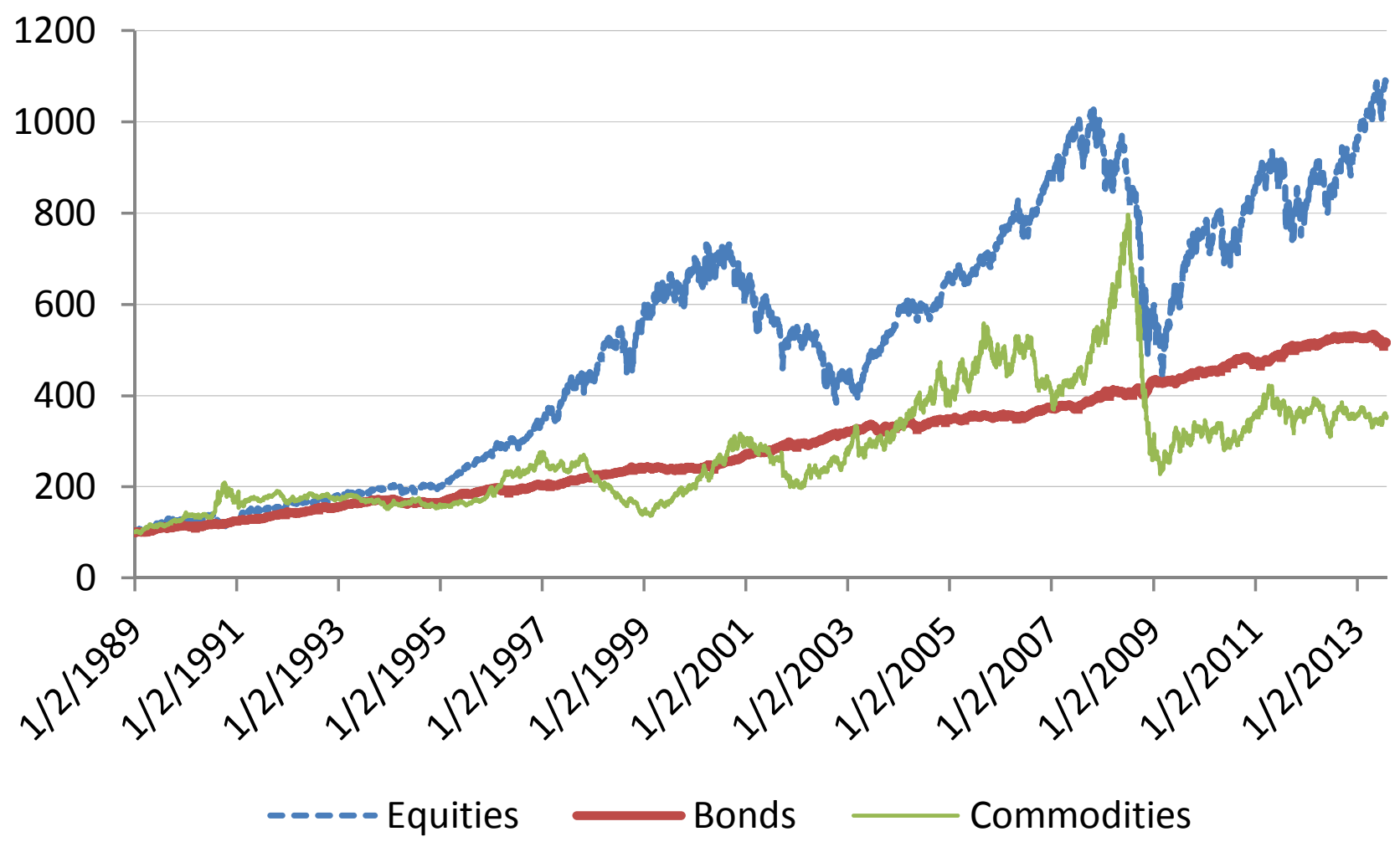

*Source: Bloomberg Finance LLP, MSCI, Russell, S&P, Worldscope, Deutsche Bank Quantitative Strategy*

Figure 2: Return, risk, and Sharpe ratio

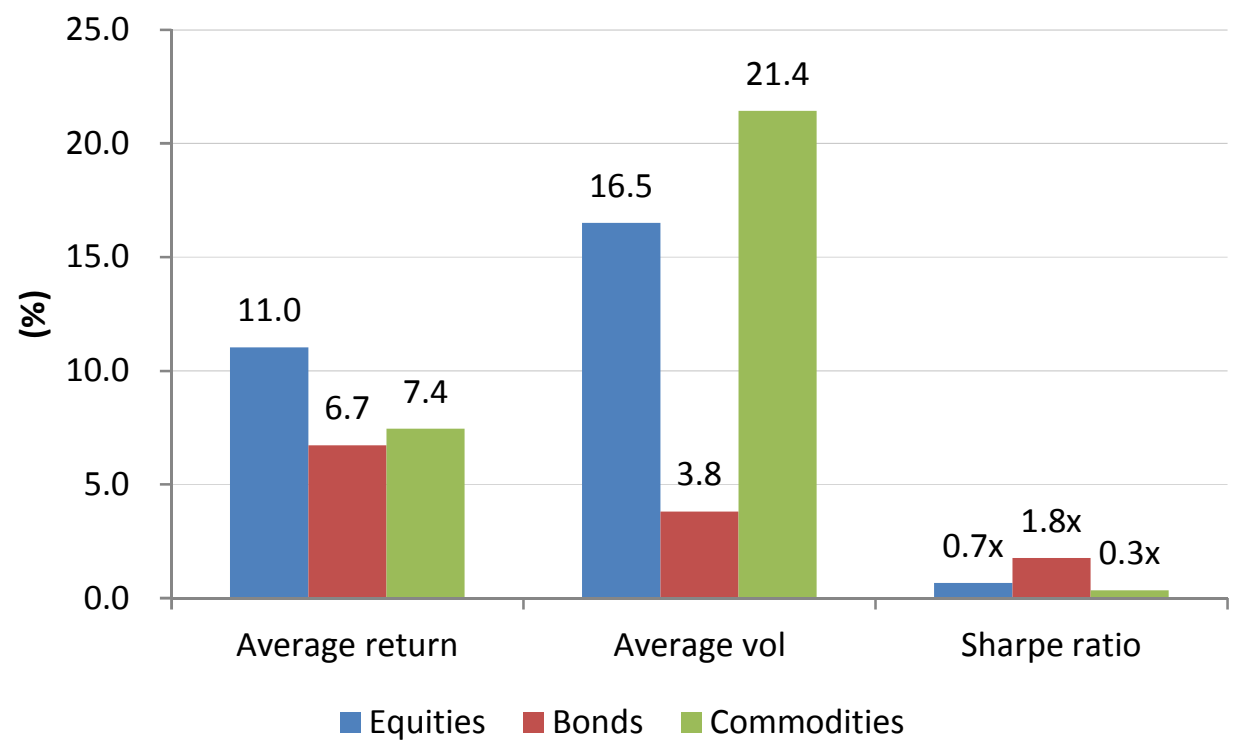

*Source: Bloomberg Finance LLP, MSCI, Russell, S&P, Worldscope, Deutsche Bank Quantitative Strategy*

Figure 3: Downside risk

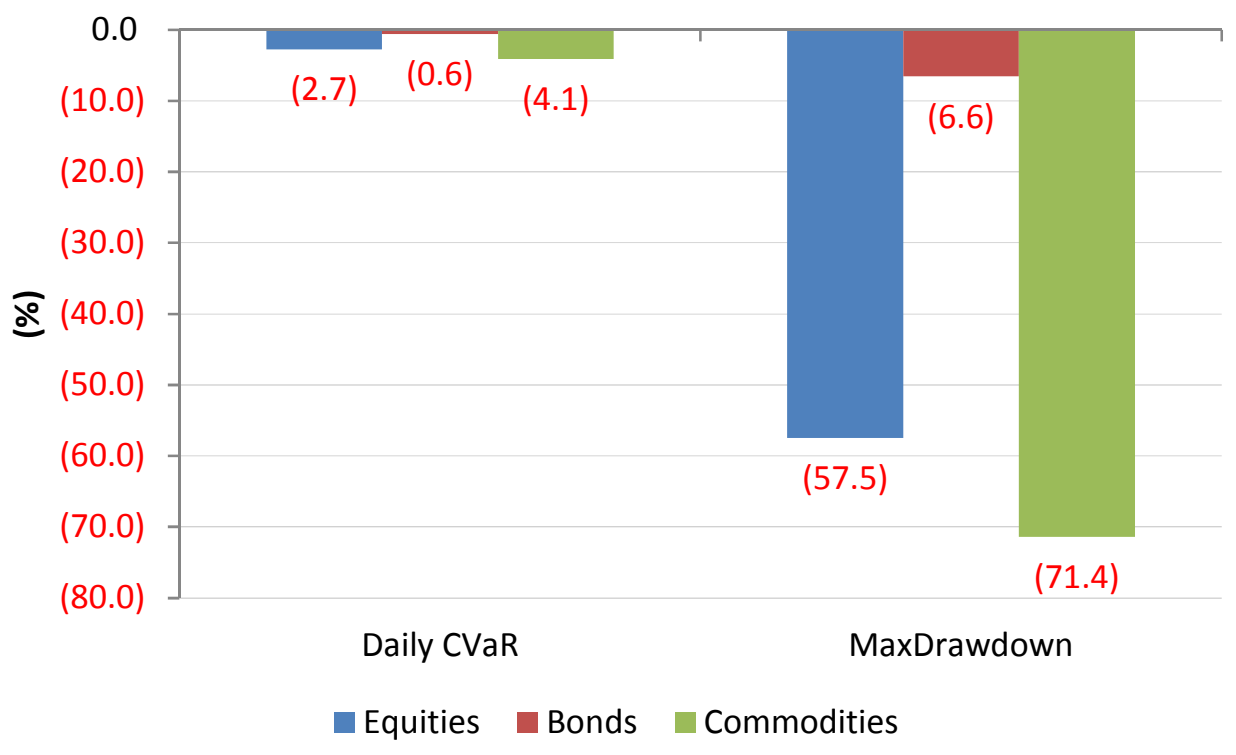

*Source: Bloomberg Finance LLP, MSCI, Russell, S&P, Worldscope, Deutsche Bank Quantitative Strategy*

Figure 4: Scatterplot of three asset classes

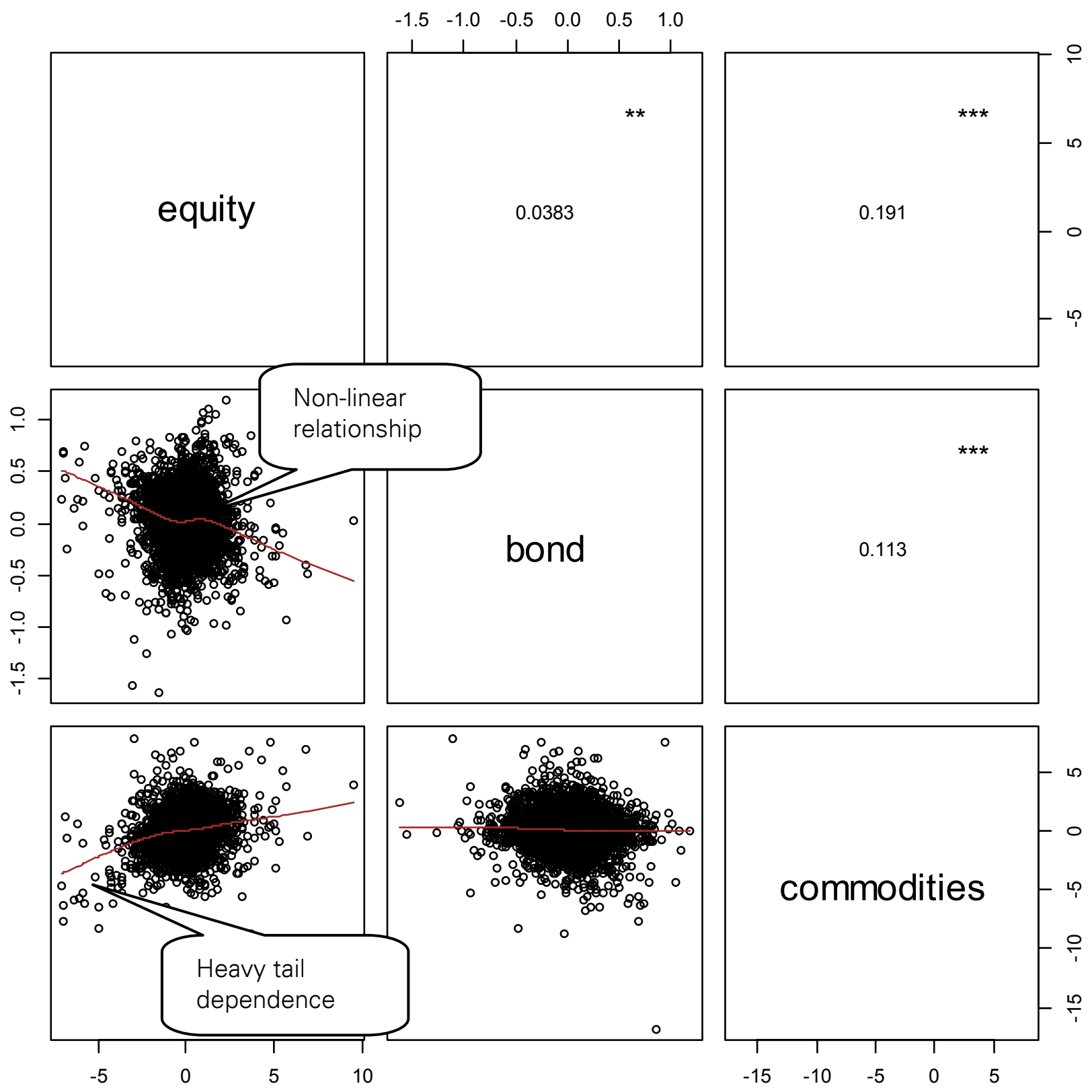

Source: Bloomberg Finance LLP, MSCI, Russell, S&P, Worldscope, Deutsche Bank Quantitative Strategy

## Estimating risk

Unlike returns, we never really know the true level of risk – not even after the fact. Actually, there isn't even a universally accepted definition of risk. In practice, risk is generally defined as volatility (standard deviation) or tail risk (e.g., value at risk or conditional value at risk). In Luo, *et al* [2013] "DB Handbook of Portfolio Construction, Part 1", we emphasized the benefit of using conditional value at risk or CVaR as a risk measure over volatility. To properly measure and predict risk, we have to take into account the stylized facts of asset returns:

- At the asset class level, asset returns typically show serial correlation
- Volatility clustering, i.e., high volatilities tend to be followed by high volatilities
- Extreme events are more common than implied by a normal distribution
- Cross asset class correlation is time varying
- Tail dependence: correlation can't capture the complete comovement among assets

In the following sections, we will discuss the above patterns of asset returns in detail. Then, we will suggest a risk model that can properly account for these patterns.

## Return serial correlation via ARMA

ACF (autocorrelations) and PACF (partial autocorrelations) graphs are useful tools to understand serial correlation. ACF measures the correlation of an asset ( $r_t$ ) with its own lagged returns ( $r_{t-k}$ ), i.e.,

$$ACF_k = \rho(r_t, r_{t-k})$$

Sometimes, we also want to measure the relationship between $r_t$ and $r_{t-k}$, after controlling the effects of other time lags $1, 2, \ldots, k-1$. That is when we use PACF. The PACF at lag $k$, $PACF_k = \phi_k$ is calculated by fitting the following linear regression:

$$r_t = \phi_0 + \phi_1 r_{t-1} + \phi_2 r_{t-2} + \cdots + \phi_k r_{t-k} + \varepsilon_t$$

At the single security level, especially for liquid assets like common stocks, we typically assume there is no return serial correlation. In other words, we tend to argue that stock returns are more like random noise, i.e., knowing yesterday's return doesn't help us predict today's return (this is also sometimes called the "weak form of the efficient market hypothesis"). For example, Figure 5 shows the ACF plot of daily IBM return. None of the first 40 lags exhibits statistically significant autocorrelation (see Figure 5).

At the asset class level, returns tend to show some moderate level of serial correlation. For example, Figure 6 and Figure 7 plot the ACF graphs for equities and bonds – in both cases, the lag-one serial correlation is clearly significant. However, the serial correlation for neither of the three asset classes is persistent, meaning that most of the higher lags are insignificant beyond lag one. Combining ACF and PACF, we can have a good understanding of the ARMA (AutoRegressive Moving Average) structure of asset returns.

The benefit of serial correlation is that it can help us to predict future returns, using past returns. The downside is that if serial correlation is not properly accounted for, the results of time series statistical models can be misinterpreted.

If we apply a simple ARMA(1,1) model[5] to daily returns of equities, the residual displays little sign of serial correlation (see Figure 9). This seems to suggest that an ARMA(1,1) model is sufficient to account for serial correlation in asset returns.

The ARMA(1,1) model can be expressed as :

$$r_t - \phi\, r_{t-1} = \varepsilon_t + \theta \varepsilon_{t-1}$$

Where,

$r_t$ is asset return at time $t$,

[5] ARMA stands for AutoRegressive Moving Average.

$\phi$ is the autoregressive (AR) parameter,

$\theta$ is the moving average (MA) parameter, and

$\varepsilon_t$ is the residual and is typically assumed to follow a known distribution

Figure 5: ACF – IBM

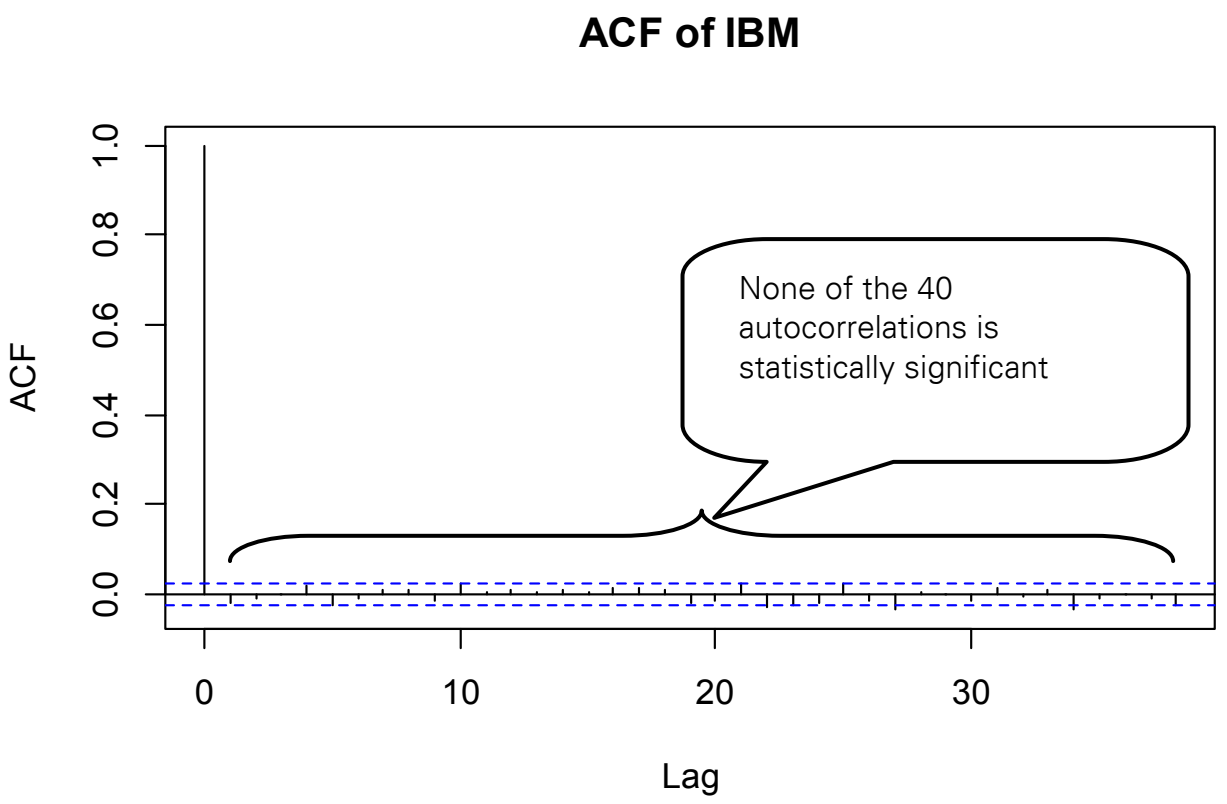

*Source: Bloomberg Finance LLP, MSCI, Russell, S&P, Worldscope, Deutsche Bank Quantitative Strategy*

Figure 6: ACF – equities

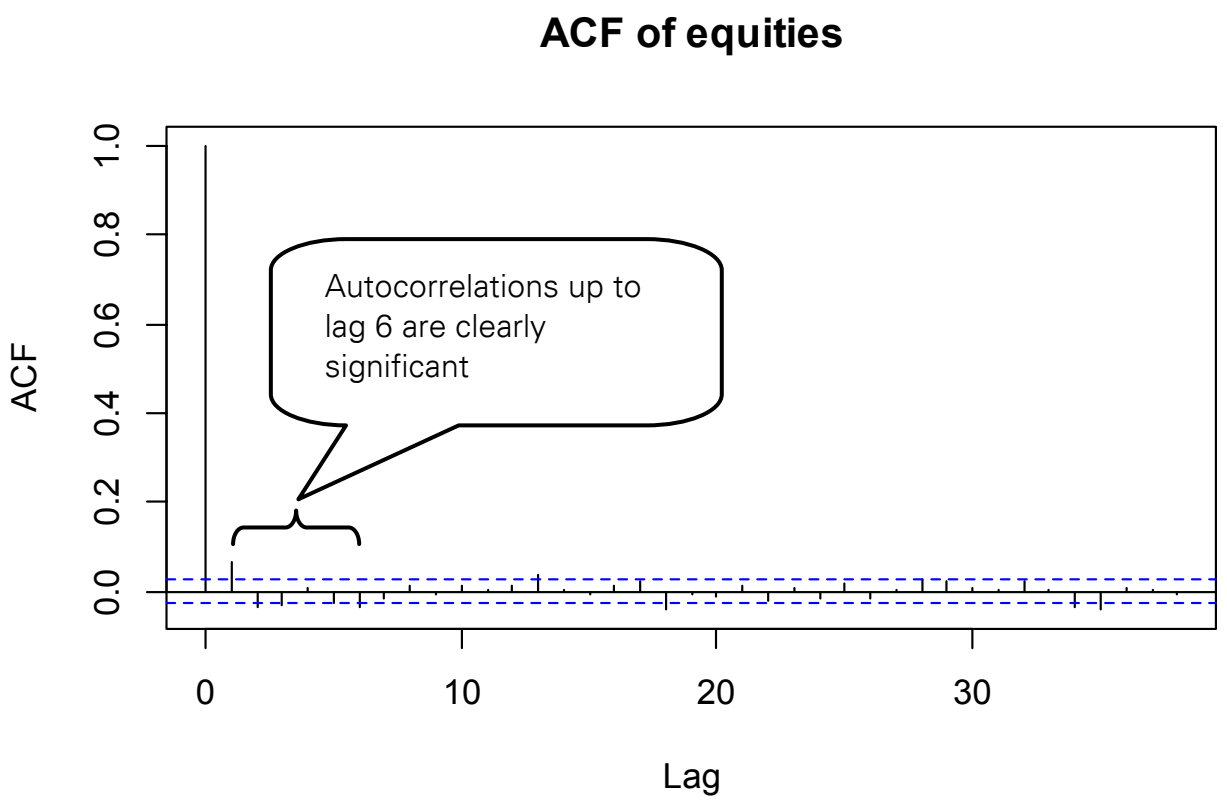

*Source: Bloomberg Finance LLP, MSCI, Russell, S&P, Worldscope, Deutsche Bank Quantitative Strategy*

Figure 7: ACF – bonds

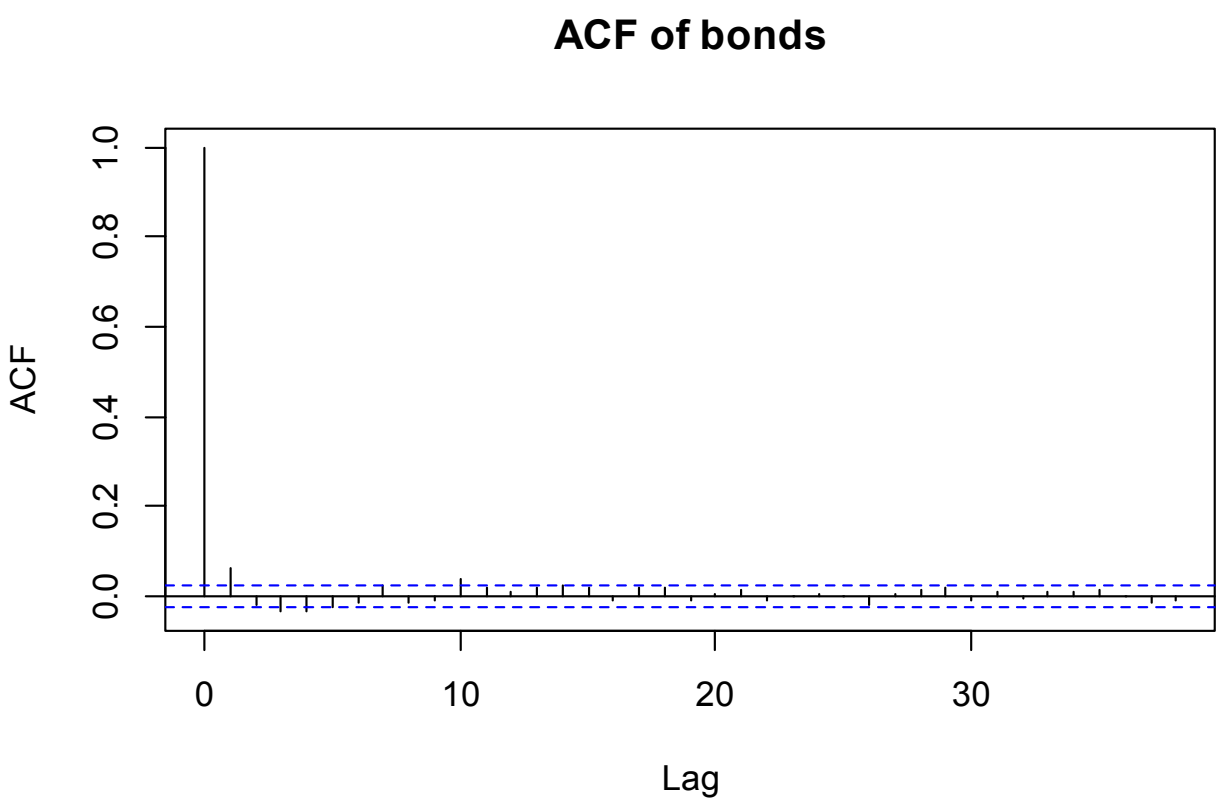

*Source: Bloomberg Finance LLP, MSCI, Russell, S&P, Worldscope, Deutsche Bank Quantitative Strategy*

Figure 8: PACF – equities

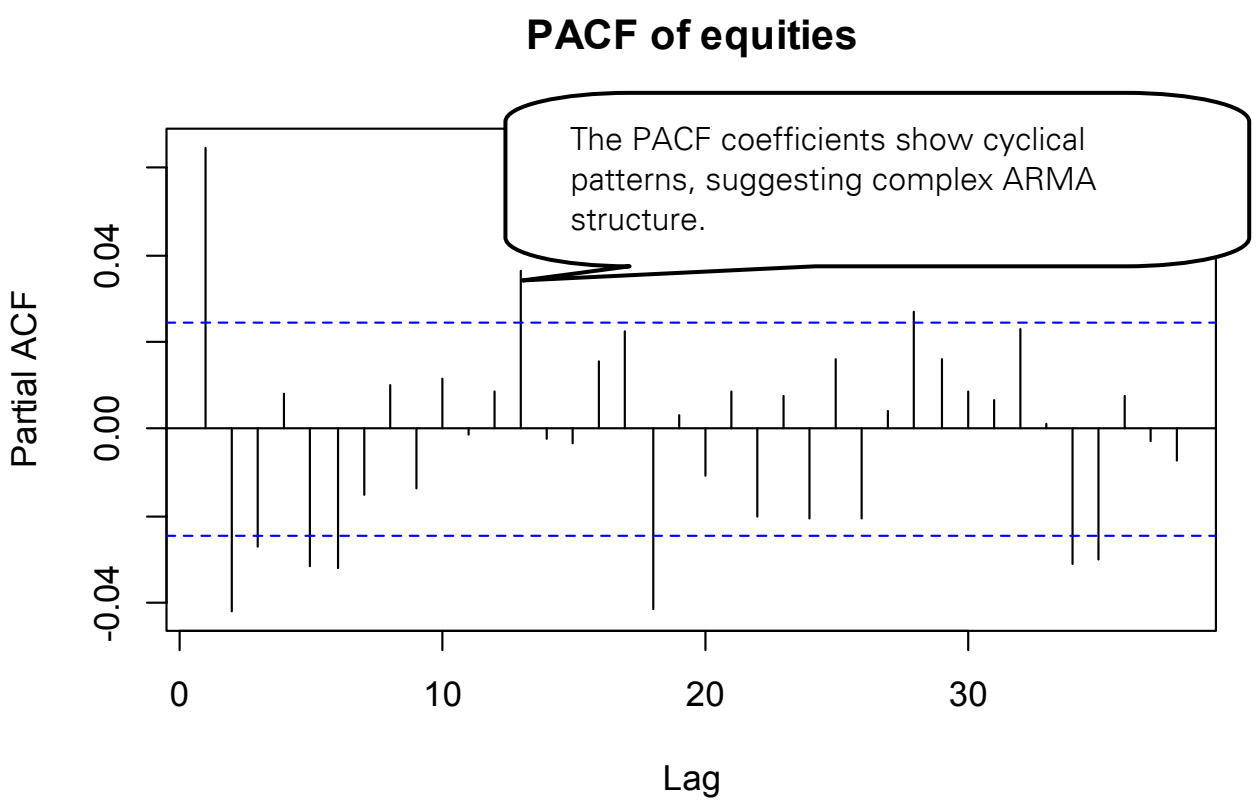

*Source: Bloomberg Finance LLP, MSCI, Russell, S&P, Worldscope, Deutsche Bank Quantitative Strategy*

Figure 9: Filtered equity return

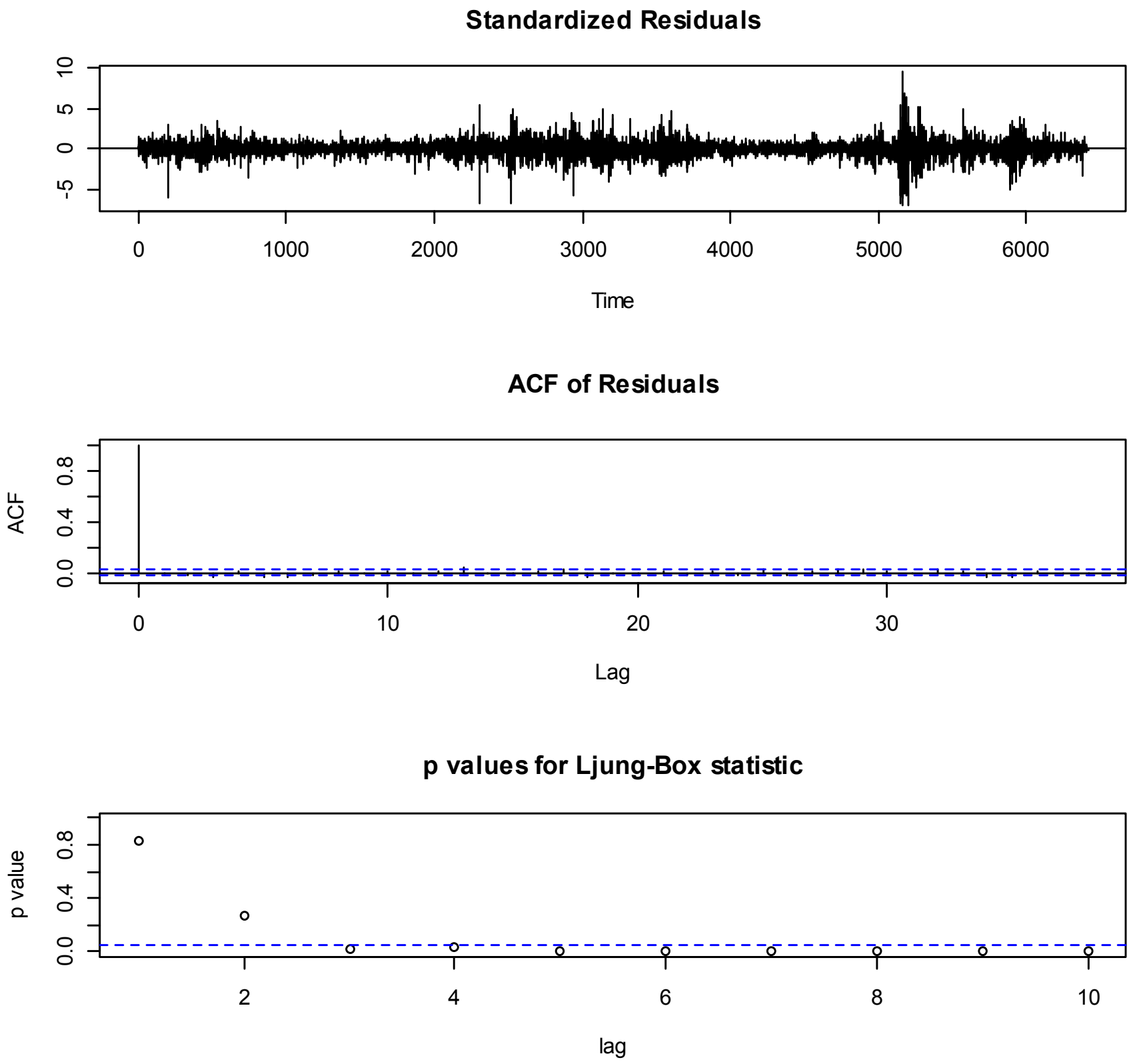

*Source: Bloomberg Finance LLP, MSCI, Russell, S&P, Worldscope, Deutsche Bank Quantitative Strategy*

## Volatility clustering and extreme outliers via GARCH

As we discussed above, asset returns typically show some modest serial correlation. On the volatility side, however, volatility tends to be highly persistent, i.e., what's known as "volatility clustering effect". In other words, a period of high volatility is likely to follow with another period of high volatility. Figure 10 and Figure 11 show the ACF and PACF plot of the squared daily returns (as a proxy for variance) of equities. Serial correlation coefficients are highly significant to higher lags for squared returns.

One interesting statistical model that accounts for the properties of asset volatility is called GARCH (Generalized AutoRegressive Conditional Heteroskedasticity), proposed by Engle [1982] and expanded by Bollerslev [1986]. Many extensions of the GARCH model have been presented in the literature since then.

Continued on the previous section, a GARCH(1,1) model can be defined as:

$$\varepsilon_t = \sigma_t z_t$$

Where $\sigma_t$ is a nonnegative process such that

$$\sigma_t^2 = \alpha_0 + \alpha_1 \varepsilon_{t-1}^2 + \beta_1 \sigma_{t-1}^2$$

$z_t$ is the residual of the GARCH model and is typically assumed to follow a known distribution, e.g., a normal distribution or a t-distribution.

The GARCH(1,1) model seems to be able to remove most of the volatility clustering effect – the residual from the GARCH(1,1) model shows little autocorrelation at higher lags (see Figure 12 and Figure 13).

More interestingly, a GARCH model has built-in ability to model heavy tails. Even if we assume $z_t$ is normally distributed, a GARCH process can still exhibit heavy tails. Because GARCH models assume the conditional variance is not constant; therefore, outliers occur when the variance is large. Sometimes, asset returns may present even heavier tails than implied by a GARCH model, and we can assume other distributions for the residual $z_t$ , e.g., a t-distribution. In practice, we find non-normal GARCH models suffer from dimensionality issues, especially when we try to fit multiple assets simultaneously.

Figure 10: ACF – squared equity returns

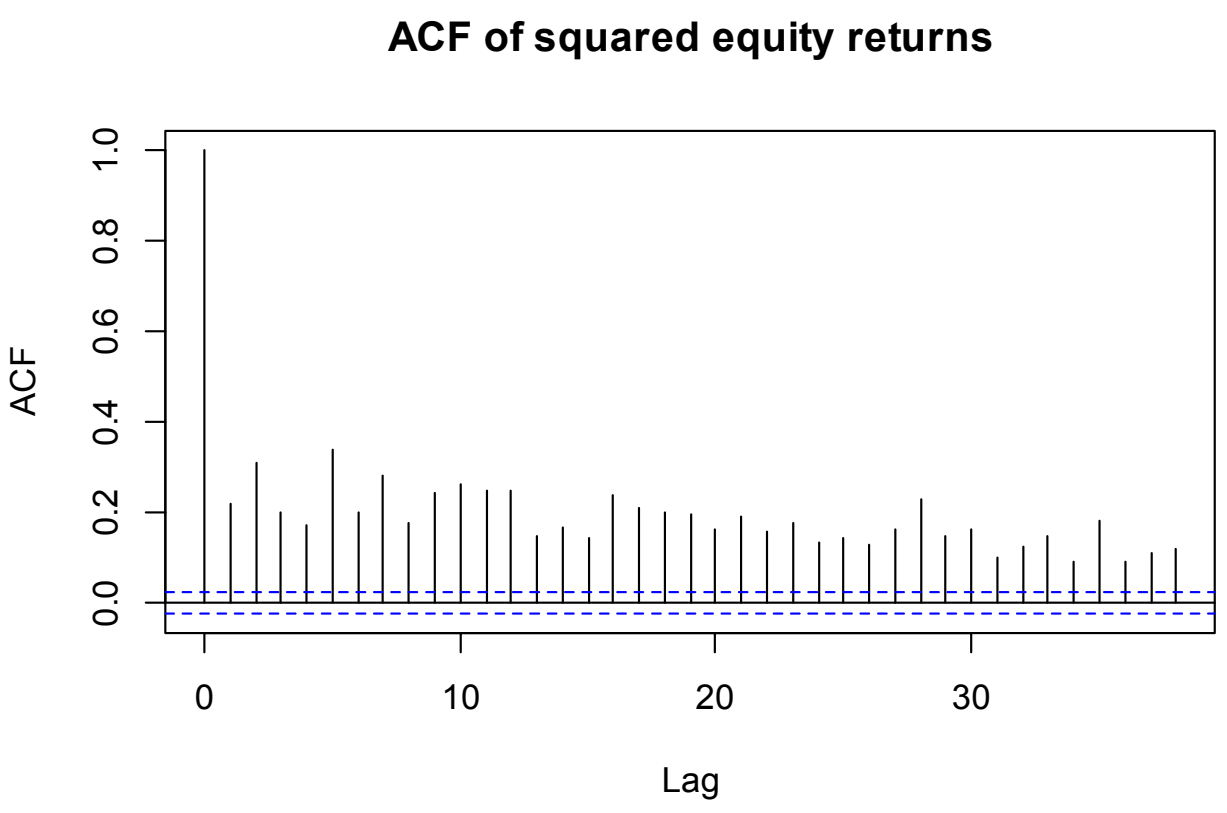

*Source: Bloomberg Finance LLP, MSCI, Russell, S&P, Worldscope, Deutsche Bank Quantitative Strategy*

Figure 11: PACF – squared equity returns

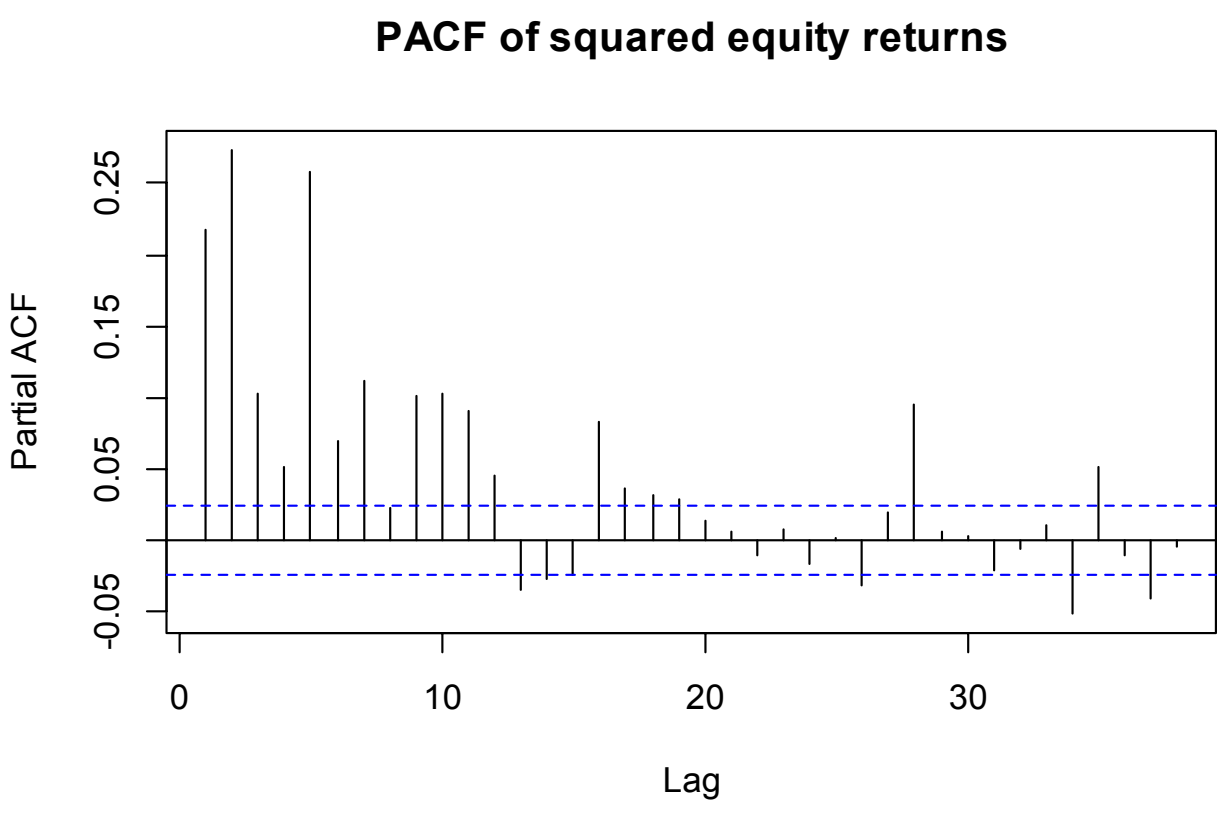

*Source: Bloomberg Finance LLP, MSCI, Russell, S&P, Worldscope, Deutsche Bank Quantitative Strategy*

Figure 12: ACF of GARCH filtered equity returns

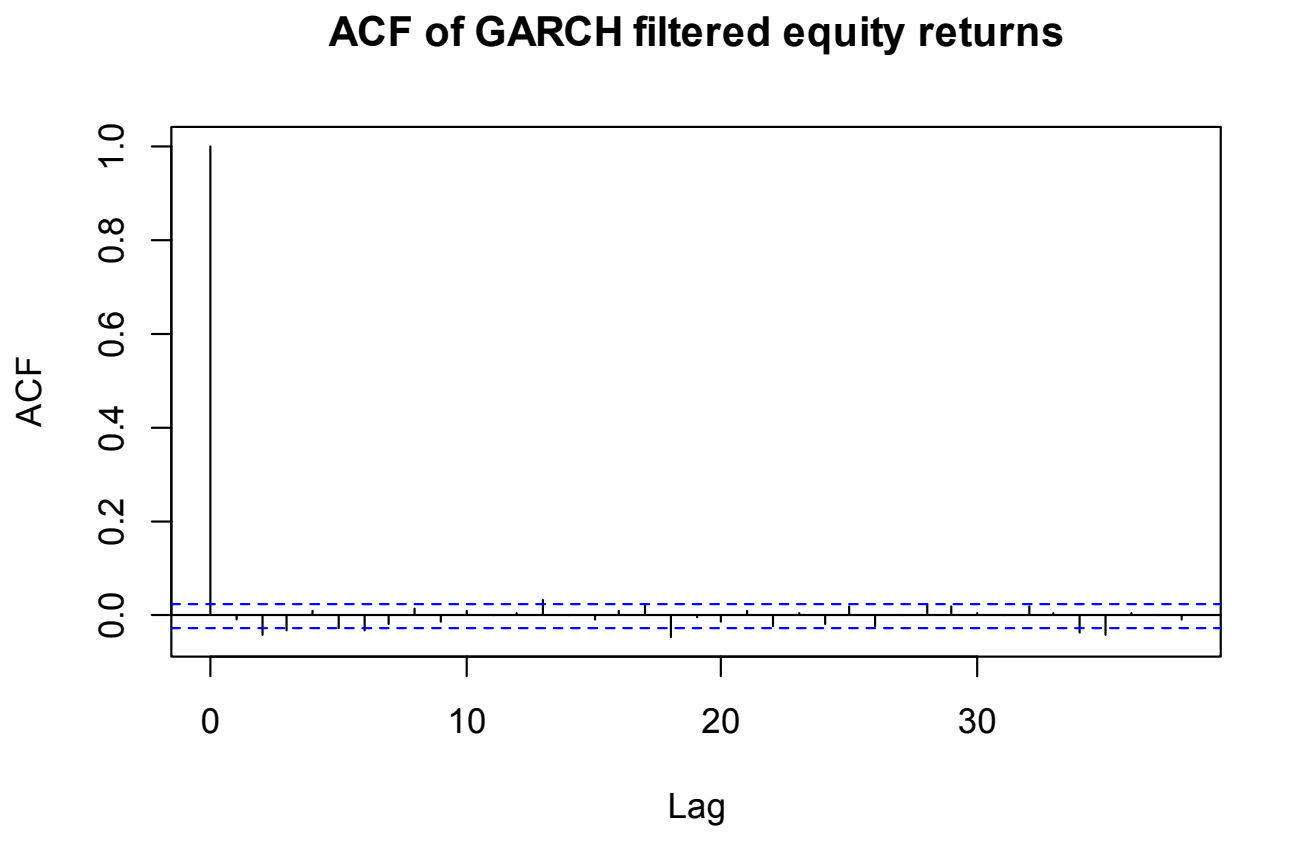

*Source: Bloomberg Finance LLP, MSCI, Russell, S&P, Worldscope, Deutsche Bank Quantitative Strategy*

Figure 13: PACF of GARCH filtered equity returns

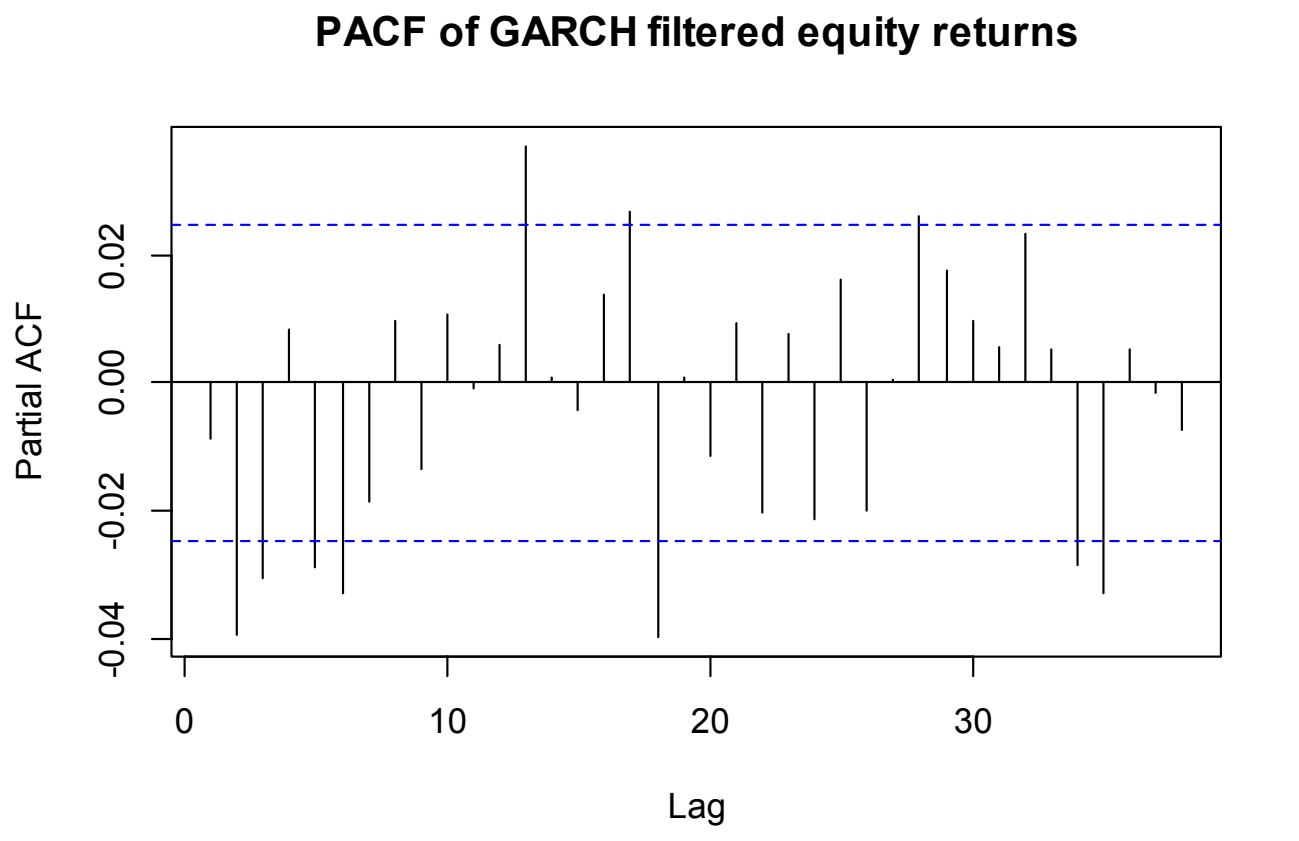

*Source: Bloomberg Finance LLP, MSCI, Russell, S&P, Worldscope, Deutsche Bank Quantitative Strategy*

## Asymmetric risk via GJR-GARCH

The so-called leverage effect (Black [1976]) refers to the well-known relationship between stock returns and volatility – volatility increases after the stock price falls. Asymmetric GARCH model, for example, the GJR-GARCH model (see Glosten, Jagannathan, and Runkle [1993]) can be used to accommodate this feature. In the GJR-GARCH(1,1) model, positive and negative shocks on the conditional variance are modeled as follows:

$$\sigma_t^2 = \alpha_0 + \left(\alpha_1 \varepsilon_{t-1}^2 + \gamma I_{t-1} \varepsilon_{t-1}^2\right) + \beta_1 \sigma_{t-1}^2$$

Where,

$I_{t-1}$ is the indicator function and takes on value of $1$ for $\varepsilon_{t-1} \le 0$ and $0$ otherwise, and

$\gamma$ is the leverage term and measures the impact of a negative shock.

We find the $\gamma$ term is highly significant for equities and insignificant for bonds and commodities (see Figure 14). It is intuitive to see that the leverage effect is positive and statistically significant for equities, as it can be justified by economic intuition. After the stock price falls, the equity value of the firm declines; therefore, the degree of leverage in its capital structure rises, producing a hike in stock volatility. The leverage coefficient for bonds is also positive, albeit insignificant. As equity and bond prices fall, panic kicks in to the markets and investors look for more protection; therefore, both realized and implied volatilities escalate.

It is also interesting to note that the $\gamma$ coefficient is negative for commodities – a phenomenon that we typically term as "inverse leverage effect". When commodity prices go up, the cost to the real economy surges, so it is negative for the economy and panic sets in.

Because the leverage coefficients are statistically insignificant for two of three assets, we decide not to include this feature in our final model.

Figure 14: Asymmetric terms

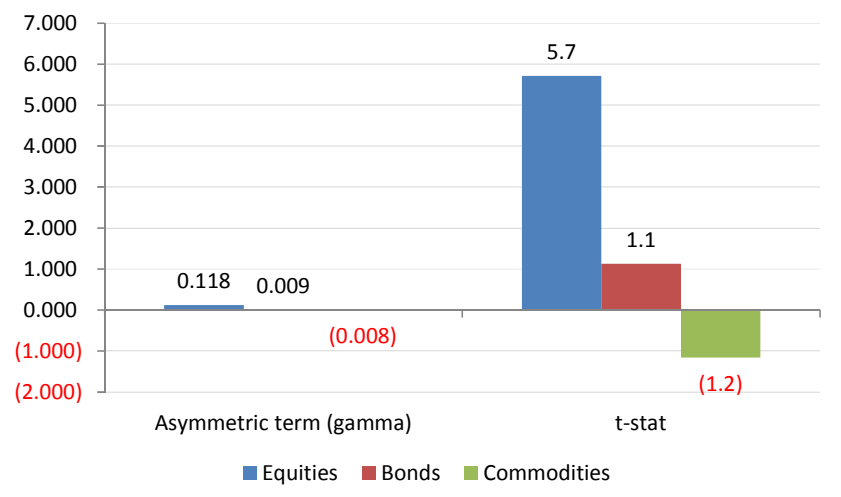

*Source: Bloomberg Finance LLP, MSCI, Russell, S&P, Worldscope, Deutsche Bank Quantitative Strategy*

## Time-varying dynamic correlation via DCC

The traditional definition of risk (i.e., variance and volatility) critically depends on the accuracy of asset-by-asset covariance matrix, which in turn, depends on asset variance/volatility and correlation. The above discussion focused on univariate estimate of asset volatility and now we would like to shift our attention to the joint distribution, or how assets move together.

The traditional multivariate GARCH model typically assumes a constant conditional correlation (CCC), mostly for dimensionality and numerical optimization concerns (see Bollerslev [1990]).

Extending the GARCH model in the previous sections to the multivariate case, we can show:

$$r_t | I_{t-1} = \mu_t + \varepsilon_t$$

Where,

$r_t$ is a $(N \times 1)$ vector of asset returns,

$I_{t-1}$ is the set of information available as of time $t-1$,

$\mu_t$ is a $(N \times 1)$ vector of conditional means, and

$\varepsilon_t$ is a $(N \times 1)$ vector of residual returns.

The residual return vector $\varepsilon_t$ can be modeled as:

$$\varepsilon_t = \Sigma_t^{1/2} z_t$$

The conditional variance matrix $\Sigma_t$ can then be modeled as follows in the CCC setup:

$$\Sigma_t = D_t R D_t$$

Where,

$D_t = diag(\sigma_{1,t}, \sigma_{2,t}, \Lambda\ , \sigma_{N,t})$ is the diagonal vector of the covariance matrix $\Sigma_t$, and

$R$ is the positive definite constant conditional correlation matrix.

Engle [2002] and Tse and Tsui [2002] introduced a dynamic structure in modeling correlation – a suite of models called dynamic conditional correlation (DCC) model, which was further extended by Cappiello *et al* [2006] and Billio *et al* [2006].

In a DCC model: $\Sigma_t = D_t R_t D_t$

where the correlation matrix $R_t$ is now allowed to be time varying. It is very challenging to estimate $R_t$ and at the same time to guarantee it to be positive definite. For interested readers, we will leave the technical details of DCC models to the above referenced papers.

The restriction of constant correlation is clearly unrealistic, as most practitioners have witnessed in real life. In fact, Figure 15 to Figure 18 show the correlation among equities, bonds, and commodities using CCC, rolling estimate, and DCC. The correlations under CCC have mostly gone down over time, as more and more data are being added to the expanding estimation window, which is clearly misleading. The rolling window sample correlation is more dynamic, but maybe too volatile, due to estimation errors. The DCC model appears to strike a good balance. We also add a line labeled "actual correlation", which is the *ex post* realized correlation using daily return data during that month[6].

## Weighted average portfolio correlation (WPC)

Another special note is how we calculate the weighted average correlation for a portfolio of assets. As shown in Luo, *et al* [2013], a portfolio's risk is not a simple weighted average of each asset's volatility. Similarly, we can't calculate a portfolio's weighted correlation as the weighted pairwise correlation, because the relationship is nonlinear.

The variance of a portfolio can be computed as:

$$\sigma_p^2 = \sum_{i=1}^{N} \omega_i^2 \sigma_i^2 + 2\sum_{i=1}^{N-1}\sum_{j>i}^{N} \omega_i \omega_j \sigma_i \sigma_j \rho_{ij}$$

where:

$\sigma_p^2$ is the variance of the portfolio

$\sigma_i$ is the volatility of asset $i$

$\omega_i$ is the weight of asset $i$

$\rho_{ij}$ is the correlation coefficient between asset $i$ and $j$

Now, let's define $\rho_{average}$ (i.e., WPC) as the weighted average pairwise correlation, then the above equation can be written as:

$$\sigma_p^2 = \sum_{i=1}^{N} \omega_i^2 \sigma_i^2 + 2\sum_{i=1}^{N-1}\sum_{j>i}^{N} \omega_i \omega_j \sigma_i \sigma_j \rho_{ij} = \sum_{i=1}^{N} \omega_i^2 \sigma_i^2 + 2\sum_{i=1}^{N-1}\sum_{j>i}^{N} \omega_i \omega_j \sigma_i \sigma_j \rho_{average}$$

Therefore,

$$\sum_{i=1}^{N-1}\sum_{j>i}^{N} \omega_i \omega_j \sigma_i \sigma_j \rho_{ij} = \sum_{i=1}^{N-1}\sum_{j>i}^{N} \omega_i \omega_j \sigma_i \sigma_j \rho_{average} = \rho_{average}\sum_{i=1}^{N-1}\sum_{j>i}^{N} \omega_i \omega_j \sigma_i \sigma_j$$

And, finally,

[6] Please note that the *ex post* "actual" correlation can't be used in constructing real life portfolios, as we don't know it at the beginning of the portfolio rebalance; rather, we can only calculate it at the end of the month. In addition, even the *ex post* "actual" correlation is still an estimate of the true correlation, which is unobservable.

$$WPC = \rho_{average} = \frac{\sum_{i=1}^{N-1}\sum_{j>i}^{N} \omega_i \omega_j \sigma_i \sigma_j \rho_{ij}}{\sum_{i=1}^{N-1}\sum_{j>i}^{N} \omega_i \omega_j \sigma_i \sigma_j}$$

Therefore, WPC is essentially the weighted pairwise correlation adjusted for asset volatility.

Figure 15: Correlation between equities and bonds

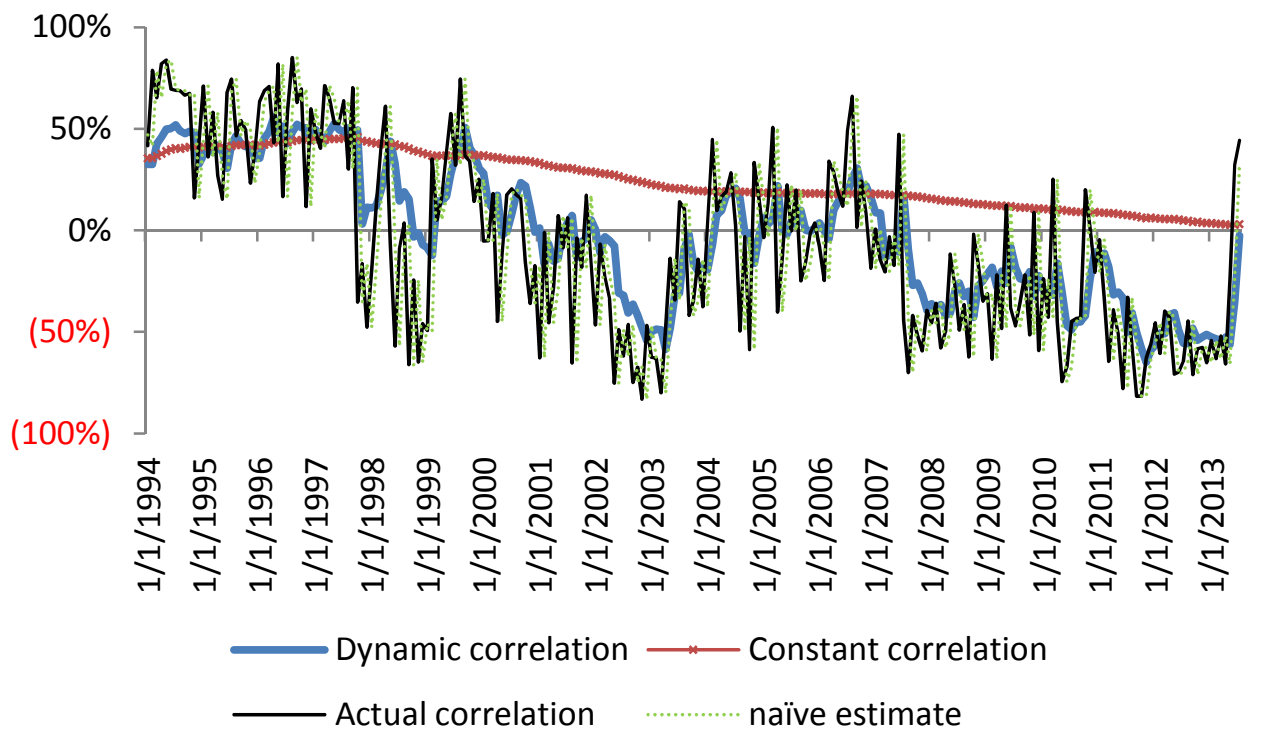

Source: Bloomberg Finance LLP, MSCI, Russell, S&P, Worldscope, Deutsche Bank Quantitative Strategy

Figure 16: Correlation between equities and commodities

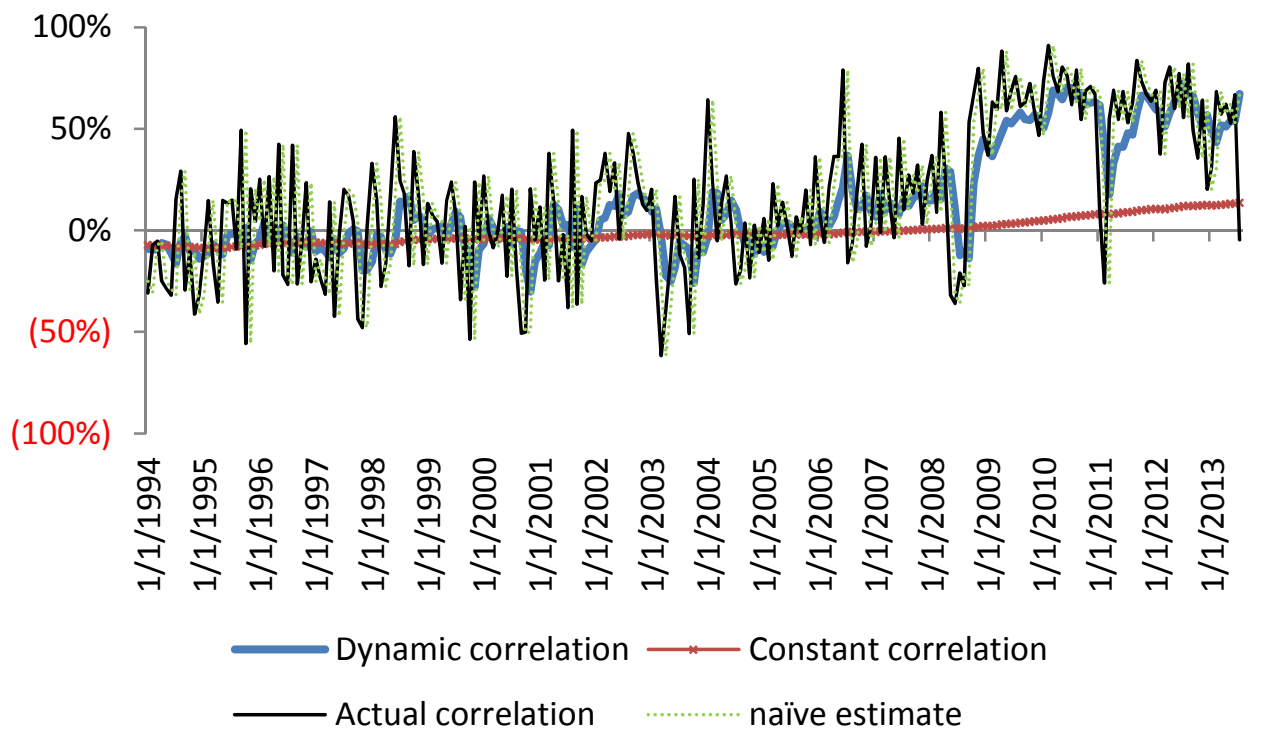

Source: Bloomberg Finance LLP, MSCI, Russell, S&P, Worldscope, Deutsche Bank Quantitative Strategy

Figure 17: Correlation between bonds and commodities

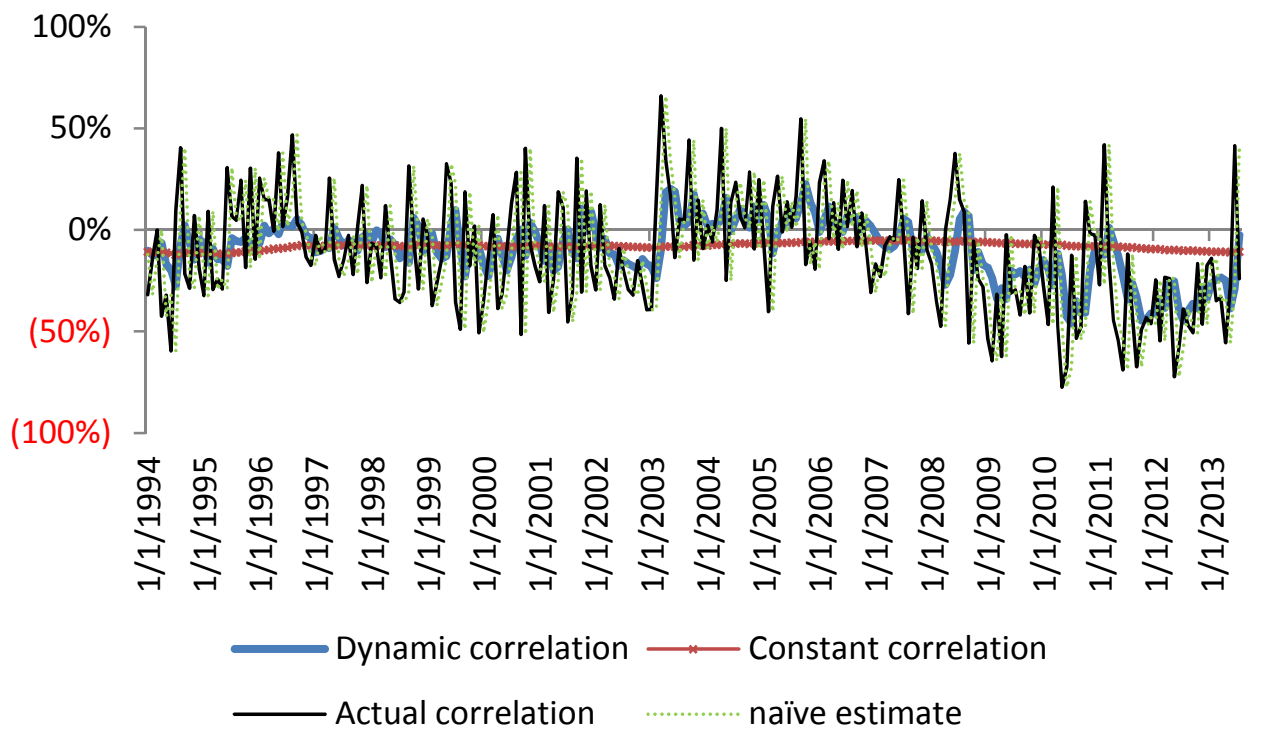

Source: Bloomberg Finance LLP, MSCI, Russell, S&P, Worldscope, Deutsche Bank Quantitative Strategy

Figure 18: Weighted average pairwise correlation

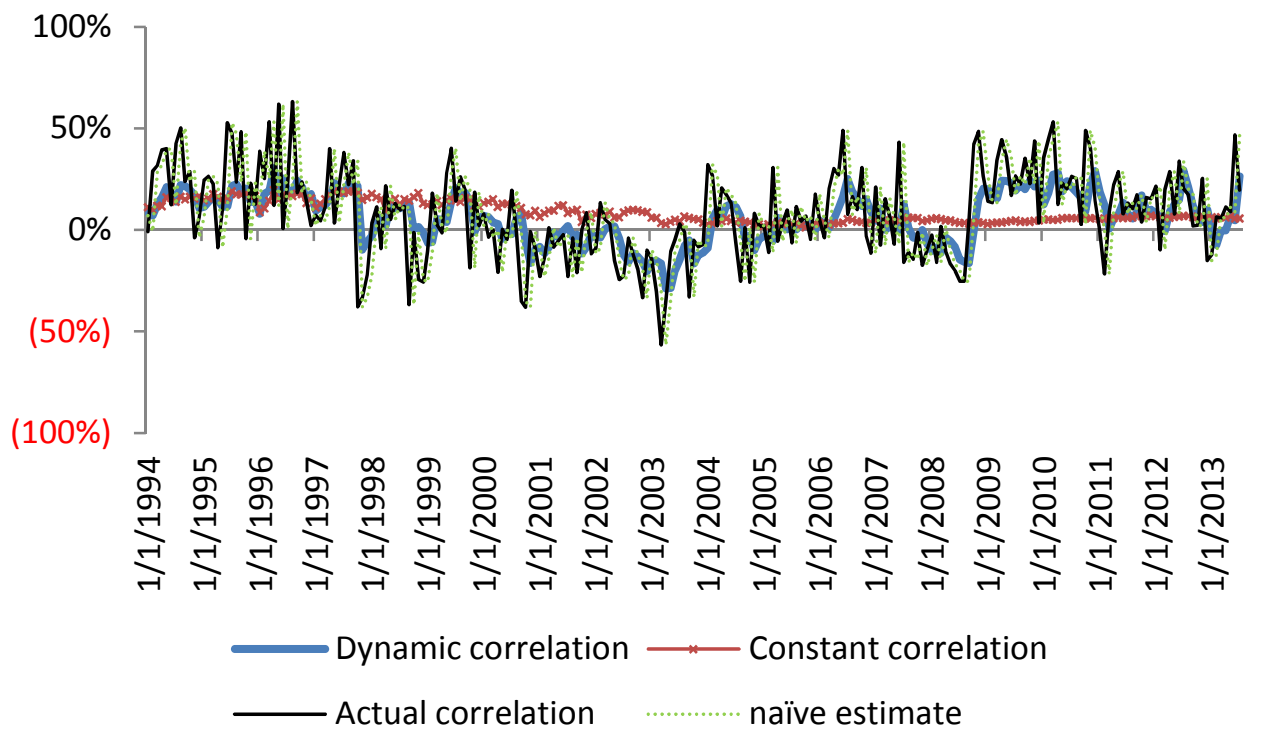

Source: Bloomberg Finance LLP, MSCI, Russell, S&P, Worldscope, Deutsche Bank Quantitative Strategy

## Tail dependence via Copula

In Luo, *et al* [2013], we emphasized the importance of incorporating tail dependence via Copula models in portfolio construction. To quickly recap, in a non-technical sense[7],

$$\text{Joint Distribution} = \text{Copula} + \text{Marginal Distribution}$$

[7] See Meucci, A. [2011], A short, comprehensive, practical guide to copulas, http://papers.ssrn.com/sol3/papers.cfm?abstract_id=1847864, for more details.

Therefore, a copula model gives us the flexibility to model joint asset return distributions. For example, we could fit an exponential GARCH model for each asset's marginal distribution, while at the same time modeling the joint distribution using a t-copula model.

Figure 19 shows the difference between Pearson correlation and Copula-based tail dependence for the three asset classes. Tail dependence coefficients among our three asset classes are clearly higher than Pearson correlation coefficients – as expected, assets are more likely to fall at the same time than the average. Some differences are strikingly large. For example, the Pearson correlation between equities and commodities is only 19%, but the tail dependent coefficient is 28% – almost 44% higher. Furthermore, the (on average) negative correlation between equities and bonds turns out to be positively related at the tail level.

Figure 19: Correlation versus tail dependence (lower triangle = correlation/upper triangle = copula tail dependence)

| | Equities | Bonds | Commodities |
|---|---|---|---|
| Equities | 100% | 8% | 28% |
| Bonds | -4% | 100% | 4% |
| Commodities | 19% | -11% | 100% |

*Source: Bloomberg Finance LLP, MSCI, Russell, S&P, Worldscope, Deutsche Bank Quantitative Strategy*

To visually examine how Pearson's correlation coefficient underestimates the true dependence, let's compare the theoretical bivariate normal distribution between equities and commodities (see Figure 20), with the empirical distribution (see Figure 21). The empirical distribution clearly has much heavier tails, i.e., the probabilities of these two assets both move higher or fall lower are much higher than other combinations.

Different from Luo, *et al* [2013], where we modeled bivariate tail dependence between a pair of assets, in this paper, we would like to jointly estimate tail dependence among three assets in a multivariate setting. As the number of assets increases, the dimensionality goes up exponentially. In Luo, *et al* [2013], we applied empirical Copula, but for multi-asset case, we need to apply a more structured model.

From Figure 21 to Figure 23, it's interesting to note that the tail dependence structures among the three asset classes are roughly symmetric, i.e., the left tail dependence isn't that different from the right tail, which suggests that a t-Copula might be sufficient to model the inter-dependence at the asset class level.

The t Copula (as defined in Embrechts, McNeil, and Straumann [2001]) can be thought of as representing the dependence structure implicit in a multivariate t distribution. Mashal and Zeevi [2002] and Breymann *et al*. [2003] have shown that the empirical fit of the t copula is generally superior to that of the Gaussian copula, where tail dependence is assumed non-existent. One reason for this is the ability of the t-Copula to better capture the phenomenon of dependent extreme values, which is often observed in financial return data.

Figure 20: Theoretical bivariate normal distribution between equities and commodities

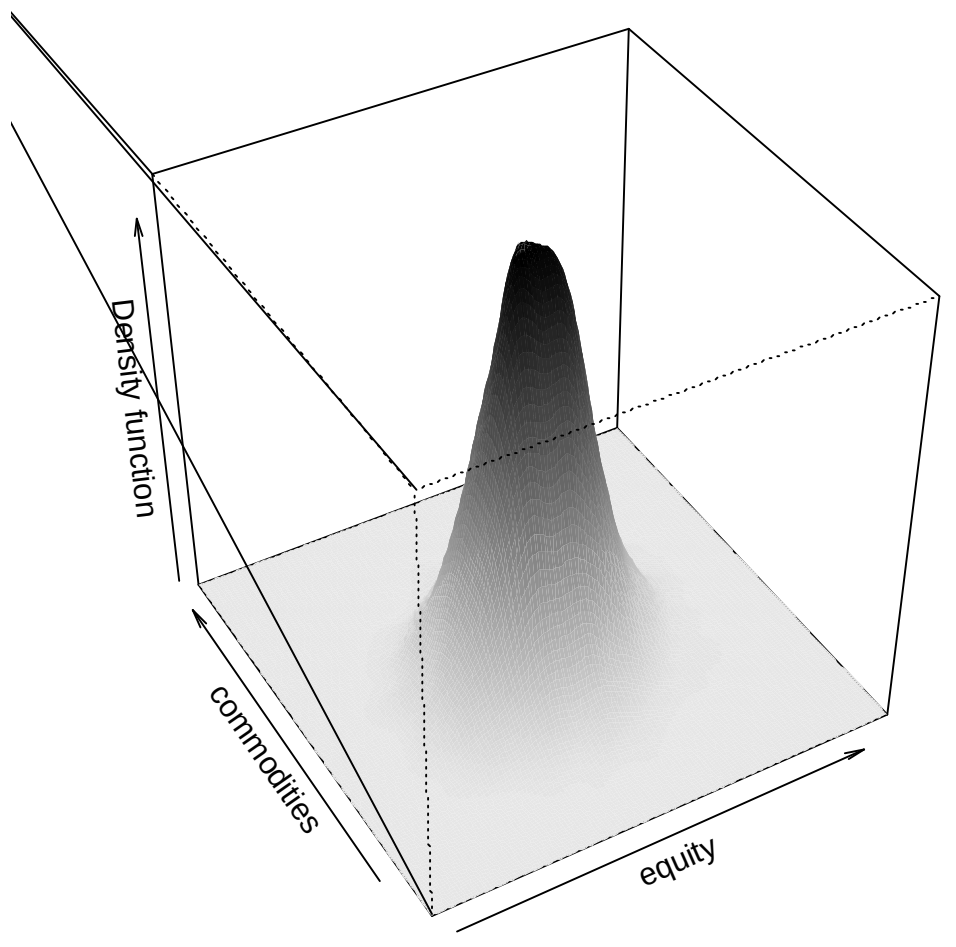

*Source: Bloomberg Finance LLP, MSCI, Russell, S&P, Worldscope, Deutsche Bank Quantitative Strategy*

Figure 21: Empirical distribution between equities and commodities

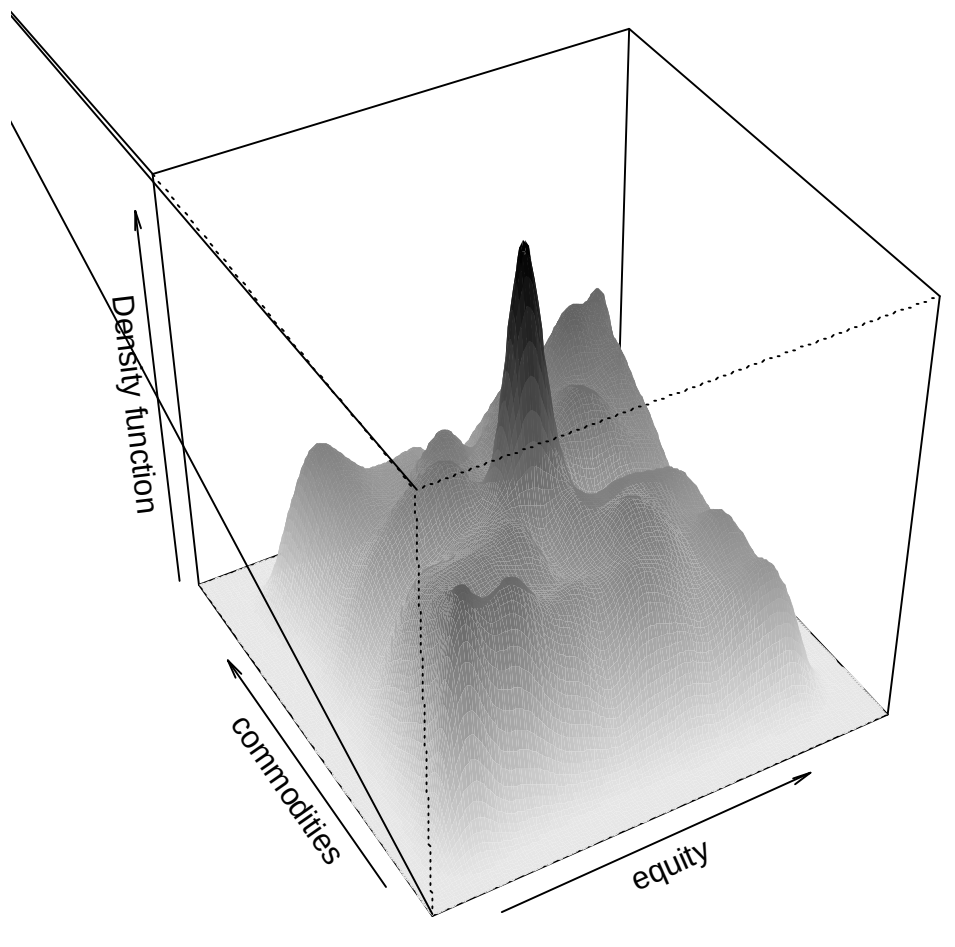

*Source: Bloomberg Finance LLP, MSCI, Russell, S&P, Worldscope, Deutsche Bank Quantitative Strategy*

Figure 22: Empirical distribution between equities and bonds

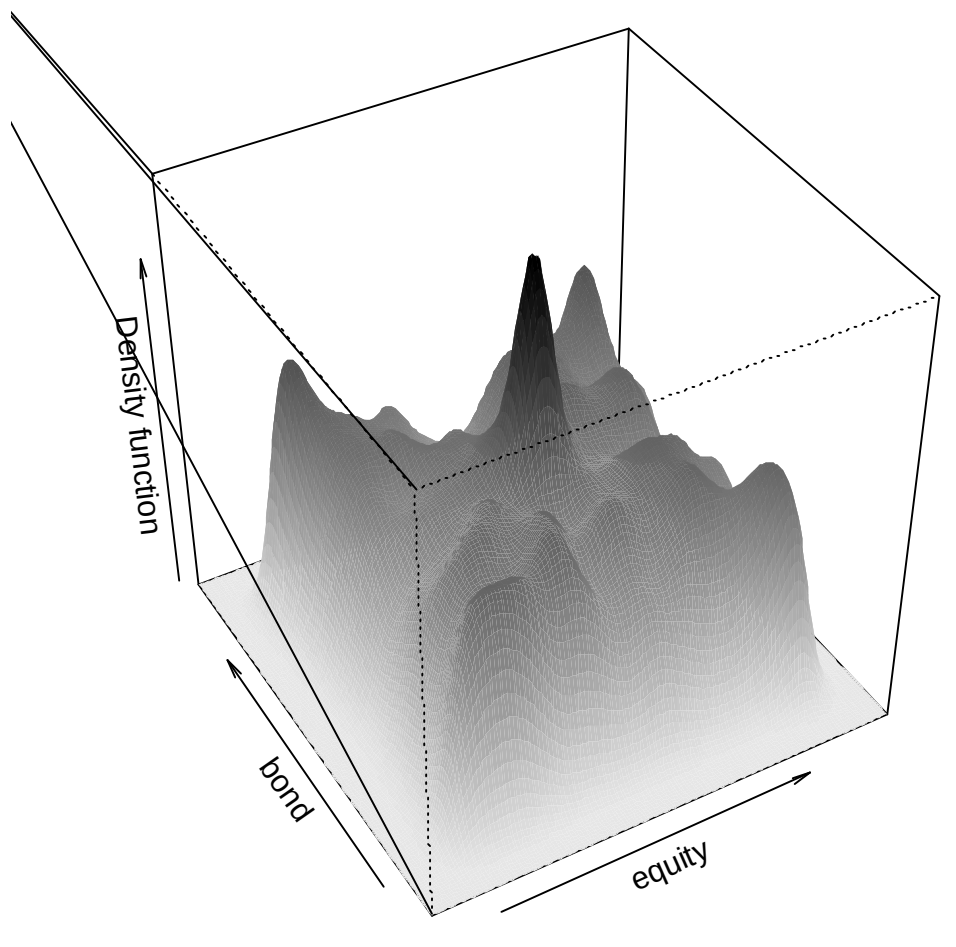

*Source: Bloomberg Finance LLP, MSCI, Russell, S&P, Worldscope, Deutsche Bank Quantitative Strategy*

Figure 23: Empirical distribution between bonds and commodities

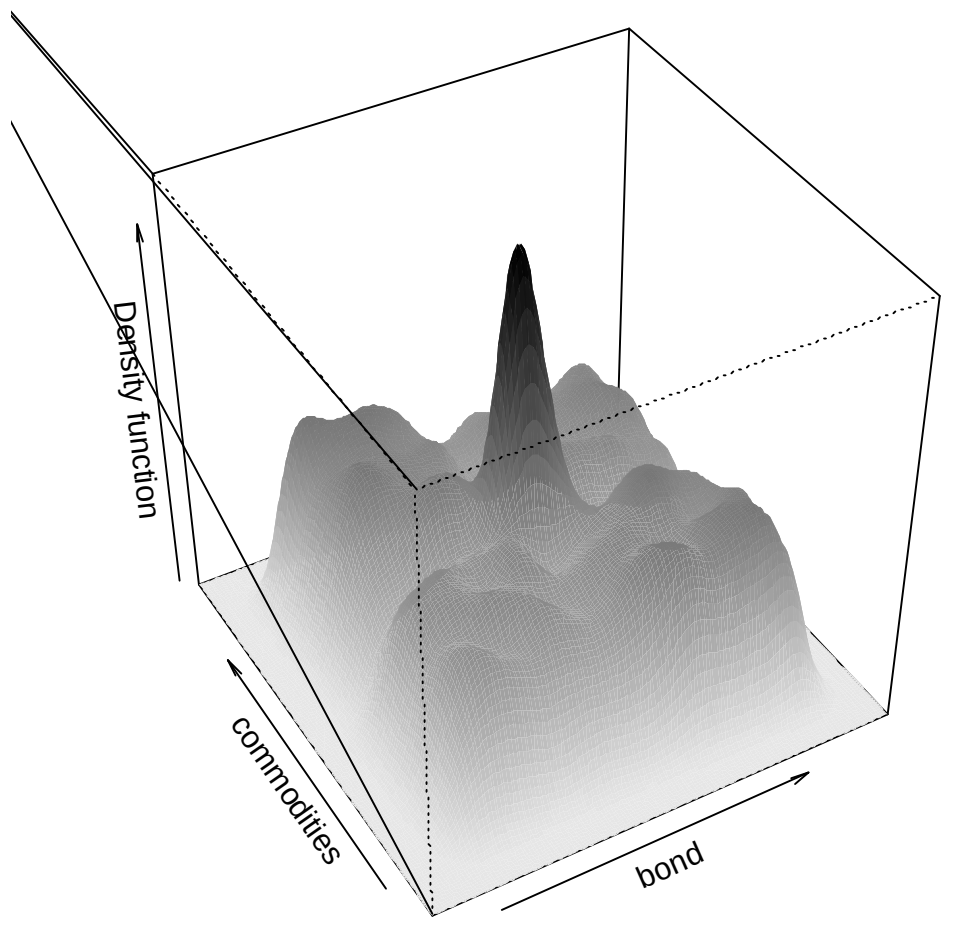

*Source: Bloomberg Finance LLP, MSCI, Russell, S&P, Worldscope, Deutsche Bank Quantitative Strategy*

# II. Predicting global financial market risk via GARCH-DCC-Copula model

Let's now review the stylized patterns of asset returns that we have identified in the previous section and suggested solutions:

- The conditional mean of asset returns using ARMA(1,1), i.e., serial correlation
- The univariate conditional variance of each asset using GARCH(1,1), i.e., volatility clustering and extreme outliers
- The dynamic conditional correlation using DCC
- The multivariate tail dependence using t-Copula

In this section, we would like to develop a statistical risk model that incorporates these features, and then use this model to predict the risk of our global capital market.

## Fitting and simulating asset returns via GARCH-DCC-Copula model

In this paper, we develop an interesting model, the GARCH-DCC-Copula model that takes into account the above four properties. It is fairly computationally intensive to fit the GARCH-DCC-Copula model for our three-asset case. Starting from January 31, 1994, we use an expanding window of daily returns for the three assets from January 2, 1989 (about five years of daily data) to fit the first model. To save computing time, we only re-estimate the model monthly, using the data available as of the time; therefore, the model is completely out-of-sample. The GARCH-DCC-Copula model is estimated as follows:

1. Each one of the three asset classes is fitted separately to an ARMA(1,1) model for conditional mean and an GARCH(1,1) model for conditional variance

2. The residuals from the ARMA(1,1)-GARCH(1,1) model are then used to fit a joint multivariate dynamic conditional correlation (DCC) model for the correlation structure

3. The residuals from the univariate ARMA(1,1)-GARCH(1,1)-DCC model are then used to fit a t-Copula model to account for tail dependence

4. The estimated parameters of the ARMA(1,1)-GARCH(1,1)-DCC-Copula model are used to simulate future returns and calculate various risk metrics

## Predicting risk and correlation

Modeling and fitting asset returns is the starting point. Now we need to use the model to predict risk. At each month end, we use the fitted GARCH-DCC-Copula model to simulate a return time series for the next 21 days for each of the three asset classes, i.e., the month-ahead return stream. Then, we can calculate the predicted return of our three-asset SAA portfolio (i.e., 50% invested in equities, 40% in bonds, and 10% in commodities) for the next 21 days. Then we repeat the same simulation 10,000 times to build a distribution of the month-ahead returns. With the 10,000 simulated return series, we can estimate both volatility and the conditional value-at-risk (CVaR) of the SAA portfolio. Figure 24 and Figure 25 show the predicted CVaR and volatility of our SAA portfolio in the past 20 years. It is intuitive to see, for example, that CVaR (and volatility) peaked during the 2008 financial crisis. We also see that the risk of bonds has increased noticeably in recent months, as a result of the potential rising interest rate concerns[8].

Since our GARCH-Copula model assumes time varying correlation, a useful side product is that we can also calculate the implied correlation of the three assets for the next 21 days (see Figure 26). The massive run-up of correlation between equities and commodities since mid-2009 is evident, which is the main driver behind the increased average implied correlation among our three asset classes. We see that bonds serve as a great diversifier over most of the 20-year history, with negative correlation with both equities and commodities. However, we can also see those negative correlations have recently come to an end.

Figure 24: Predicted risk (conditional value-at-risk)

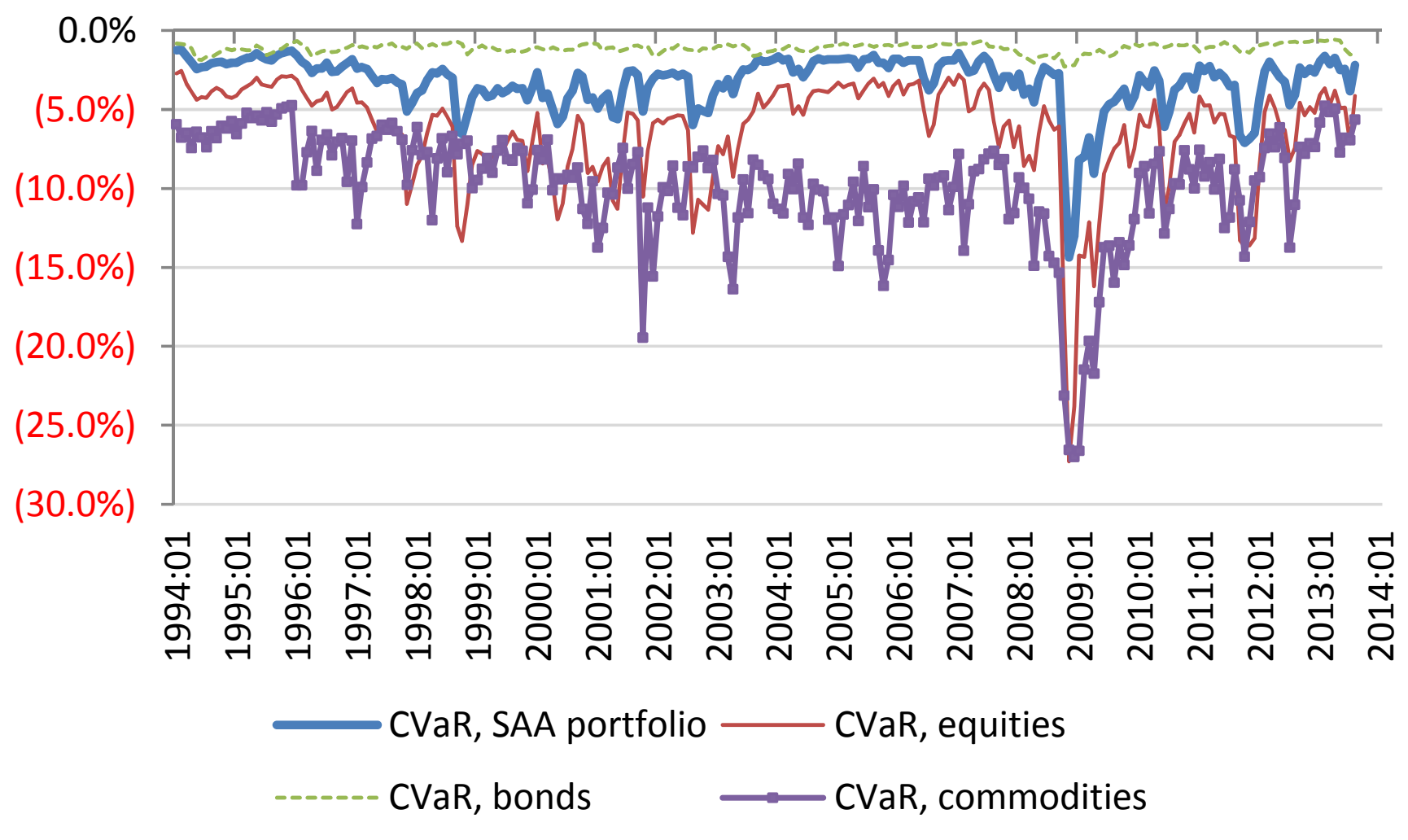

*Source: Bloomberg Finance LLP, MSCI, Russell, S&P, Worldscope, Deutsche Bank Quantitative Strategy*

[8] We will have more discussions around rising interest rates in later sections.

Figure 25: Predicted risk (volatility)

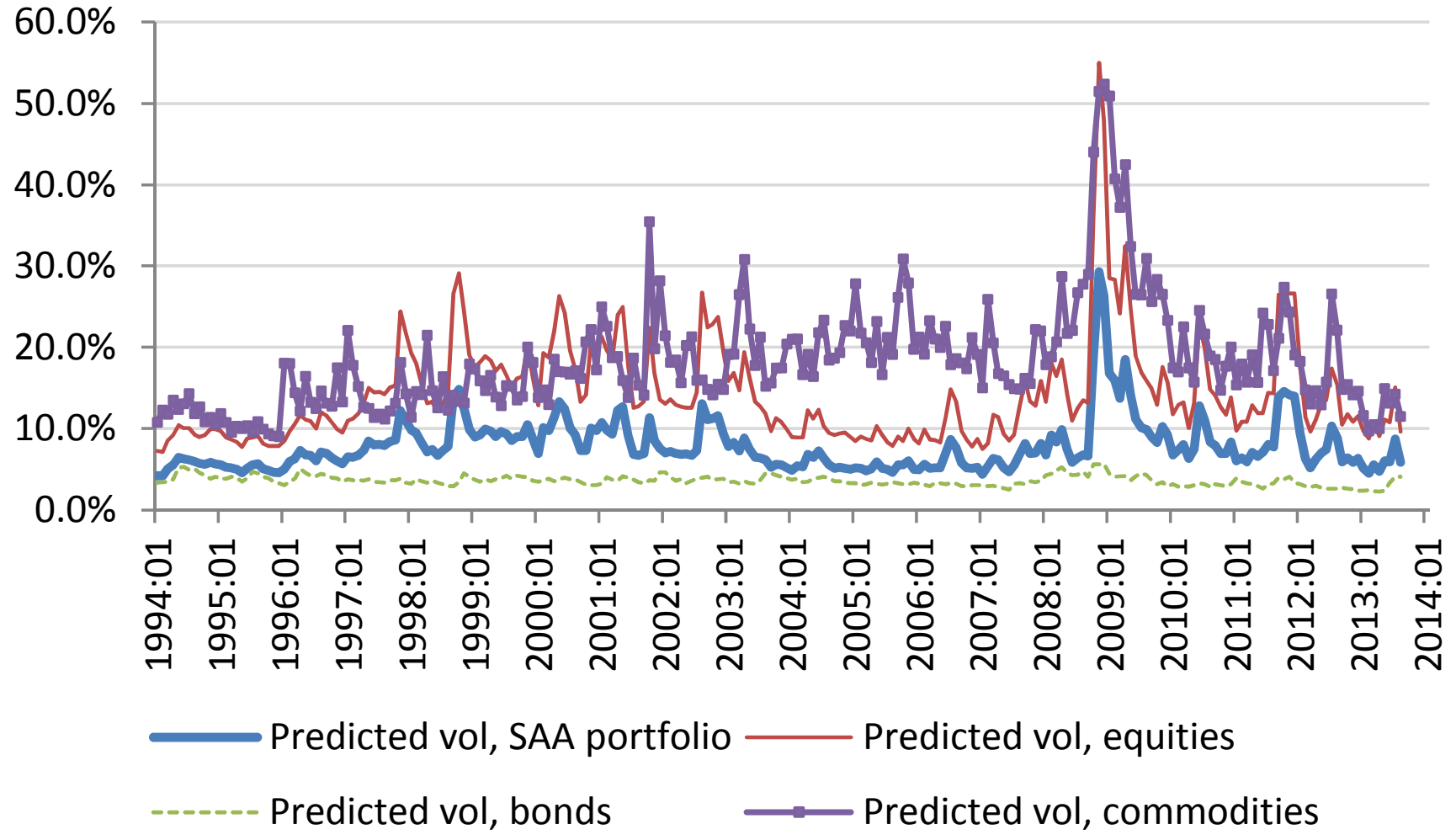

*Source: Bloomberg Finance LLP, MSCI, Russell, S&P, Worldscope, Deutsche Bank Quantitative Strategy*

Figure 26: Predicted correlation

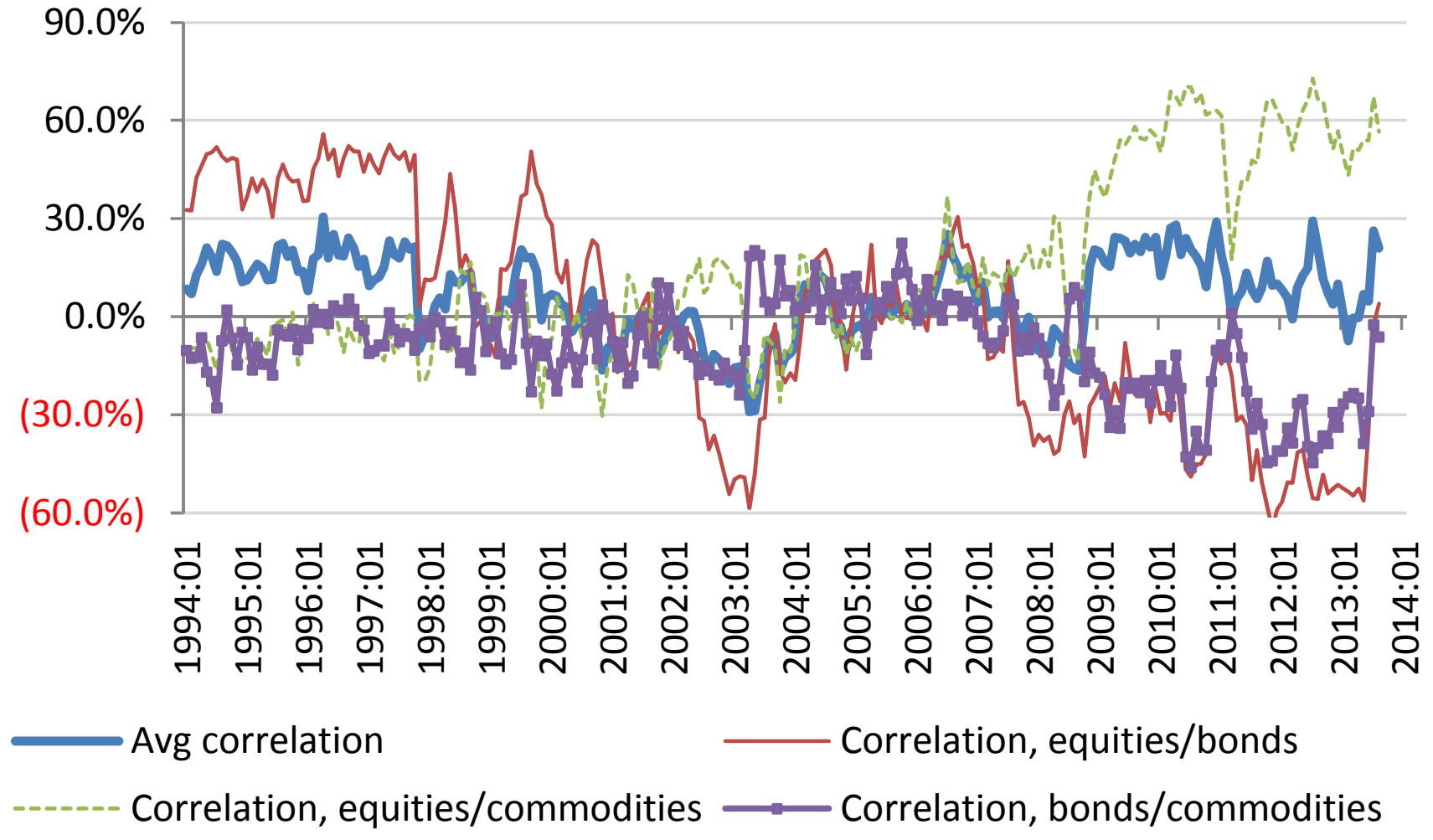

*Source: Bloomberg Finance LLP, MSCI, Russell, S&P, Worldscope, Deutsche Bank Quantitative Strategy*

## The accuracy of our risk and correlation prediction

The GARCH-DCC-Copula model does appear to model the stylized patterns of asset returns fairly well. However, the bottom line is how well our model predicts actual risk and correlation. The trick is, of course, that we never really know the true level of risk or correlation, because they are unobservable. Nonetheless, we try to use look-ahead estimates of risk and correlation as our benchmarks. The look-ahead risk is simply the realized volatility for each asset, using daily returns of that asset over the *following* calendar month. Similarly, the look-ahead correlation is computed using daily returns of a pair of assets over the *next* calendar month. These look-ahead risk and correlation can't be used in real-time portfolio management, because we can't calculate them until the following month end, but they do use forward-looking information and serve as sensible benchmarks.

A much simpler way to predict risk is to use realized volatility, e.g., realized volatility calculated with previous month's daily returns – hereafter called "naïve" estimate. Figure 27 shows the time series IC, i.e., the correlation between our predicted risk (using either naïve estimates or our GARCH-DCC-Copula model) with the "actual" volatility (using realized volatility with daily returns of the following month). Our GARCH-Copula model demonstrates higher predictive power than the naïve model for all three asset classes, but particularly well for the most volatility asset class – commodities.

To assess the ability to predict correlation, as shown in Figure 28, we compare our GARCH-DCC-Copula model with GARCH-CCC-Copula (i.e., the constant correlation model) and naïve estimate (calculating correlation using previous month's daily returns). Our GARCH-DCC-Copula model clearly dominates the other two models, as measured by time series IC (i.e., the correlation between our model prediction and "actual" correlation[9]).

A more accurate risk model does not always guarantee a better portfolio. In the next section, we try to apply our GARCH-DCC-Copula model to a few real-life asset allocation strategies (both risk-based and alpha-based) to show whether it actually improves portfolio performance.

Figure 27: Predicting risk, correlation between estimates and "actual" risk

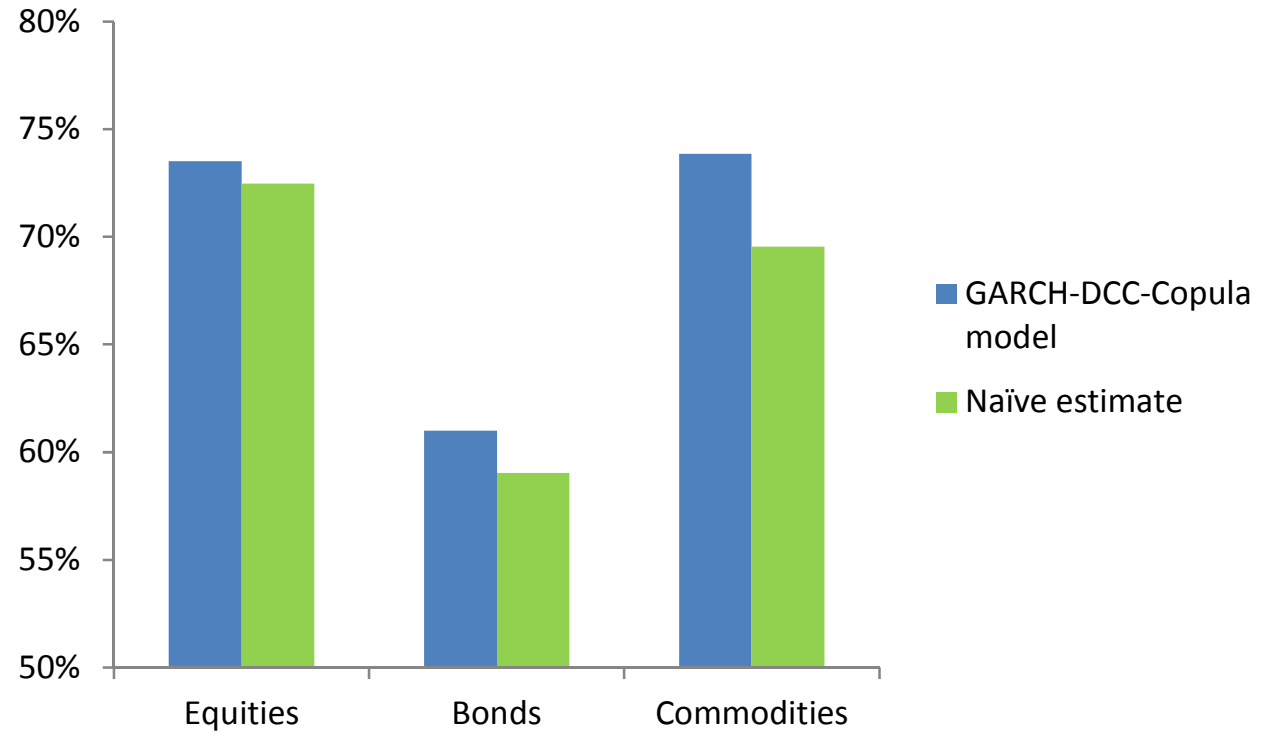

Source: Bloomberg Finance LLP, MSCI, Russell, S&P, Worldscope, Deutsche Bank Quantitative Strategy

Figure 28: Predicting correlation, correlation between estimates and "actual" correlation

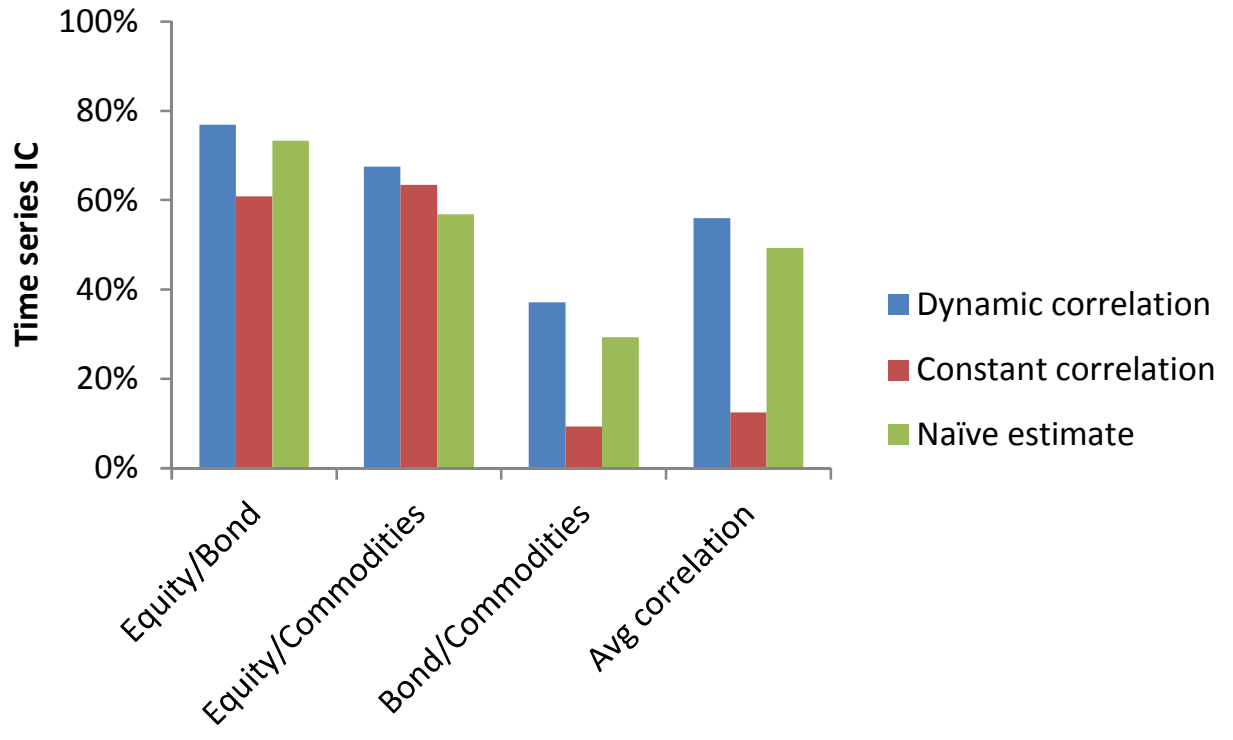

Source: Bloomberg Finance LLP, MSCI, Russell, S&P, Worldscope, Deutsche Bank Quantitative Strategy

[9] As a reminder, the "actual" correlation is calculated using the following month's daily returns.

# III. GARCH-DCC-Copula model for asset allocation

In this section, we extend our GARCH-DCC-Copula model from a three-asset case to a more real-life 11-asset application. The purpose is to assess:

- Whether our GARCH-DCC-Copula model fits real asset return data well
- Whether we can build better risk models using our GARCH-DCC-Copula model
- More importantly, whether our GARCH-DCC-Copula can help us to build more efficient portfolios

## Investment universe

If our GARCH-Copula model indeed estimates risk and correlation better than traditional risk models using sample data, it should help us make better asset allocation decisions. In this section, we use a real-life asset allocation strategy with 11 asset classes[10]:

1. US large cap equity (Russell 1000)
2. US small cap equity (Russell 2000)
3. International equity (MSCI EAFE)
4. Emerging markets equity (MSCI EM)
5. REITs (S&P Global REITs)
6. US treasuries (Deutsche Bank US All Treasuries Index)
7. US high yield bonds (Deutsche Bank US High Yield Index)
8. Investment Grade Sovereign (Deutsche Bank USD Investment Grade Sovereign Bond Index)
9. EM credit (Deutsche Bank Emerging Markets Bond USD Index)
10. Commodities (S&P/GSCI)
11. Gold (front-end gold futures)

Our benchmark is a traditional 60-40 allocation: 30% in US large-cap equities, 30% in EAFE equities, 20% in US treasuries, and 20% in investment grade sovereign bonds[11].

[10] To be consistent, we use the same 11 asset classes as in Luo, et al [2013].
[11] The benchmark is again the same as in Luo, et al [2013].

As shown in Figure 29, since 2000, gold and REITs have delivered the highest cumulative returns, while EAFE equities and commodities have produced the lowest returns. US treasuries, investment grade sovereign bonds, and US high yield bonds tend to be much less risky than EM equities and commodities (see Figure 30). In terms of Sharpe ratio, EM credit, investment sovereign bonds, and US treasuries have delivered the best performance, while commodities and EM equities lagged at the bottom (see Figure 31).

All equity indices are highly correlated (see Figure 32), while bonds, EM credit, and commodities seem to be good hedges. Commodities and gold reveal strong nonlinear relationships with equities.

Figure 29: Cumulative returns of the 11 asset classes

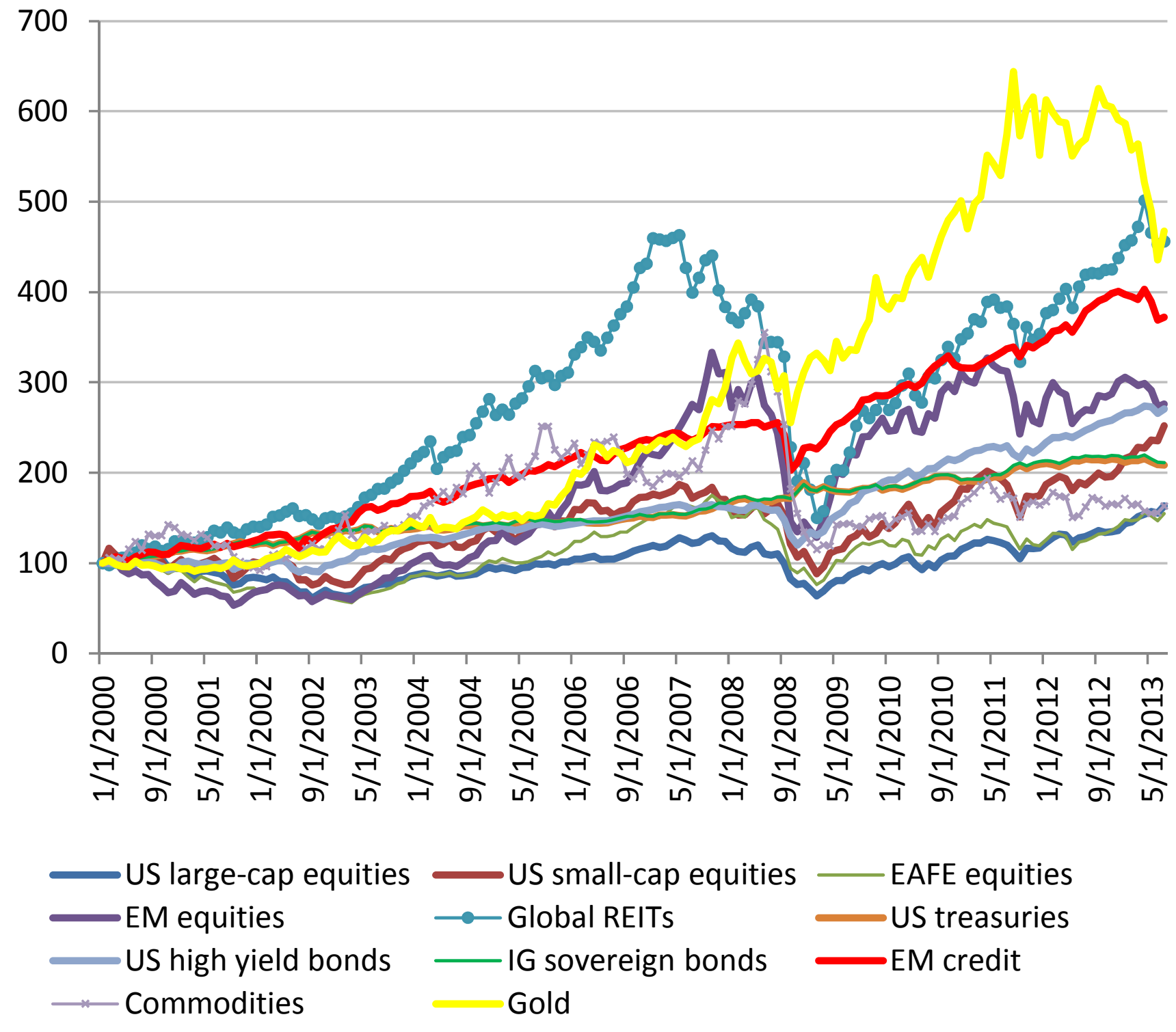

*Source: Bloomberg Finance LLP, MSCI, Russell, S&P, Worldscope, Deutsche Bank Quantitative Strategy*

## Figure 30: Historical average return and volatility

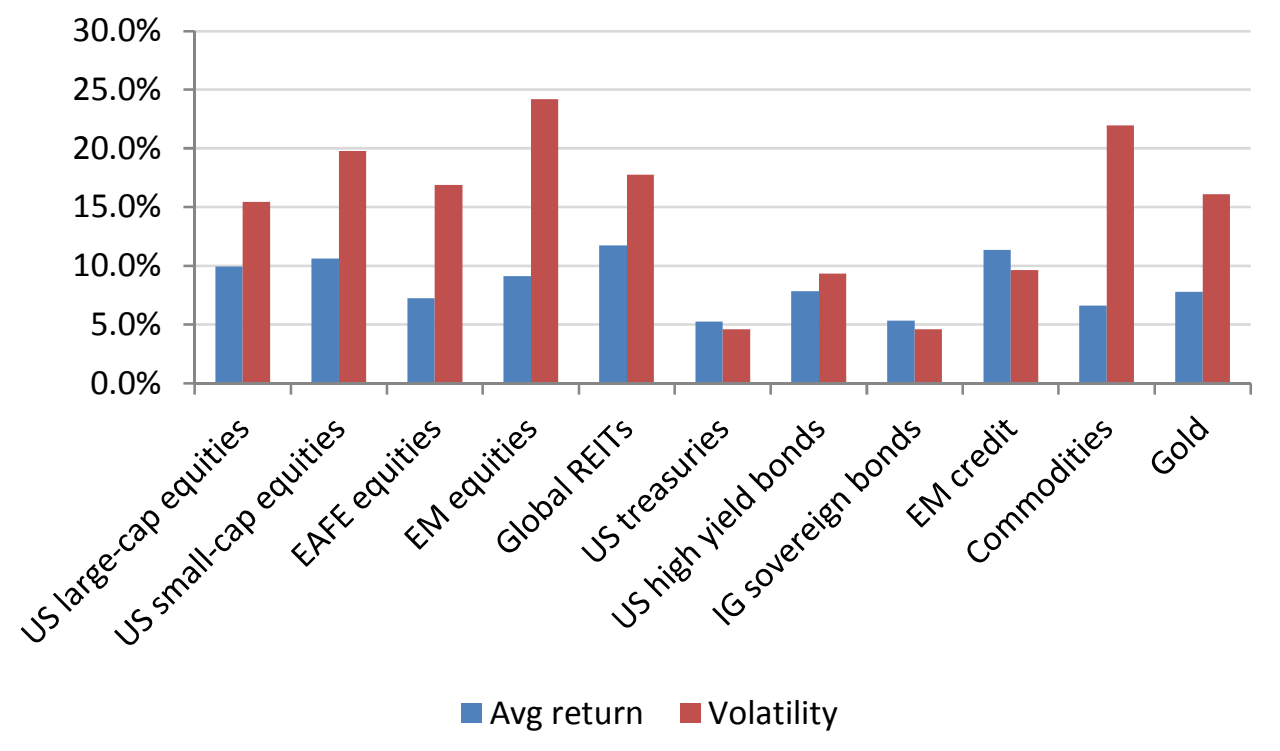

*Source: Bloomberg Finance LLP, MSCI, Russell, S&P, Worldscope, Deutsche Bank Quantitative Strategy*

## Figure 31: Historical Sharpe ratio

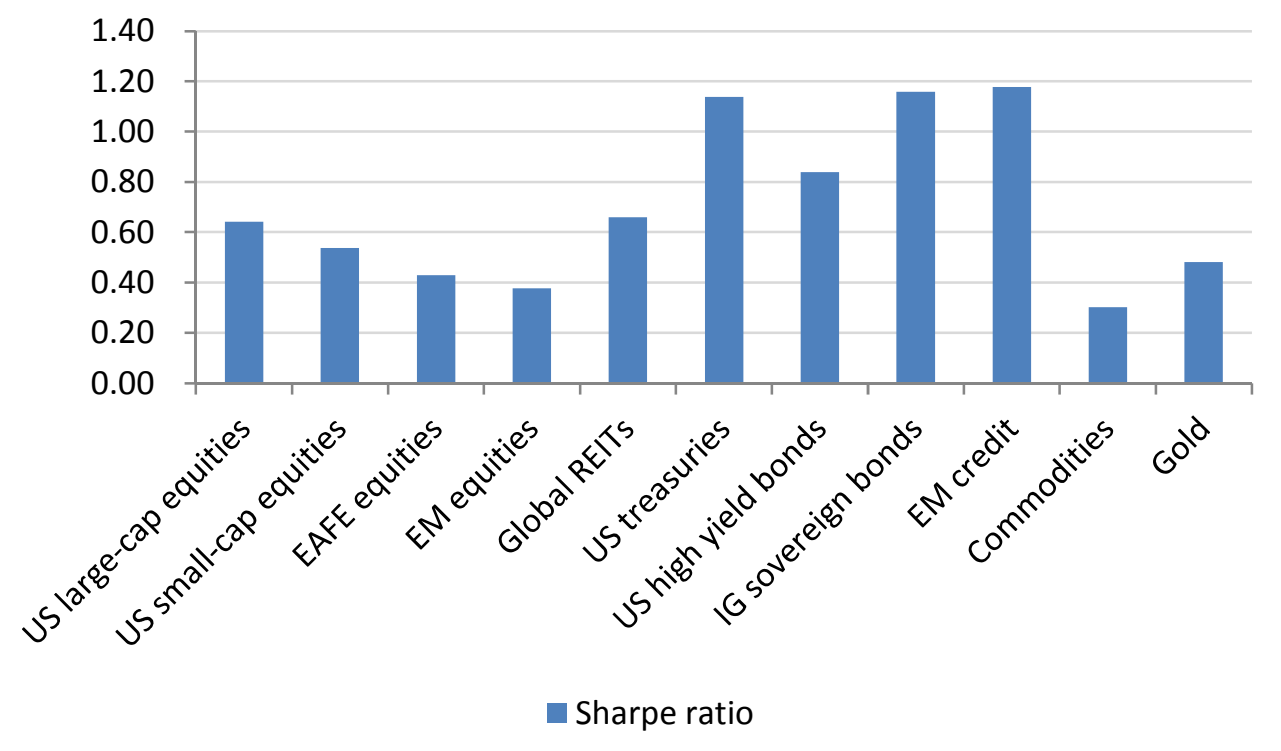

*Source: Bloomberg Finance LLP, MSCI, Russell, S&P, Worldscope, Deutsche Bank Quantitative Strategy*

## Figure 32: Scatterplot of 11 asset classes

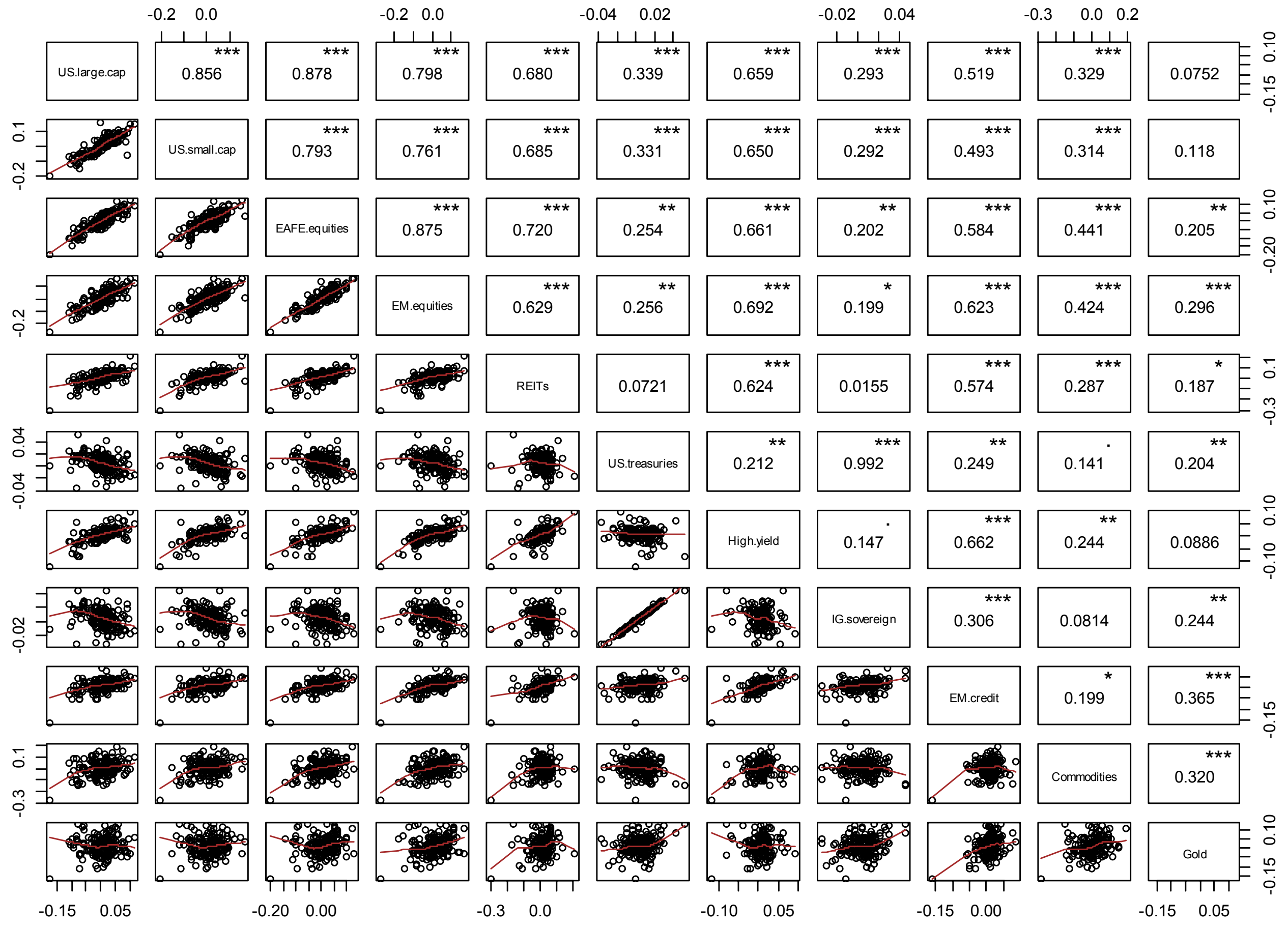

*Source: Deutsche Bank*

## Portfolio construction techniques

We then build our asset allocation portfolio with the following techniques[12]:

- Benchmark: a traditional 60-40 portfolio, i.e., 30% in US large-cap equities, 30% in EAFE equities, 20% in US treasuries, and 20% in investment grade sovereign bonds.
- Naïve diversification strategies
  - Equally weighted portfolio (EquallyWgted), by equally weighting all assets
  - Inverse volatility portfolio (InvVol), by weighting each assets by the inverse of their volatilities
  - Risk parity or equal risk contribution (RiskParity), by allocating equal risk budget to each asset, fully accounting for the covariance matrix
- Sophisticated diversification strategies
  - Maximum diversification portfolio (MaxDiversification), by maximizing diversification ratio[13]
  - Minimum tail dependence portfolio (MinTailDependence), by minimizing weighted average tail dependence among all assets
- Risk minimization strategies
  - Global minimum variance portfolio (GlobalMinVar), by minimizing the portfolio variance
  - Minimum variance-tail dependence (MinVarTail), by minimizing the variance-tail dependence matrix
  - Minimum CVaR (MinCVaR), by minimizing the portfolio conditional value-at-risk

## Three different risk models

We then compare the performance of the same suite of risk-based allocations as in Luo, *et al* [2013], using three sets of risk models:

- Rolling window: risk metrics are estimated using trailing one-year actual returns
- Expanding window: risk models are estimated using an expanding window (minimum five years of daily data)
- GARCH-DCC-Copula risk model

The trade-off between the two sample-based risk models is that the rolling window estimation tends to use more timely information, while the expanding window uses more data, which helps reduce estimation errors.

---

[12] The benchmark, naïve and sophisticated diversification strategies, and risk minimization strategies are fully defined in Luo, et al [2013].

[13] Please note that Maximum Diversification Portfolio (MDP®) and Diversification Ratio (DR®) are trademarks of TOBAM.

To be robust, we decide not to fine tune the “best” statistical model. Rather, we use exactly the same GARCH-DCC-Copula setup as in the previous three-asset example. We estimate our GARCH-DCC-Copula model as follows:

1. Each one of the 11 asset classes are fitted separately to an ARMA(1,1)-GARCH(1,1) model

2. The residuals from the ARMA(1,1)-GARCH(1,1) model are then used to fit a joint multivariate DCC model for the correlation structure and t-Copula for tail dependence

3. The estimated parameters of the ARMA(1,1)-GARCH(1,1)-DCC-Copula model are used to simulate daily returns

4. The simulated daily returns are used as the estimation universe for the risk models applied to the various risk-based allocations

## Performance comparison

First of all, we would like to point out that the performance of the benchmark and EquallyWgted portfolios does not depend on any risk models. We include them in the analysis below for simple comparison purpose.

From a return aspect (see Figure 33), portfolios constructed using our GARCH-DCC-Copula model produce higher *ex post* returns consistently than the two other sample-based risk models. Expanding window risk model tends to perform slightly better than rolling window risk model.

More interestingly, from an *ex post* risk perspective (see Figure 34), our GARCH-DCC-Copula risk model successfully reduces both variance-based risk (e.g., GlobalMinVar) and tail risk (e.g., MinTailVar and MinCVaR).

Our GARCH-DCC-Copula risk model appears to improve portfolio Sharpe ratio across the board (see Figure 35) and reduce downside risk even further (see Figure 36). As a side product, we once again verify that all risk-based allocations outperform the benchmark, regardless of the risk models and portfolio construction techniques.

Lastly, portfolios constructed using our GARCH-DCC-Copula risk model are also more diversified (see Figure 37) and less likely to be crowded (see Figure 38).

Figure 33: Realized return comparison

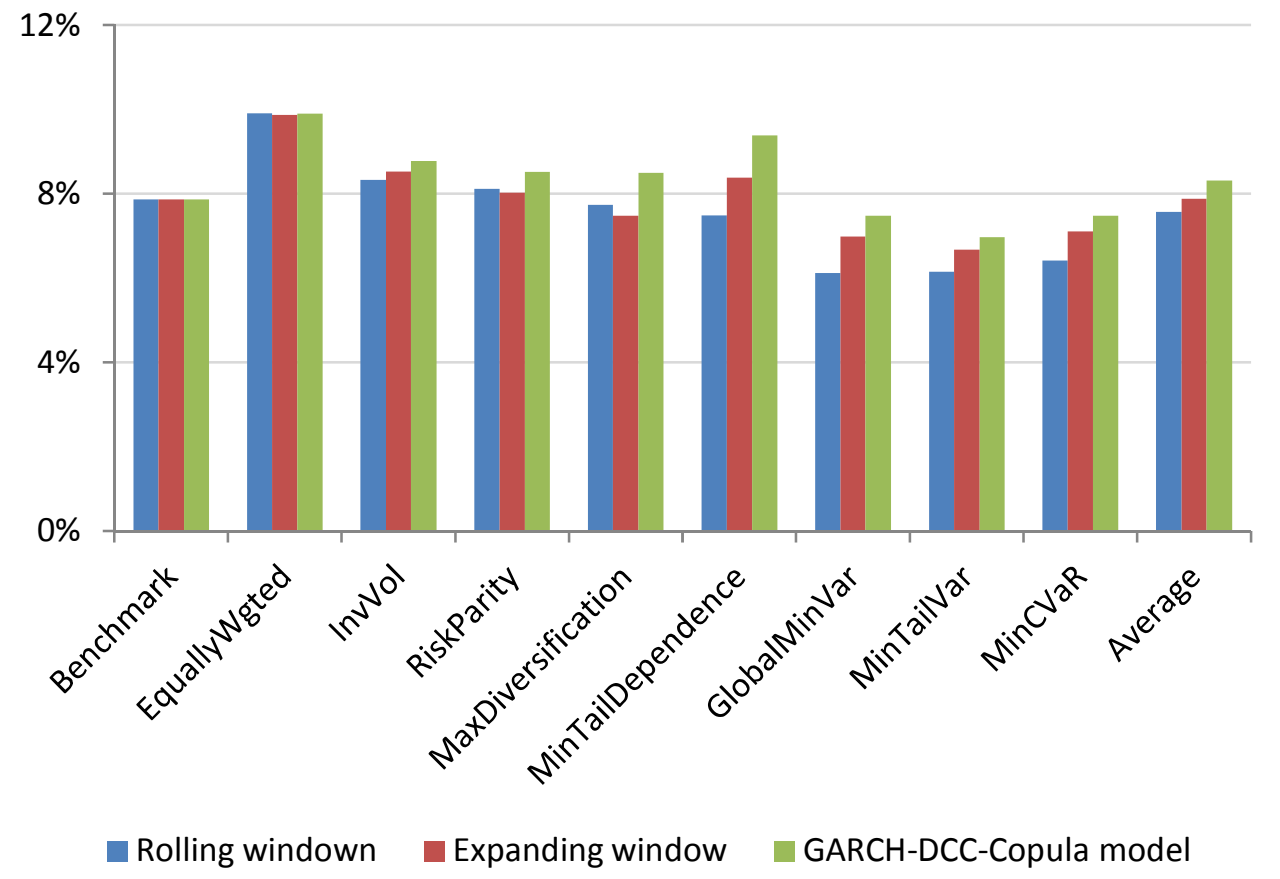

*Source: Bloomberg Finance LLP, MSCI, Russell, S&P, Worldscope, Deutsche Bank Quantitative Strategy*

Figure 34: Realized volatility comparison

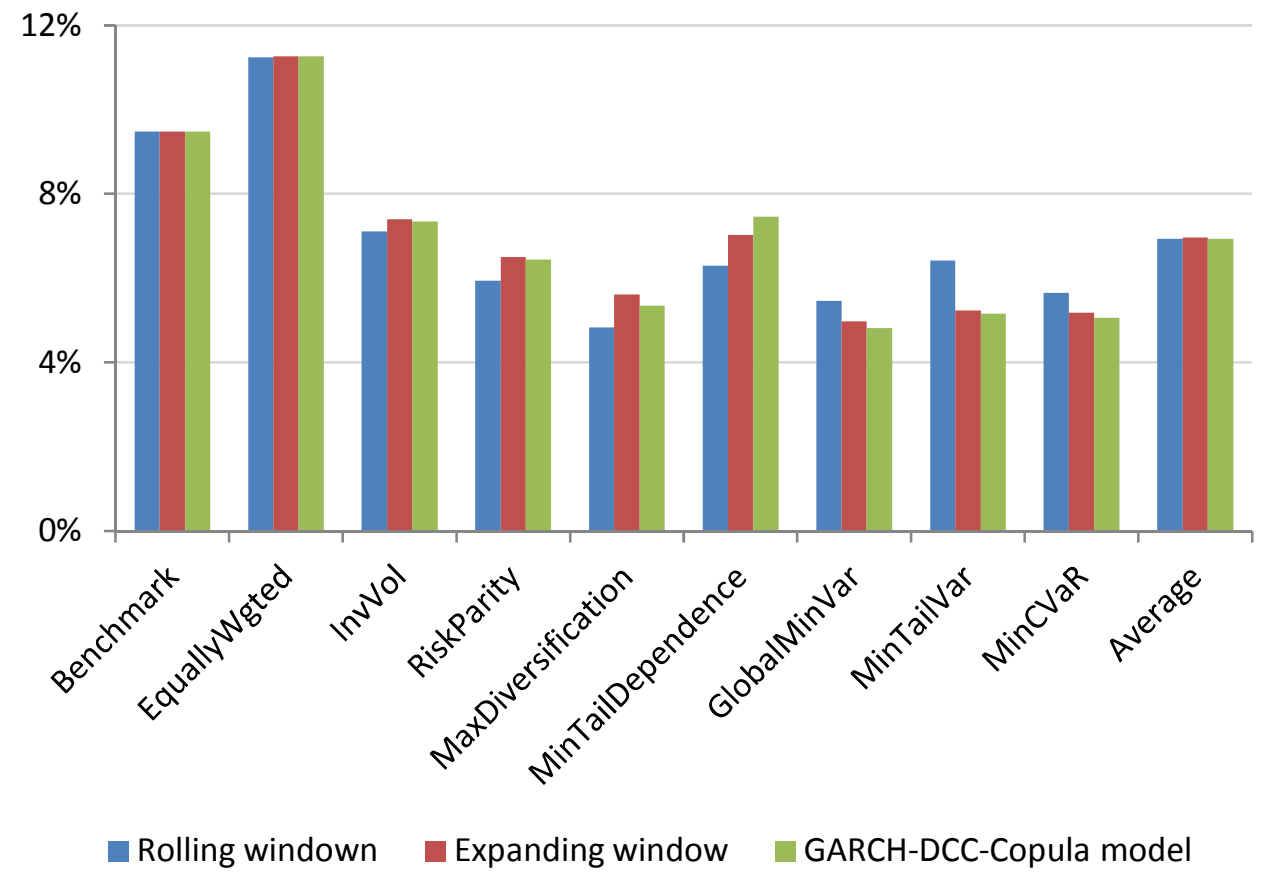

*Source: Bloomberg Finance LLP, MSCI, Russell, S&P, Worldscope, Deutsche Bank Quantitative Strategy*

Figure 35: Sharpe ratio comparison

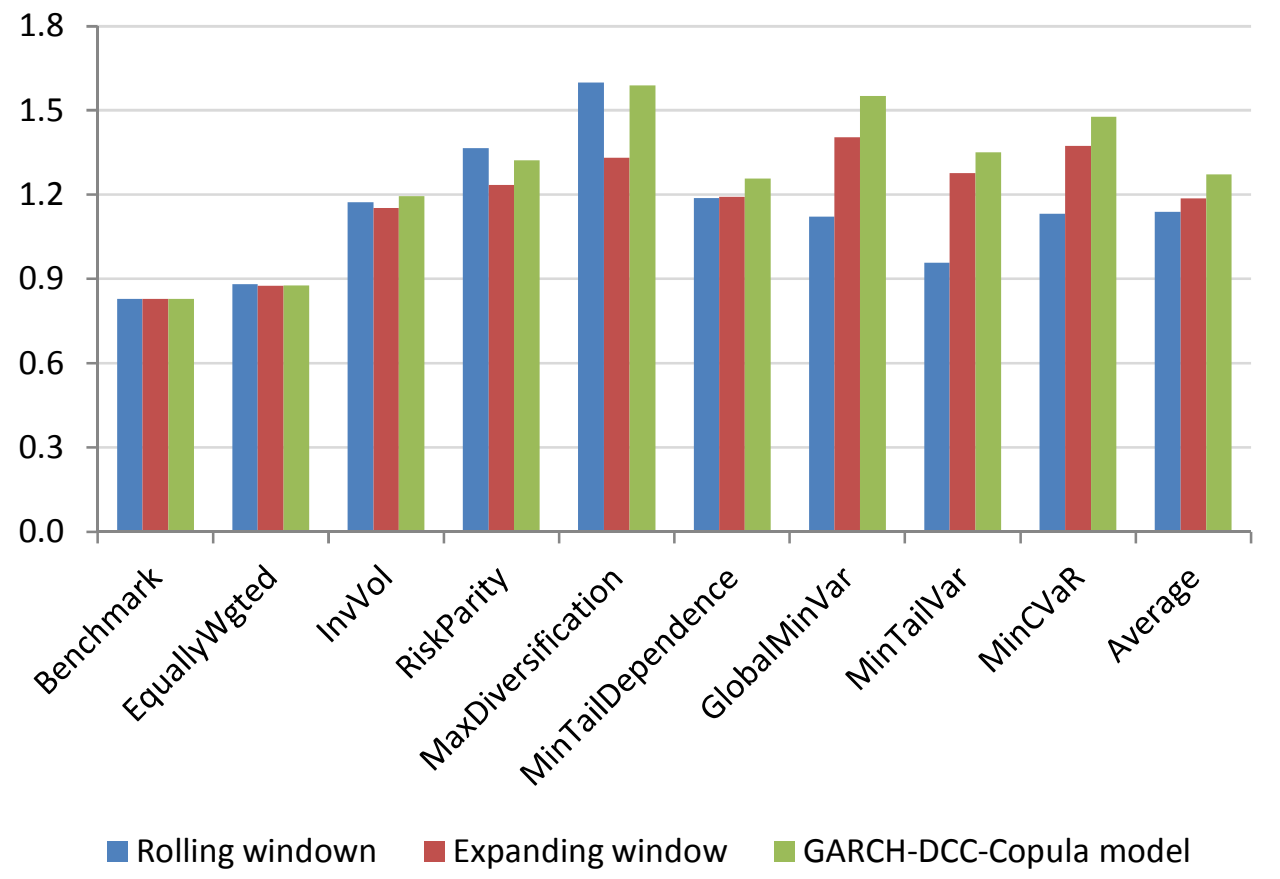

*Source: Bloomberg Finance LLP, MSCI, Russell, S&P, Worldscope, Deutsche Bank Quantitative Strategy*

Figure 36: Downside risk (maximum drawdown)

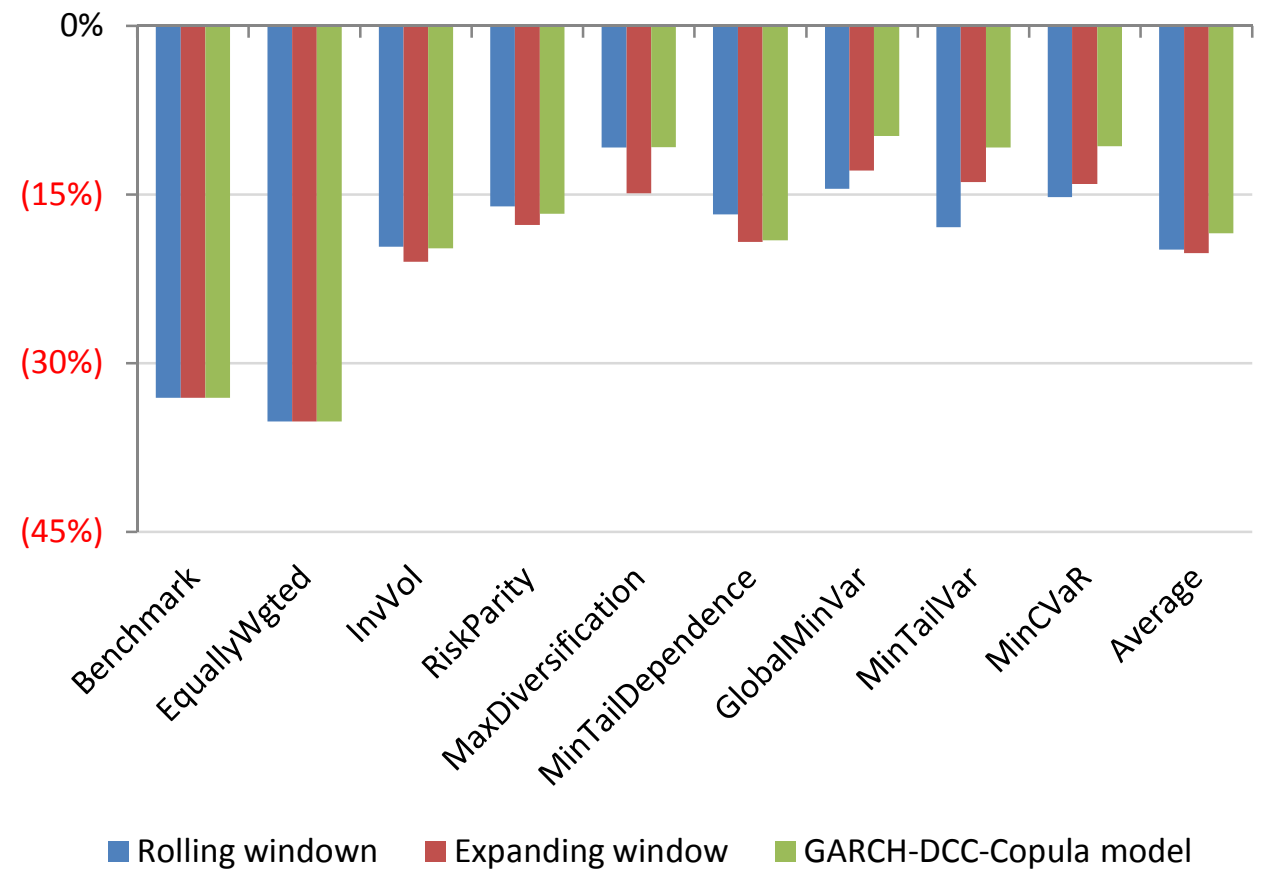

*Source: Bloomberg Finance LLP, MSCI, Russell, S&P, Worldscope, Deutsche Bank Quantitative Strategy*

Figure 37: Diversification ratio comparison

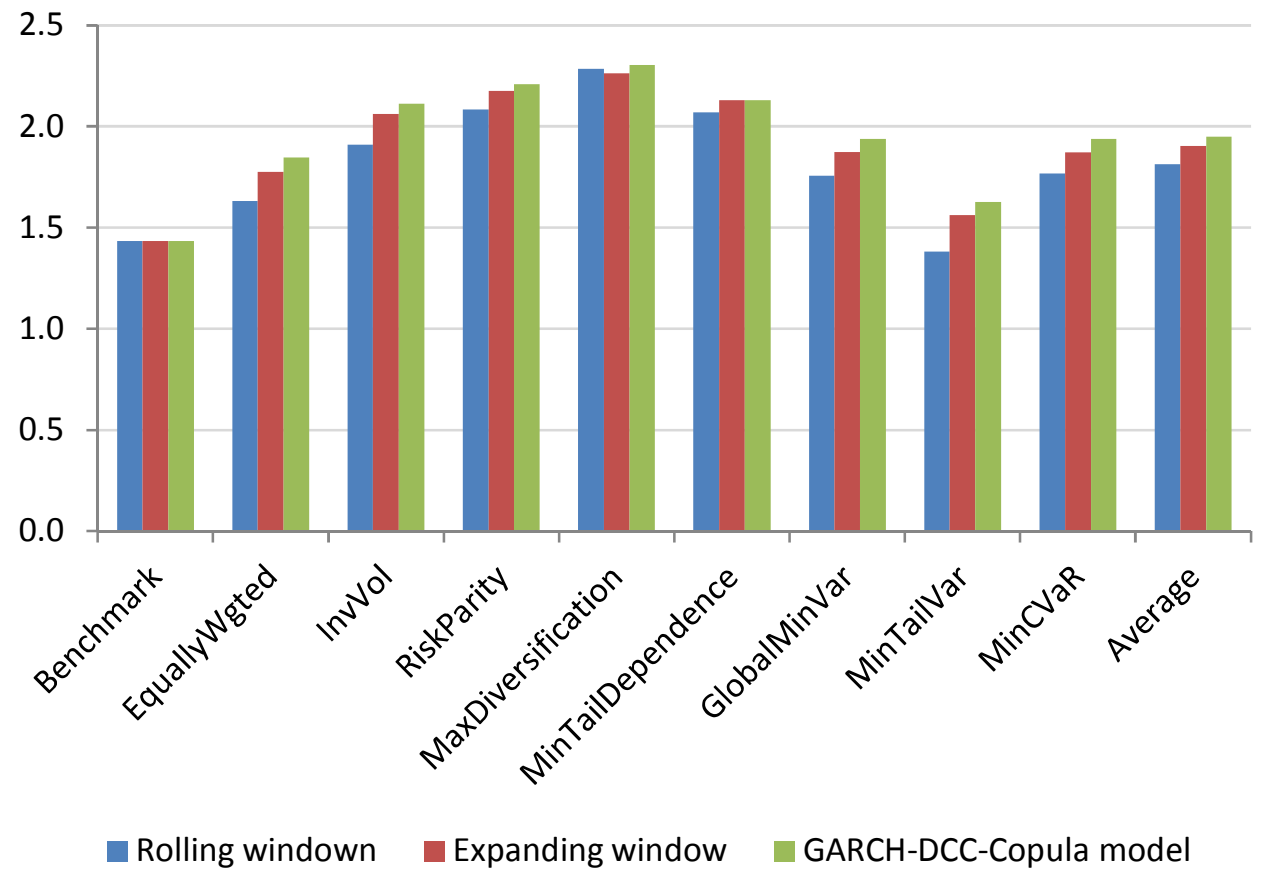

*Source: Bloomberg Finance LLP, MSCI, Russell, S&P, Worldscope, Deutsche Bank Quantitative Strategy*

Figure 38: Weighted portfolio tail dependence

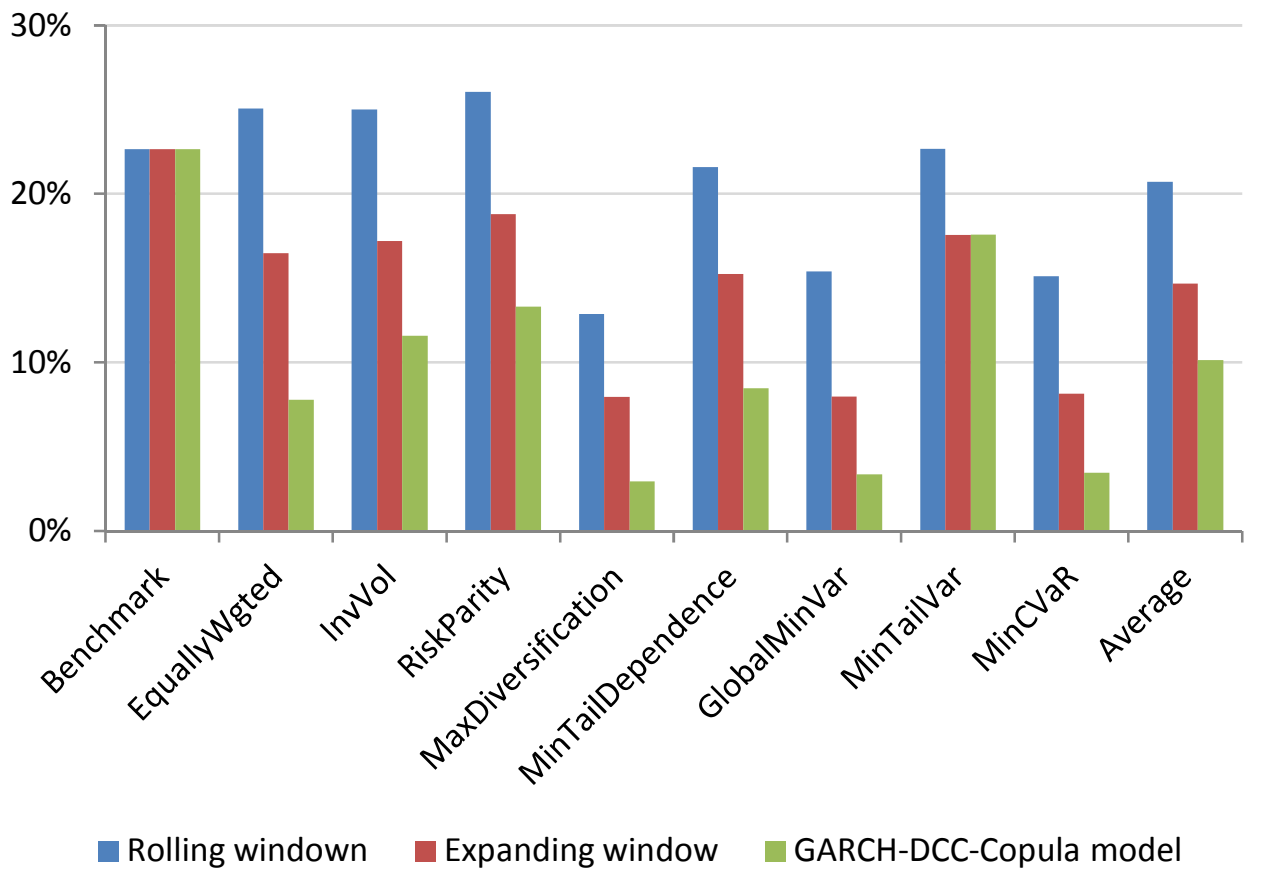

*Source: Bloomberg Finance LLP, MSCI, Russell, S&P, Worldscope, Deutsche Bank Quantitative Strategy*

## A case study – rising interest rate

Let's use a recent example to demonstrate how our improved risk model can better capture the fast moving economic environment and the implications for asset allocation decisions.

One of the major concerns in asset allocation today is that interest rates in the US and most developed countries are close to record low (see Figure 39); therefore, the upside for fixed income securities is likely to be limited, but downside risk can be significant[14]. Most risk-based asset allocation strategies are still heavily overweighting fixed income. On the other hand, although the economic recovery in the US appears to be strong, Eurozone, Japan, and EM still seem to be questionable. If another major financial or geopolitical crisis hits the global economy, interest rates may not rise as fast as investors have feared. In that case, bonds may still provide attractive returns and hedges. The recent concerns on the Fed's tapering of quantitative easing and crisis in Syria further intensify the debate. Figure 40 shows our GARCH-DCC-Copula model predicted CVaR for bonds. The current predicted downside risk for bonds is actually close to 20-year highs.

If history gives us any guidance, we can look at past rate hikes. From December 2008 to June 2009, US 10-year treasury yield rose from 2.21% to 3.53% – an increase of almost 50% in six months. Not surprising, US treasuries (along with investment grade sovereign bonds, and REITs) fell over -5% during the same period, while EM equities rallied over 36% (see Figure 41). Our GARCH-DCC-Copula model correctly predicted a heightened downside risk for bonds (see Figure 42), much more so than the rolling window or expanding window risk models suggested. As a result, our GARCH-DCC-Copula model underweighted bonds, compared to risk models estimated using either a rolling window or an expanding window. As shown in Figure 43 and Figure 44, we use RiskParity allocation as an example. The rolling window risk model would allocate around 10%, 80%, and 10% to equities, bonds, and commodities, while our GARCH-DCC-Copula model would construct a portfolio with over 20% equities, less than 70% bonds, and around 10% in commodities during the same period. As a result, as shown in Figure 45, our GARCH-DCC-Copula model delivered a return of 5.2%, compared to 4.2% under the rolling window risk model (almost 25% higher).

[14] This is the so-called the "Great Rotation", i.e., investors are likely to rotate out of bonds into stocks. As quants, we would like to let our model speak; rather than coming up with bold statements.

As a final example, Figure 46 shows the predicted risk for emerging markets equities – another asset class that has seen great selling pressure and volatility in recent months. Our GARCH-DCC-Copula model again seems to be able to pick up the rising risk faster than the sample-based risk models.

Figure 39: Wealth curve – US treasuries versus interest rate

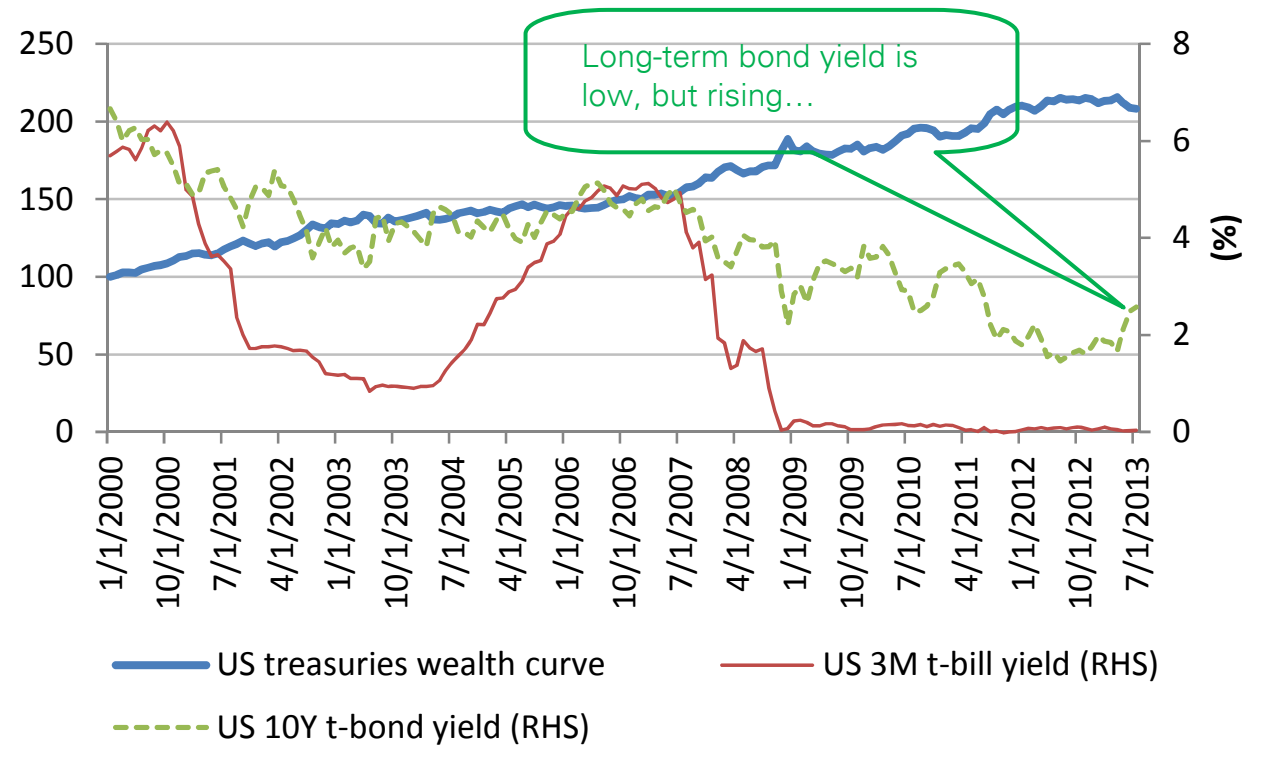

Source: Bloomberg Finance LLP, MSCI, Russell, S&P, Worldscope, Deutsche Bank Quantitative Strategy

Figure 40: Predicted CVaR – fixed income

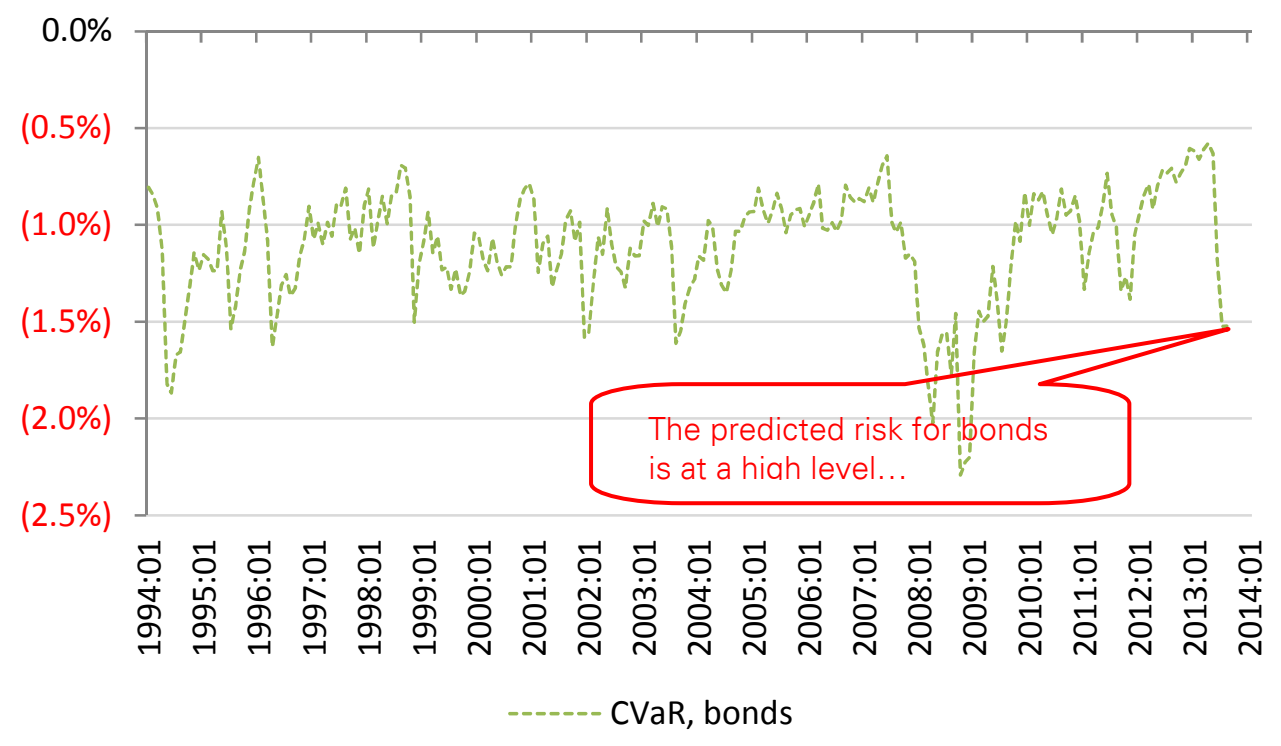

Source: Bloomberg Finance LLP, MSCI, Russell, S&P, Worldscope, Deutsche Bank Quantitative Strategy

Figure 41: Asset returns, 12/31/2008 to 6/30/2009

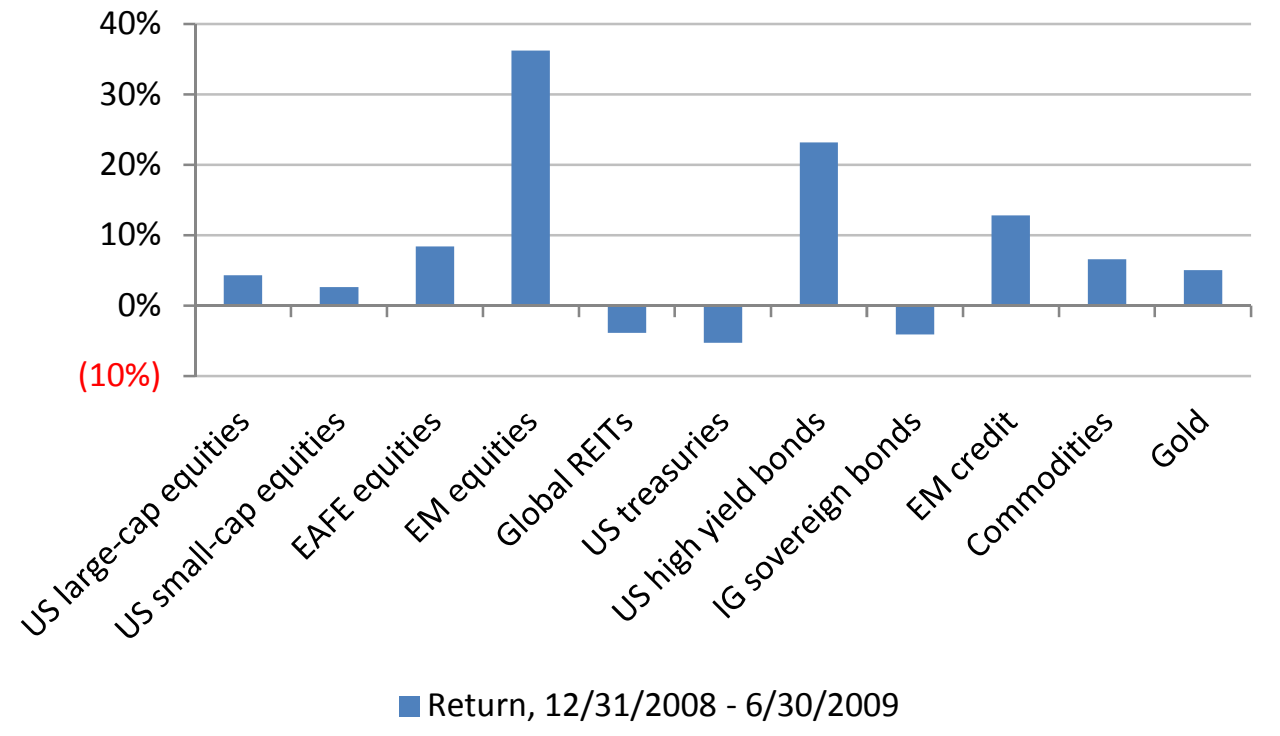

Source: Bloomberg Finance LLP, MSCI, Russell, S&P, Worldscope, Deutsche Bank Quantitative Strategy

Figure 42: Predicted risk for US treasuries

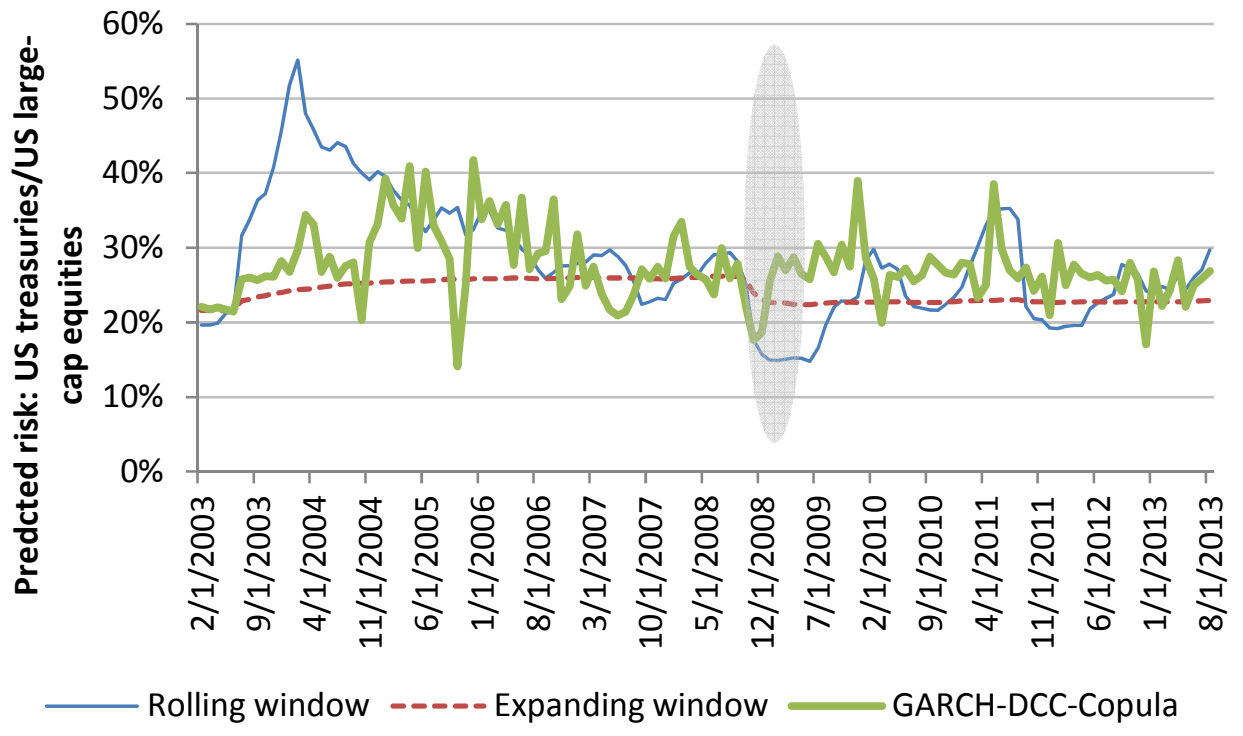

Source: Bloomberg Finance LLP, MSCI, Russell, S&P, Worldscope, Deutsche Bank Quantitative Strategy

Figure 43: Plain vanilla risk parity asset weight

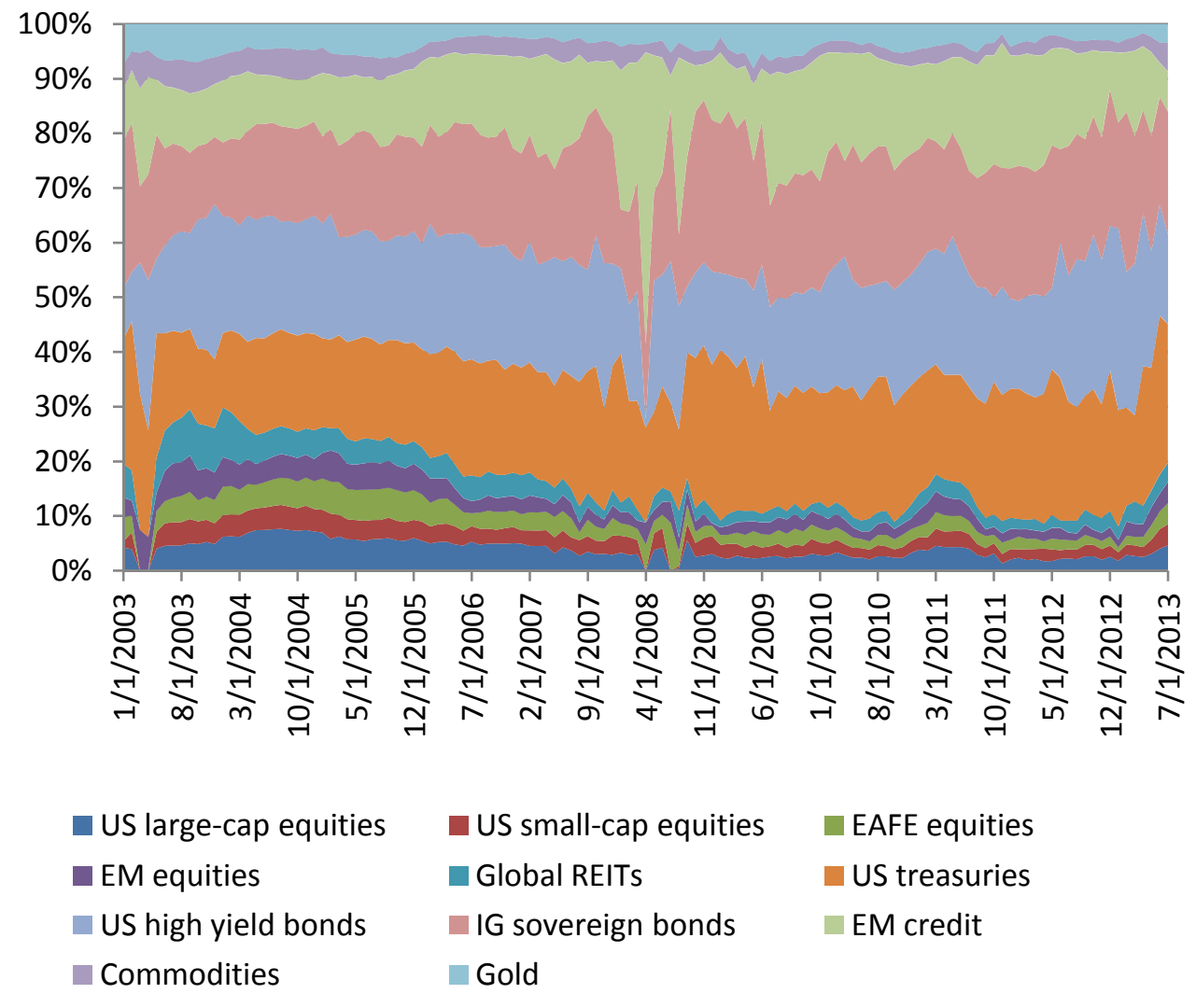

*Source: Bloomberg Finance LLP, MSCI, Russell, S&P, Worldscope, Deutsche Bank Quantitative Strategy*

Figure 44: GARCH-DCC-Copula risk parity asset weight

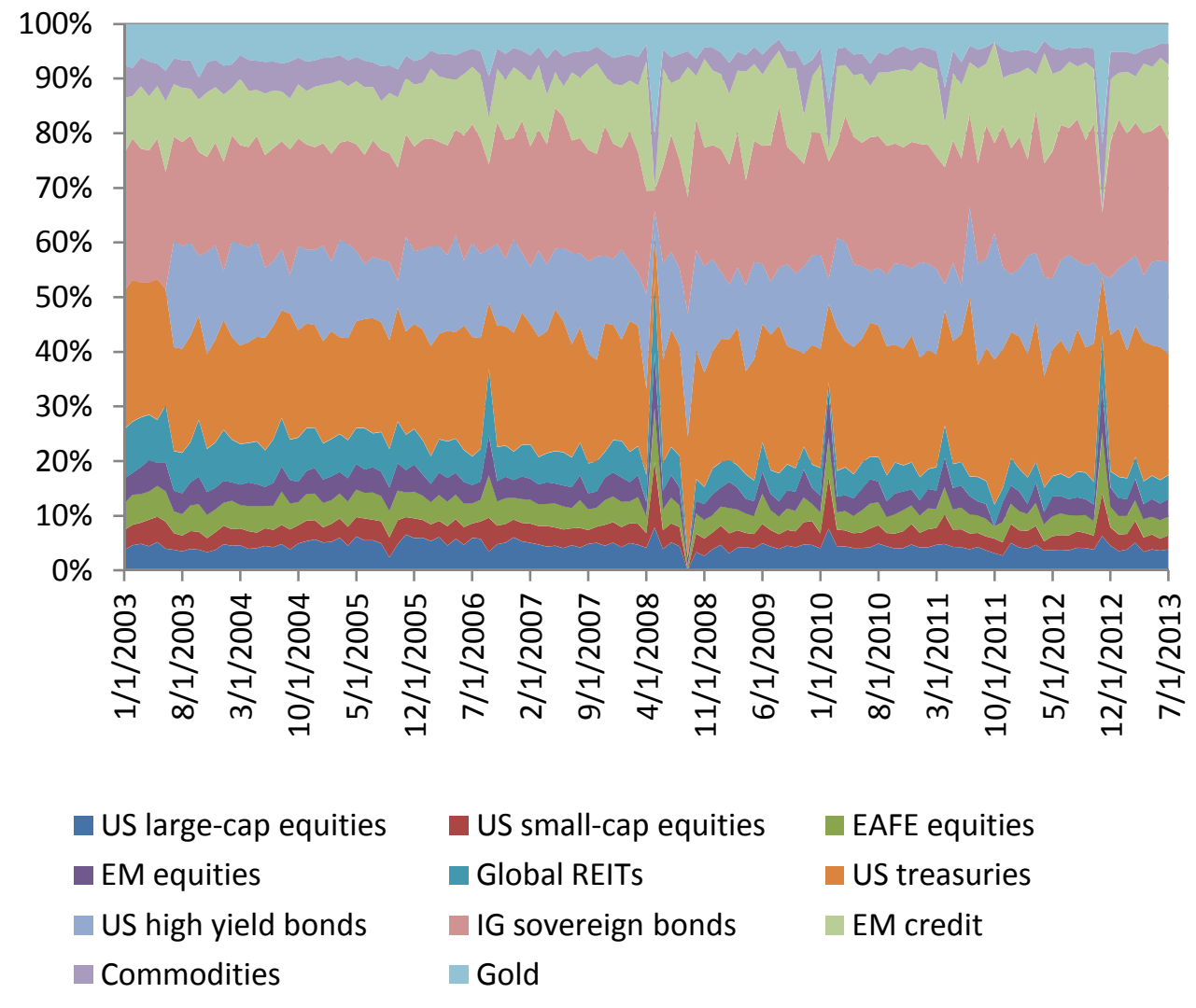

*Source: Bloomberg Finance LLP, MSCI, Russell, S&P, Worldscope, Deutsche Bank Quantitative Strategy*

Figure 45: Cumulative performance

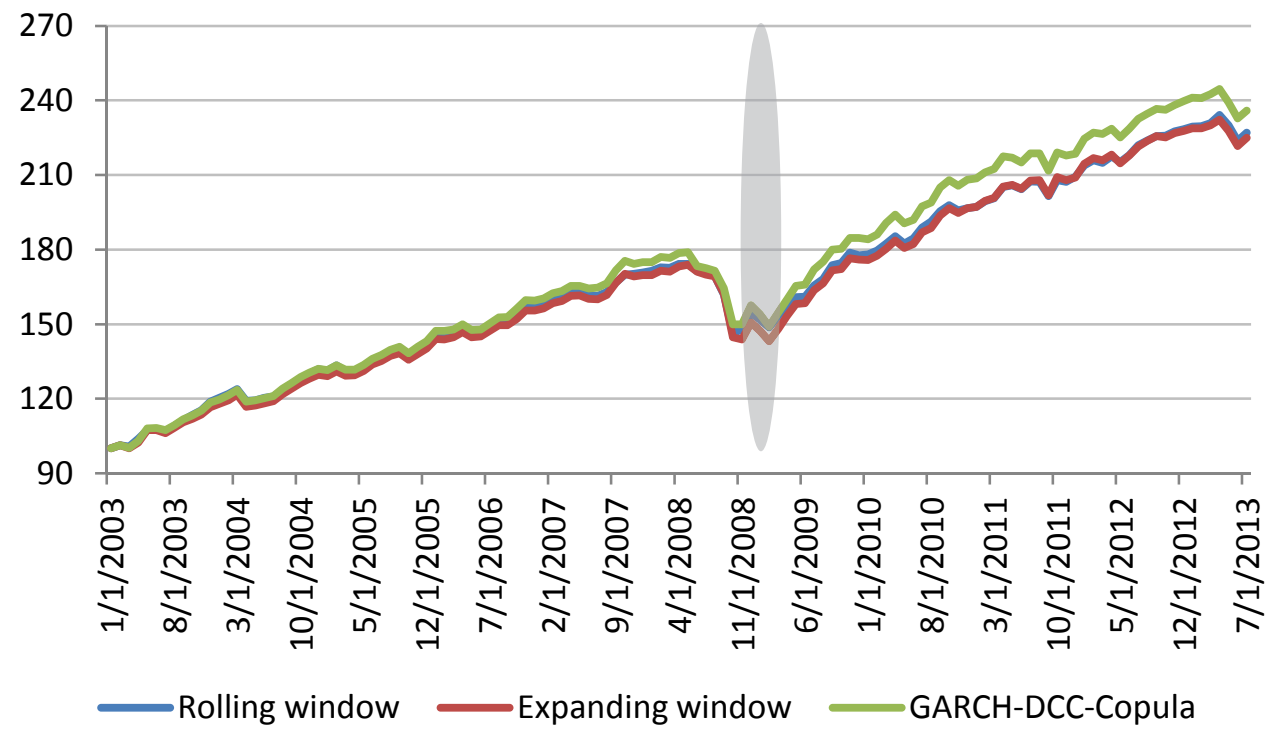

*Source: Bloomberg Finance LLP, MSCI, Russell, S&P, Worldscope, Deutsche Bank Quantitative Strategy*

Figure 46: Predicted risk for EM equities

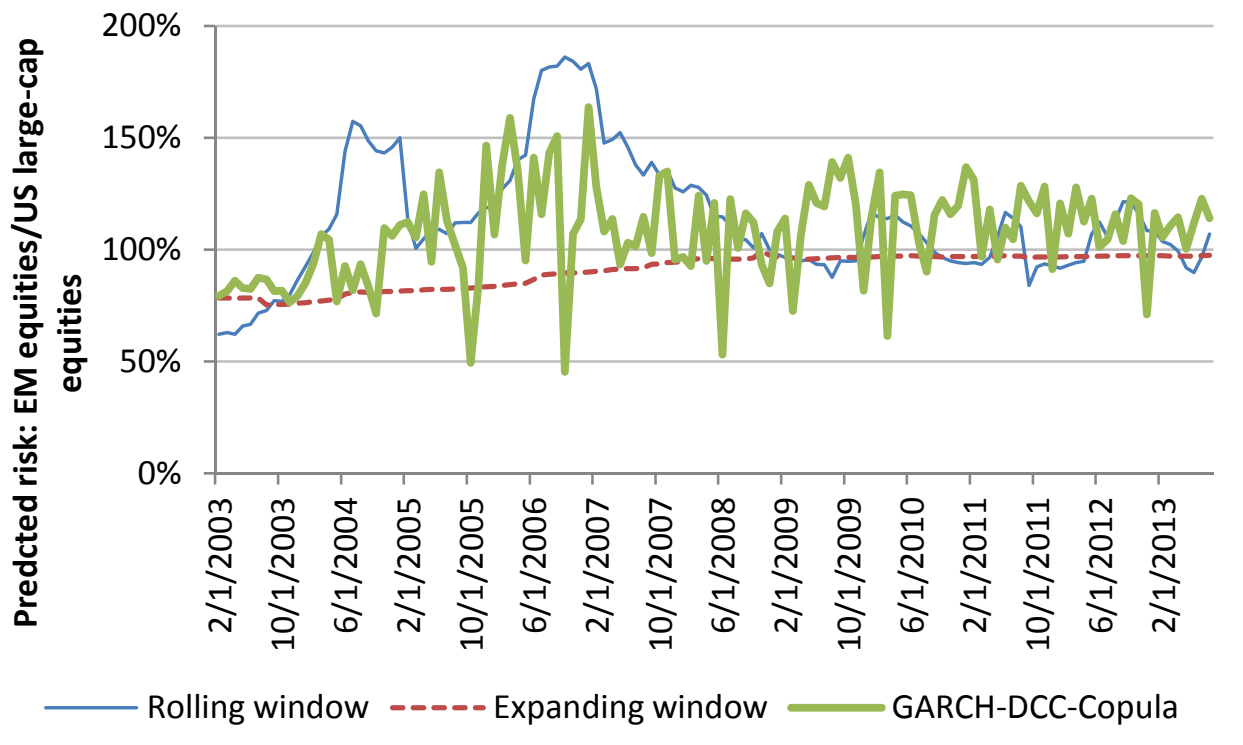

*Source: Bloomberg Finance LLP, MSCI, Russell, S&P, Worldscope, Deutsche Bank Quantitative Strategy*

# IV. GTAA strategies

The strategies in the previous section and the strategies defined in Luo, *et al* [2013] are purely risk based, i.e., the only input in constructing these portfolios is the estimated risk, while risk refers to variance-covariance, correlation, tail dependence, variance-tail dependence, or conditional value at risk.

Risk is generally considered to be easier to predict than returns and has fewer estimation errors, and that's why we have focused on risk-based allocations to establish our portfolio construction techniques.

## Return prediction

In real investing, many managers also have return prediction, based on either systematic quantitative models or discretionary qualitative process. Dynamic asset allocations with explicit return prediction are often referred to as GTAA or global tactical asset allocation strategies.

In this paper, to demonstrate how our GARCH-DCC-Copula model can help GTAA managers; we use three sets of return prediction models as simple examples. The purpose of this research is not to build the best possible alpha model; rather the focus is how risk and portfolio construction can supplement our alpha models. Therefore, the three alpha models are for demonstration purpose only:

- **Naïve alpha – short-term average.** The first forecast of asset returns is basically just the rolling one-year average of daily returns. We are essentially assuming a medium horizon momentum effect.
- **Naïve alpha – long-term average.** In the second naïve alpha model, our predicted return (i.e., alpha) for each asset is based on that asset's long-term return. Figure 47 shows our long-term return assumptions. Please note that this is a model that suffers from look-ahead bias, as we use the entire history to calculate the long-term average returns.
- **GTAA model using VRP.** As shown in Luo, *et al* [2011] "Quant TAA" and Luo, *et al* [2012] "New insights in country rotation", VRP or variance risk premium has shown great predictive power of multiple asset classes. In this example, we use a 60-month rolling window to fit and then predict the return of each asset classes, conditional on VRP.

Figure 47: Long-term average asset returns

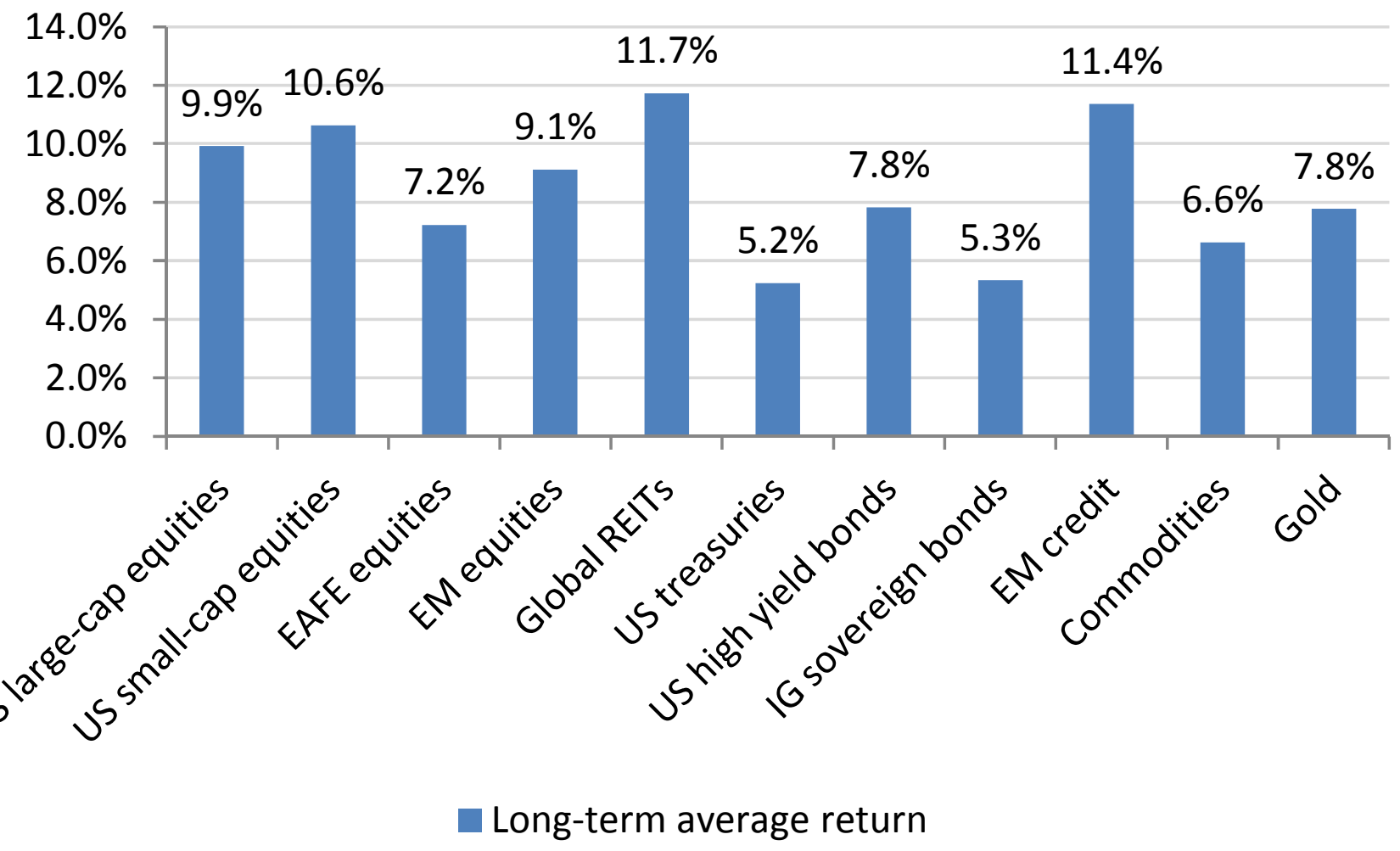

*Source: Bloomberg Finance LLP, MSCI, Russell, S&P, Worldscope, Deutsche Bank Quantitative Strategy*

## A quick review of VRP indicator

As defined in Luo, *et al* [2011], VRP or variance risk premium is more of a fear index, measuring the difference between market implied risk and realized risk. A simple example is the VRP for US equities. For US equities, the implied risk is simply the CBOE VIX index. We prefer to work in variance terms, so the implied risk is actually the squared VIX. The realized risk is the realized variance of actual returns. To get a more accurate estimate of realized variance, we use high frequency five-minute returns. Therefore, VRP is defined as:

$$VRP_{US,Equities} = VIX^2 - \sum r^2_{S\&P500}$$

When VRP is high and positive, it means that the market is in a 'panic mode', i.e., implied risk is higher than what the risk model suggests. During those phases, investors are reluctant to hold risky assets; therefore, risky assets are depressed. We find, especially at extreme levels, market tends to overreact; therefore, we should buy risky assets at extremely high VRP levels, and vice verse for periods when VRP is low and negative.

Figure 48 plots the lagged one-month VRP and the S&P 500 total return index[15]. It appears that VRP leads the US equity market – typically, when VRP falls sharply, US equity market also plunges afterwards; vice verse, when VRP rallies, equity market also recovers subsequently. Figure 50 shows the correlation between one-month lagged VRP and forward one-month returns of the 11 asset classes. The correlation coefficients for most asset classes are highly significant. It is also clear that risky assets (e.g., equities, REITs, high yield bonds, and to a lesser extent, commodities) are likely to outperform defensive assets (e.g., investment grade sovereign bonds and to a lesser extent, gold) after VRP reaches to high levels.

[15] Please note that this graph is only to show some simple intuition. It is not for the purpose of formal statistical analysis – VRP is a stationary time series, while S&P 500 index is clearly an I(1) process.

Figure 48: VRP versus S&P 500

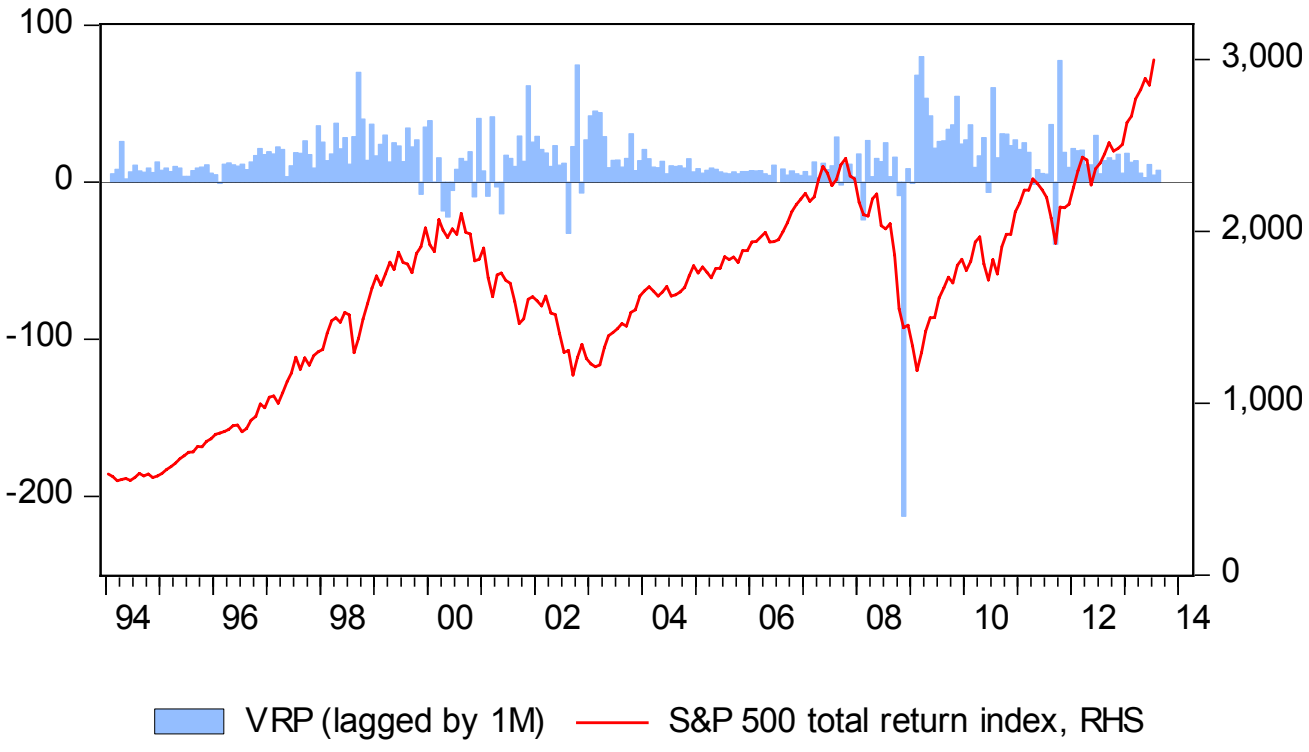

Source: Bloomberg Finance LLP, MSCI, Russell, S&P, Worldscope, Deutsche Bank Quantitative Strategy

Figure 49: One-month ahead predictive power of VRP

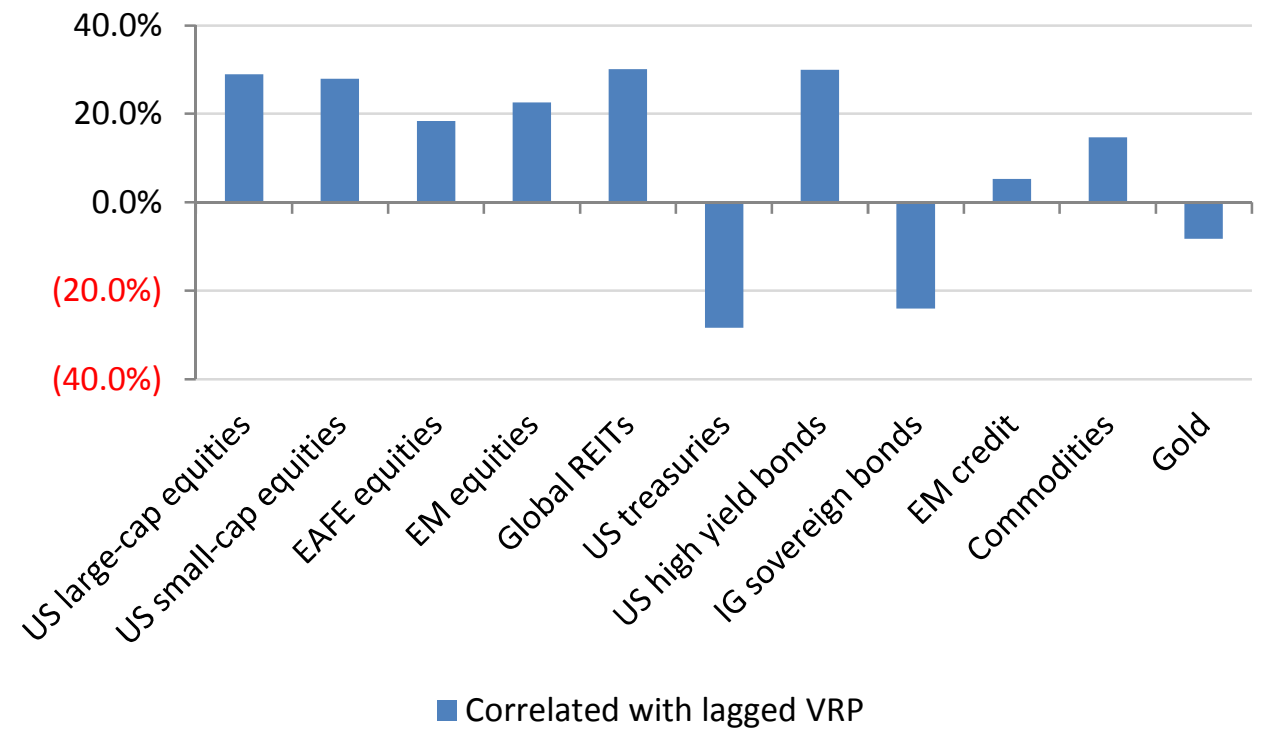

Source: Bloomberg Finance LLP, MSCI, Russell, S&P, Worldscope, Deutsche Bank Quantitative Strategy

## Forecasting asset returns with VRP

A simple forecasting model based on linear regression can be constructed:

$$r_t = \hat{\beta}_{0,t} + \hat{\beta}_{1,t} VRP_{t-1} + \varepsilon_t$$

Where,

$r_t$ is a time series of asset returns,

$\hat{\beta}_{0,t}$ and $\hat{\beta}_{1,t}$ are the OLS estimated regression intercept and slope coefficients,

$VRP_{t-1}$ is the lagged variance risk premium as of time $t-1$, and

$\varepsilon_t$ is the regression residual

We need to perform the above OLS regression for each asset at each month end. The estimated coefficients $\hat{\beta}_{0,t}$ and $\hat{\beta}_{1,t}$ can then be used in predicting asset return for time $t+1$:

$$\hat{r}_{t+1} = \hat{\beta}_{0,t} + \hat{\beta}_{1,t} VRP_t$$

To measure the performance of our GTAA model, we rely on IC or information coefficient, which is computed as the (cross-sectional) correlation coefficient between our predicted returns for each assets ($\hat{r}_{t+1}$) and actual realized returns in the subsequent period ($r_{t+1}$), at a given point of time. As shown in Figure 50, despite the recent downgrade in performance, the long-term predictive power of our simple VRP-based GTAA model is reasonable[16].

---

[16] In a later section, we will show that our global risk indicator can be used as an additional predictor of asset returns. The GTAA model based on both VRP and our risk regime indicator will be used as our final GTAA model.

Figure 50: Information coefficient (IC) of GTAA model using VRP signal

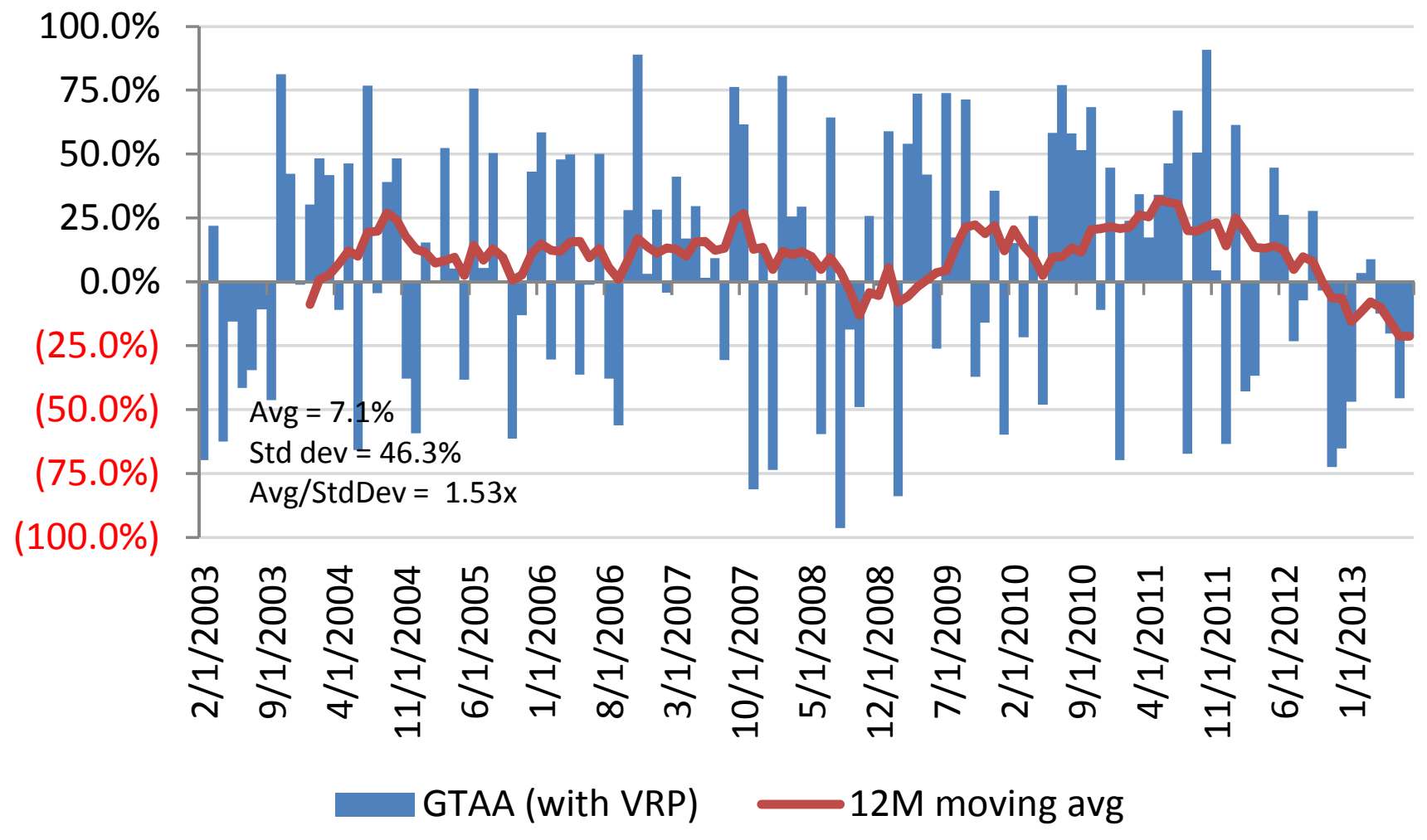

Source: Bloomberg Finance LLP, MSCI, Russell, S&P, Worldscope, Deutsche Bank Quantitative Strategy

## Portfolio construction techniques for alpha strategies

The classic portfolio construction technique with return prediction is mean-variance optimization (MVO). In particular, a simple form of MVO is the maximum Sharpe ratio portfolio (MaxSharpe) – this is the portfolio on the efficient frontier with the highest *ex ante* Sharpe ratio[17]. Of course, we don't have to pick either the GlobalMinVar or MaxSharpe portfolios – indeed, we could choose any portfolio on the efficient frontier, depending on our overall risk budget and willingness to take leverage. The MaxSharpe portfolio can be structured as:

$$\arg\max_{\omega} \left(\omega_t' \hat{r}\right) \Big/ \sqrt{\omega_t' \Sigma_t \, \omega_t}$$

Subject to:

$$\omega_t' \iota = 1$$

$$\omega_t \geq 0$$

where,

$\hat{r}_t$ is the vector of predicted asset returns at time $t$,

$\omega_t$ is the vector of asset weights at time $t$,

$\iota$ is a vector of 1's, and

[17] This is the portfolio with the highest *ex ante* or expected Sharpe ratio. Of course, there is no guarantee that this is also the portfolio with the highest *ex post* or realized Sharpe ratio. Because most return forecast models have relatively low predictive power (or high estimation error), compared to risk prediction, as we will show shortly, in most cases, this is not the portfolio with the highest *ex post* Sharpe ratio.

$\Sigma_t$ is the asset-by-asset covariance matrix at time $t$.

Figure 51 shows an example of mean-variance efficient frontier with our 11 asset classes.

Figure 51: Mean-variance efficient frontier

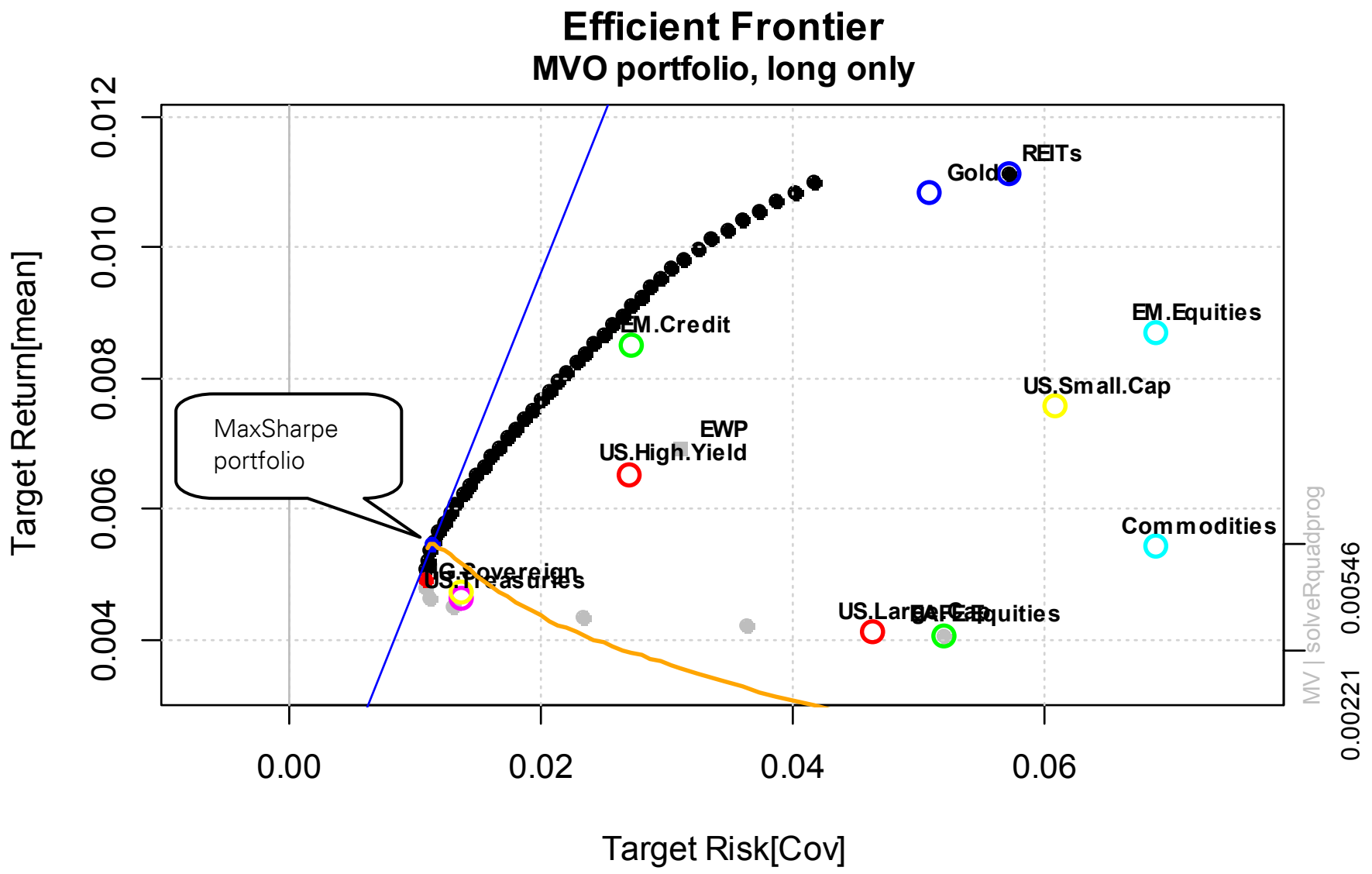

Source: Bloomberg Finance LLP, MSCI, Russell, S&P, Worldscope, Deutsche Bank Quantitative Strategy

As discussed in Luo, *et al* [2013], we have shown reasons why mean-variance framework might not be ideal, given that mean and variance can't fully account for the non-normal distribution patterns in asset returns. An interesting alternative is the so-called mean-CVaR optimization (MCVaR), where we use conditional value-at-risk as the trade-off to expected returns.

Similar to the mean-variance efficient frontier, we can also plot the mean-CVaR efficient frontier (see Figure 52 for an example). Also, similar to the MaxSharpe portfolio, we can also construct the maximum expected return/CVaR portfolio (MaxReturnCVaR). On the mean-CVaR efficient frontier, this is the portfolio with the highest *ex ante* expected return/CVaR ratio. Similar to MVO portfolios, the MinCVaR and MaxReturnCVaR are two extreme examples. We could well choose any other portfolios between these two extremes, depending on our risk tolerance and leverage constraint. We only need to make small modifications to the derivations of minimum CVaR (MinCVaR) portfolio introduced in Luo, *et al* [2013], as follows:

$$\arg\max_{\omega} = \frac{\omega' r}{\gamma + \frac{1}{(1-\alpha)S}\sum_{s=1}^{S}\left(f(\omega, r_s) - \gamma\right)^{+}}$$

Figure 52: Mean-CVaR efficient frontier

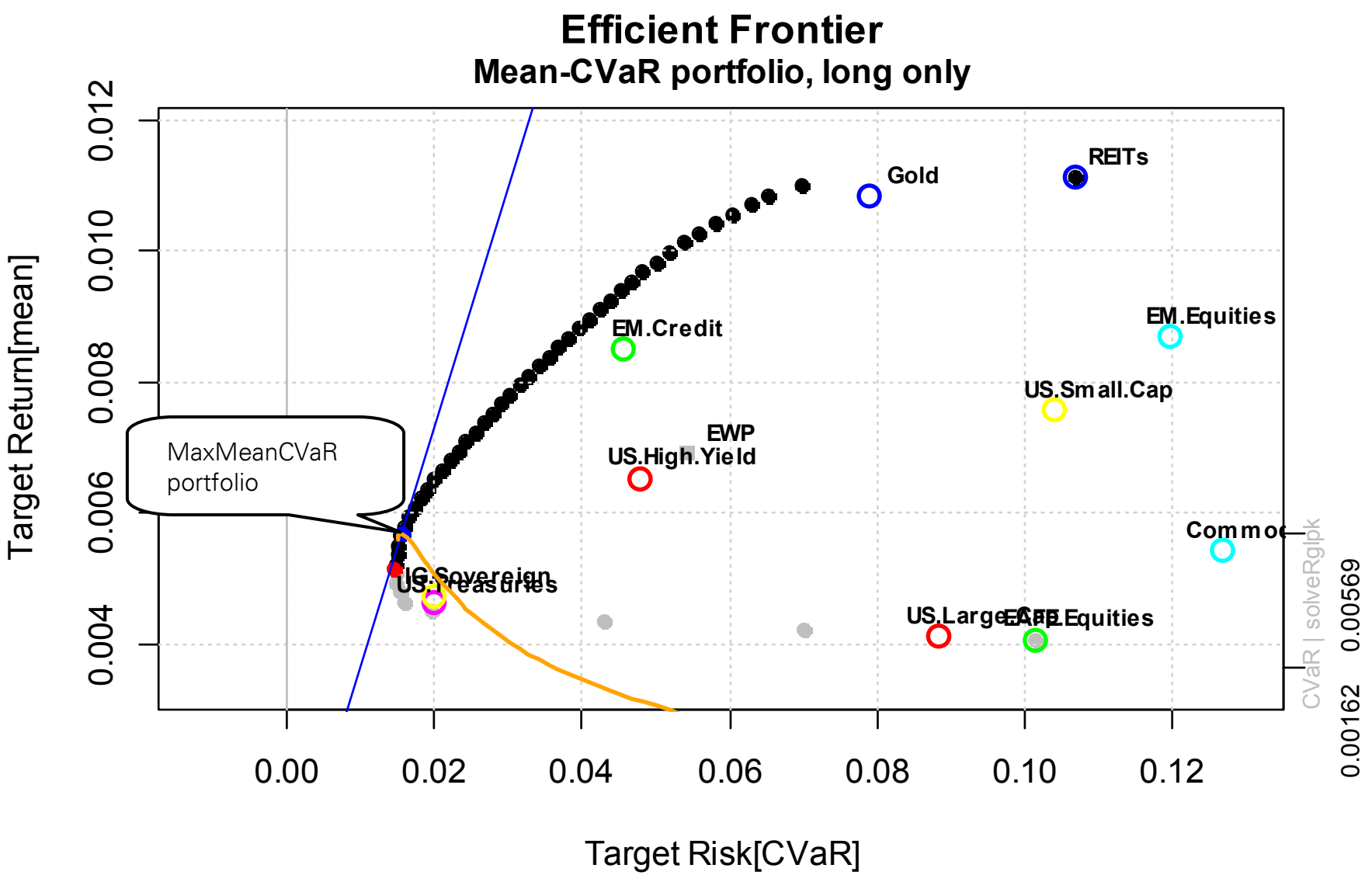

*Source: Bloomberg Finance LLP, MSCI, Russell, S&P, Worldscope, Deutsche Bank Quantitative Strategy*

## Scenario-based optimization via fractional programming

As we try to solve the MaxSharpe or MaxReturnCVaR problems, we are essentially trying to maximize a ratio. In mathematical optimization, fractional programming is one of the tools for this purpose. Fractional programming is a generalization of linear-fractional programming. The objective function in a fractional program is a ratio of two functions that are in general nonlinear. The ratio to be optimized often describes some kind of efficiency of a system. Interested readers can consult Charnes and Cooper [1962] and Stoyanov, Rachev, Fabozzi [2007] for technical details.

# V. Predicting market risk regime

Since Hamilton's seminal paper (see Hamilton [1989]) on Markov regime switching, there has been a large number of papers suggesting that nonlinear modeling can be quite useful in macroeconomic analysis with regime change (see, for example Ang and Bekaert [2002], [2004] and our own work in style rotation, Luo, *et al* [2010]).

## Theory of Markov switching model

In this paper, we do not intend to provide a thorough coverage of the theory behind regime switching models; rather, we only try to outline the main ideas and focus on applications. Interested readers please refer to Hamilton [1994].

For a random variable, $y_t$, we assume it depends on the value of an unobserved discrete state (or regime) variable, $s_t$. We further assume there are $M$ possible states (or regimes). In practice, $M$ is typically set as a small number, with two being the most common, because the numerical optimization is very difficult to converge with more than two states. The switching model further assumes that there is a different model for each regime. For given independent variables (or regressors) $X_t$ and $Z_t$, the conditional mean of $y_t$ in regime $m$ is:

$$\mu_{t,m} = X_t' \beta_m + Z_t' \gamma$$

Where $\beta_m$ and $\gamma$ are $k_X$ and $k_Z$ vectors of regression coefficients and $\beta_m$ depends on regimes, while $\gamma$ does not.

For the regression error terms, we typically assume they follow a normal distribution. Furthermore, variance may depend on regime too. Therefore,

$$y_t = \mu_{t,m} + \sigma_m \varepsilon_t$$

$\varepsilon_t$ is typically assumed as $iid$ standard normally distributed.

In Markov switching models, we assume the regime probabilities follow a first-order Markov process, meaning the probability of being in a regime $i$ depends only on the previous regime; therefore:

$$P(s_t = j \mid s_{t-1} = i) = p_{ij}(t)$$

### Dynamic regime switching model

In this paper, we also incorporate one of the latest developments in Markov switching modeling – the dynamic regime switching model. In the dynamic regime switching model, we further allow serially correlated errors, which is quite important, as we are modeling risk (measured by conditional value-at-risk) and correlation – both show strong serial correlation (see Figure 53 and Figure 54). Again, we will emphasize the practical application rather than theories – interest readers could consult Fruhwirth-Schnatter [2006] for technical details.

Figure 53: ACF of predicted CVaR

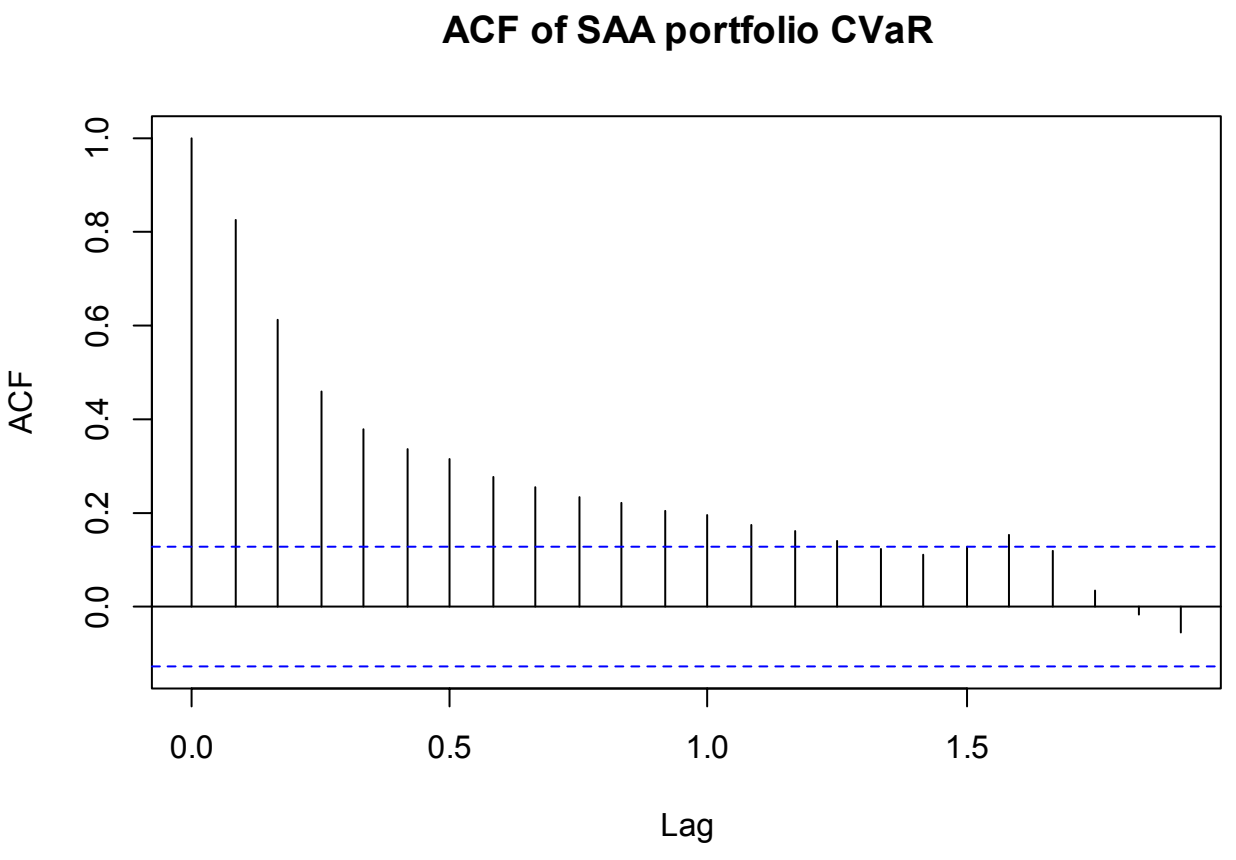

*Source: Bloomberg Finance LLP, MSCI, Russell, S&P, Worldscope, Deutsche Bank Quantitative Strategy*

Figure 54: ACF of predicted portfolio correlation

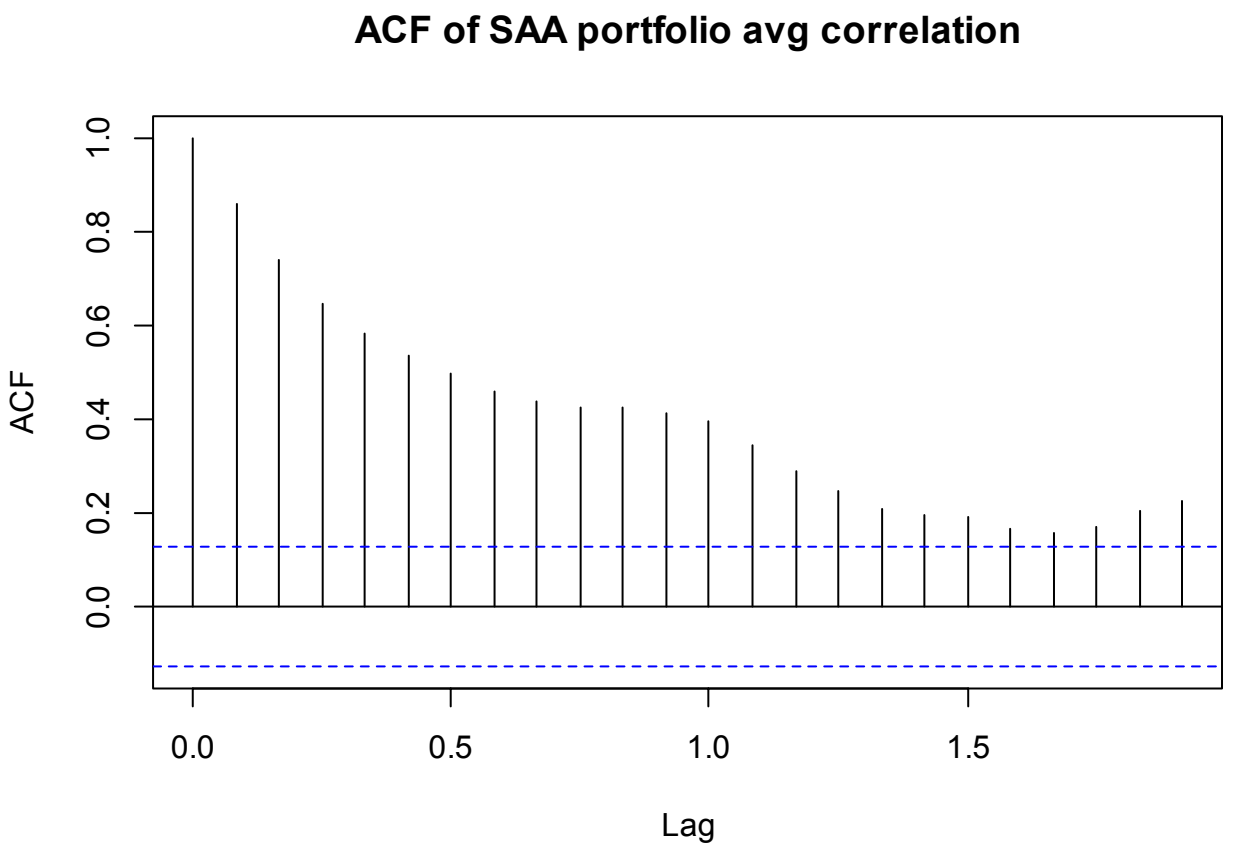

*Source: Bloomberg Finance LLP, MSCI, Russell, S&P, Worldscope, Deutsche Bank Quantitative Strategy*

### Time-varying transitions and regime heteroskedasticity

In this paper, we borrow the time-varying transition matrix approach by Kim and Nelson [1999], where the transition probability $p_{ij}(t)$ is assumed time varying, depending again on our favorite VRP indicator.

We also find Kim and Nelson's [1999] regime heteroskedasticity applies. In other words, the variance of our regression residuals also varies over time. Again, this is quite intuitive. During volatile times, the volatility of our CVaR should be higher than that during quiet times, i.e., the vol-on-vol effect.

## Global financial risk (and correlation) regime switching model

Therefore, to summarize, our regime switching model applied to the SAA portfolio risk (measured by CVaR)[18] is set up as:

$$CVaR_t = \mu_t(s_t) + \phi_1(CVaR_{t-1} - \mu_{t-1}(s_t)) + \sigma(s_t)\varepsilon_t$$

$$\mu_t(s_t) = \beta_0(s_t) + \beta_1(s_t)VRP_{t-1}$$

$$P(s_t = j \mid s_{t-1} = i) = p_{ij,t}(VRP_{t-1})$$

In this setup, VRP plays two important roles. First of all, VRP helps us to better predict risk regimes via the mean function, because the levels of VRP[19] are very different in high and low risk regimes (see Figure 55 and Figure 56). On the other hand, most of the nice fit between VRP and CVaR seems to be determined by a single data point in November 2008, where the predicted risk

[18] We also estimate a similar regime switching model for our predicted correlation.
[19] In order to use VRP in real-time prediction, we lag VRP by one month.

was all-time high, while VRP (in October 2008[20]) was also at all-time record low.

Secondly, VRP also influences the transition probability from low-risk regime to high-risk regime and vice versa. The VRP coefficient on high risk regimes is higher than that on low risk regimes, suggesting that a higher reading of VRP tends to be associated with a higher probability of high risk regime in the following month (see Figure 57).

Figure 55: SAA portfolio CVaR versus VRP

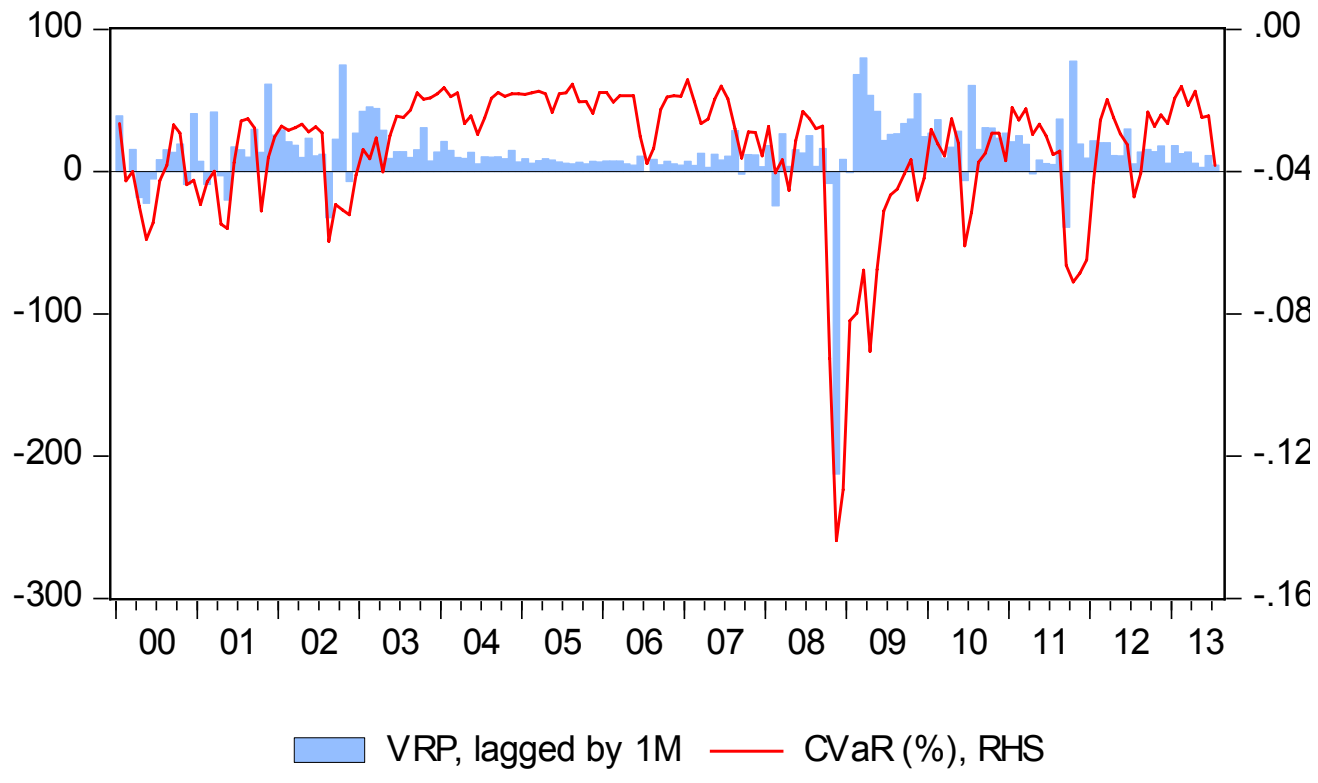

*Source: Bloomberg Finance LLP, MSCI, Russell, S&P, Worldscope, Deutsche Bank Quantitative Strategy*

Figure 56: Scatterplot of CVaR and VRP

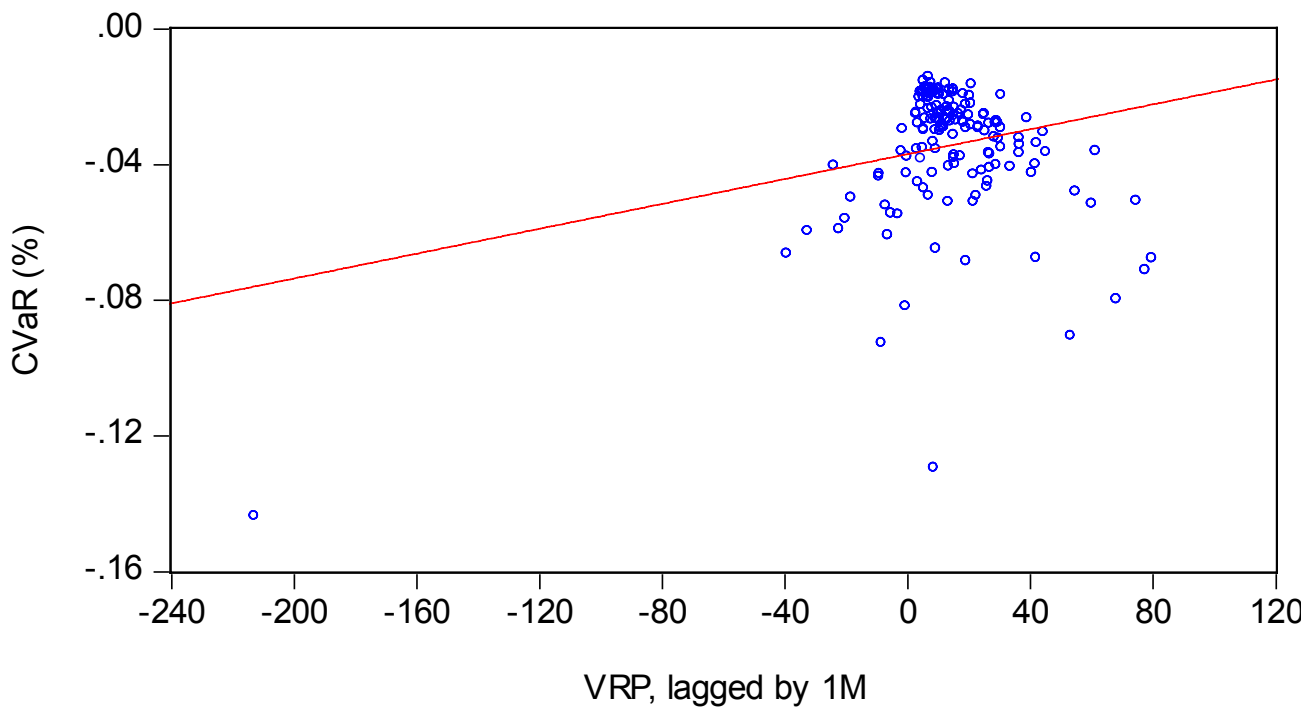

*Source: Bloomberg Finance LLP, MSCI, Russell, S&P, Worldscope, Deutsche Bank Quantitative Strategy*

[20] Remember that October-November 2008 was at the deepest of the global financial crisis.

Figure 57: Time-varying Markov transition probabilities

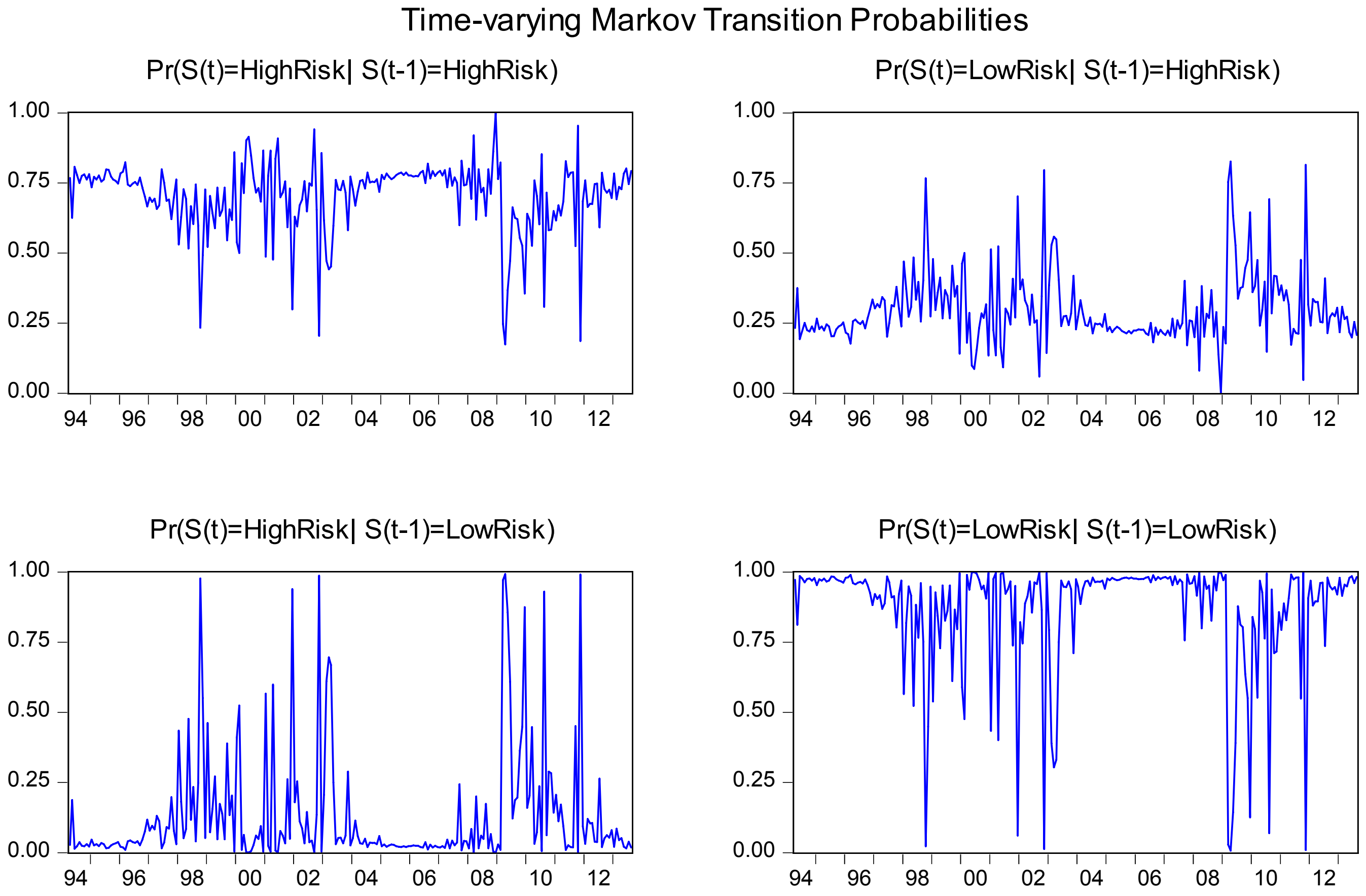

*Source: Bloomberg Finance LLP, MSCI, Russell, S&P, Worldscope, Deutsche Bank Quantitative Strategy*

## Four kinds of regime probabilities

In standard academic literature, there are three kinds of regime switching probabilities: one-step ahead prediction, filtered, and smoothed. The one-step ahead predicted probability uses information (or data) as of a given time $t$ to predict the regime probability in time $t+1$, while the filtered probability uses data as of $t$ to estimate the likelihood in time $t$: therefore, the filtered probability is more accurate than the one-step ahead prediction, but it suffers from look-ahead bias. Finally, there is also the smoothed estimate of regime probabilities, which means that we use all data in the entire sample to estimate the regime probability at time $t$. We can't use either filtered or smoothed probabilities in real-time forecast, but they are useful to understand the past.

In almost all academic papers that we have reviewed, the authors most likely use the one-step ahead prediction and claim that's the out-of-sample prediction. The reality is that it's not. The model is estimated using all data to get the model parameters, then the predictions are made using data only as of time $t$. The prediction is out-of-sample, but model parameters are not.

Therefore, in this paper, we would like to introduce the fourth kind of regime probability – true out-of-sample regime probability. In the true out-of-sample setting, we would re-estimate the Markov regime switching model every period, using data only available as of that time. Then we forecast the regime probabilities, using the model parameters and data available as of the time. We use the last period predicted regime probability as our predicted regime probability for next period. The same procedure will be repeated every period. This process ensures that we truly use data only available as of the time to make predictions. For the rest of this paper, unless it's otherwise indicated, all regime probabilities are truly out-of-sample.

The true out-of-sample probability (see Figure 58) is more volatile than the one-step ahead predicted probability (see Figure 59). As shown in Figure 63, in our example, the correlation between true out-of-sample forecast and this one-step-ahead prediction is only 68%. The difference is especially significant at the beginning of the sample, where we had less data to estimate the model. This once again highlights the importance of true out-of-sample backtesting and the impact of look-ahead bias in the investment decision process.

Figure 60 and Figure 61 show the filtered and smoothed probabilities, which appear to be more stable. As we use more and more information in determining regimes backwards, we can be more precise. Figure 62 plots all four probabilities in one chart – it's clear that, although four probabilities are highly correlated, the difference is also noticeable, especially in early samples.

Based on the predicted probability, we then further define two regimes: high risk and low risk (or high correlation and low correlation). We take a simple rule that when the probability in one regime is higher than 50%, we set it at that regime.

We find our real-time global financial risk regime indicator has considerable predictive power of asset returns and quantitative stock-selection factor returns. It can also help us to select the best portfolio construction techniques under different market environment. We will discuss the applications and implications of our risk regimes in the next few sections.

Figure 58: True out-of-sample regime probability (in high risk regime)

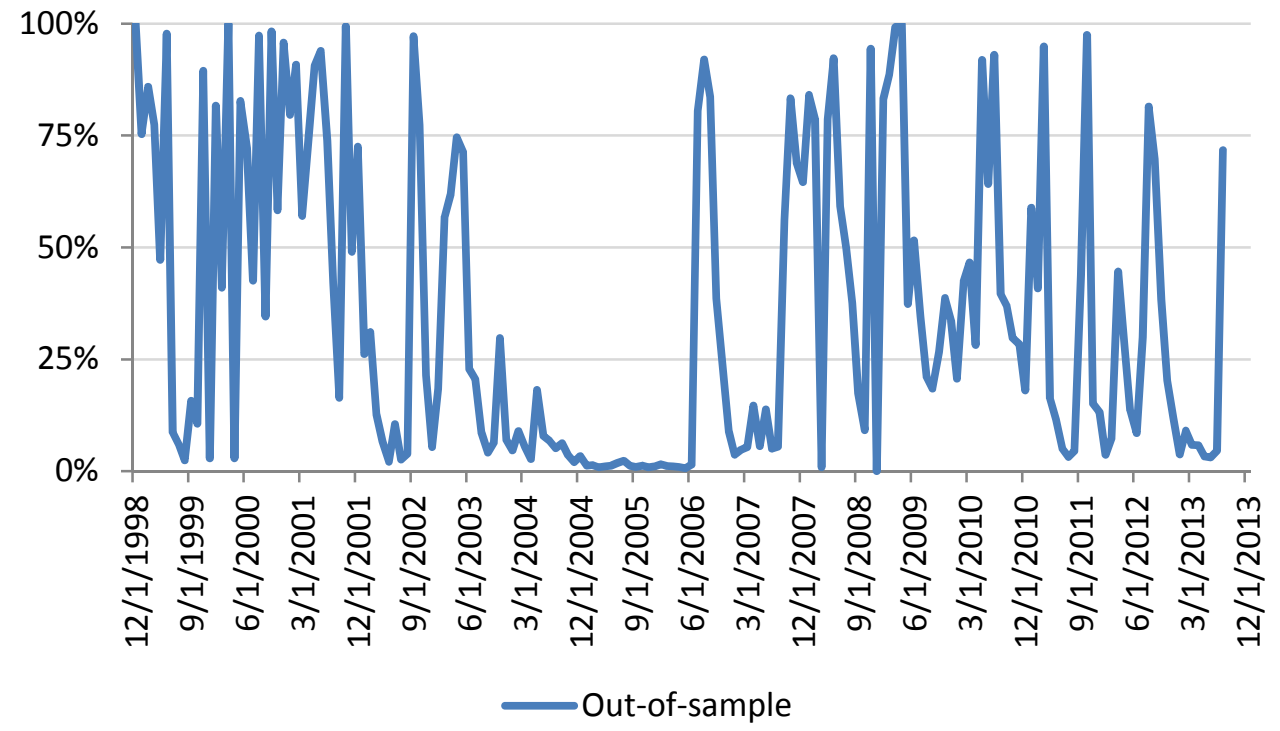

Source: Bloomberg Finance LLP, MSCI, Russell, S&P, Worldscope, Deutsche Bank Quantitative Strategy

Figure 59: One-step ahead predicted probability of high risk regime

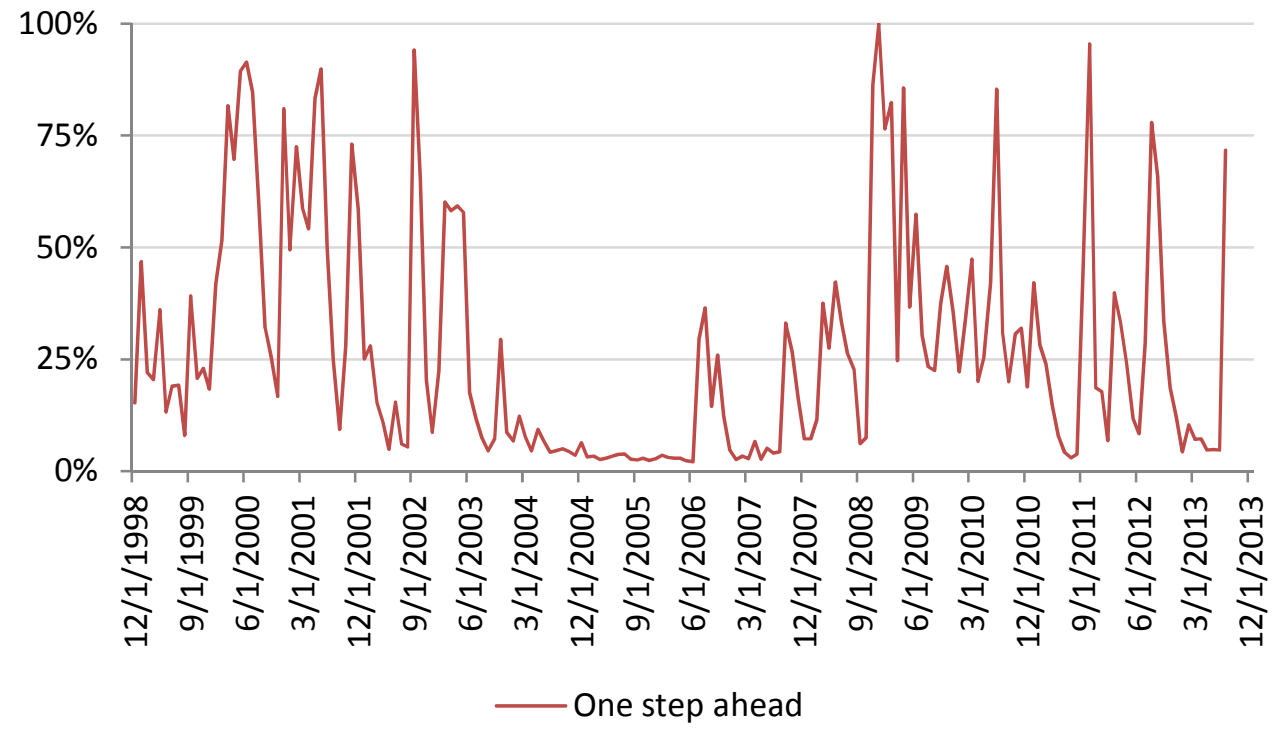

Source: Bloomberg Finance LLP, MSCI, Russell, S&P, Worldscope, Deutsche Bank Quantitative Strategy

Figure 60: Filtered probability in high risk regime

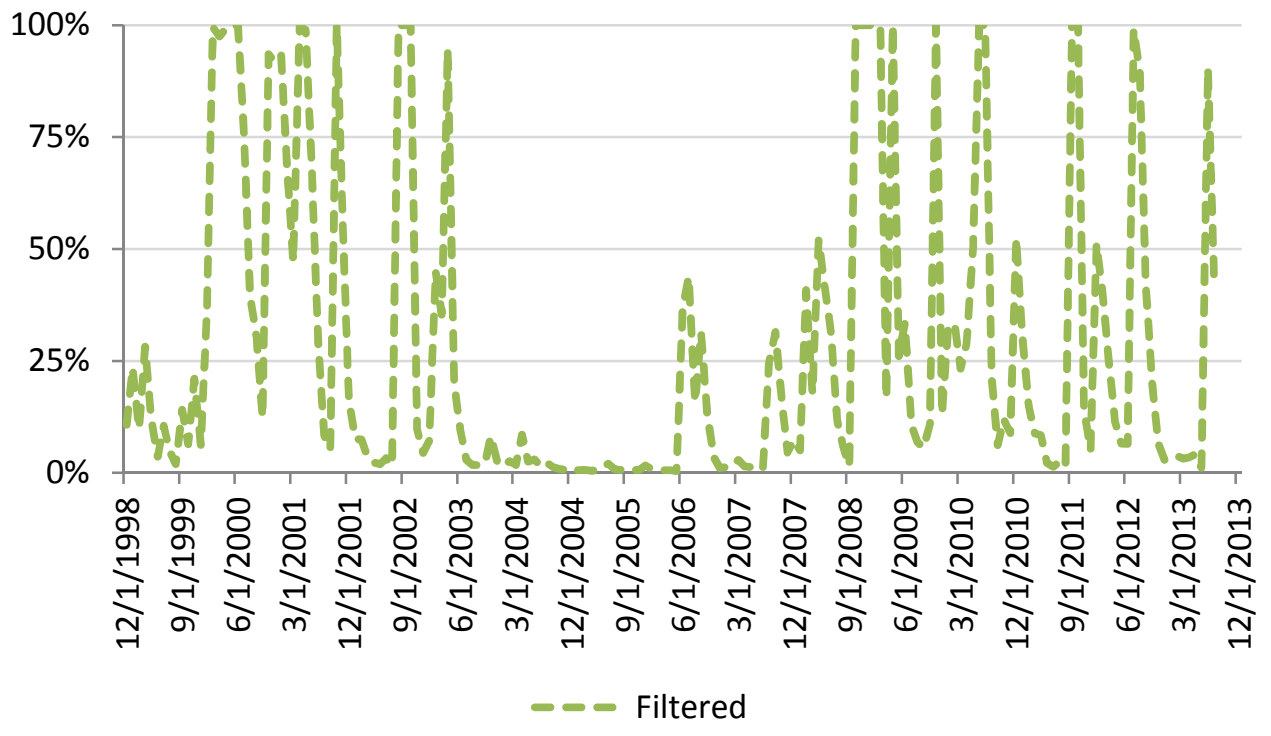

*Source: Bloomberg Finance LLP, MSCI, Russell, S&P, Worldscope, Deutsche Bank Quantitative Strategy*

Figure 61: Smoothed probability in high risk regime

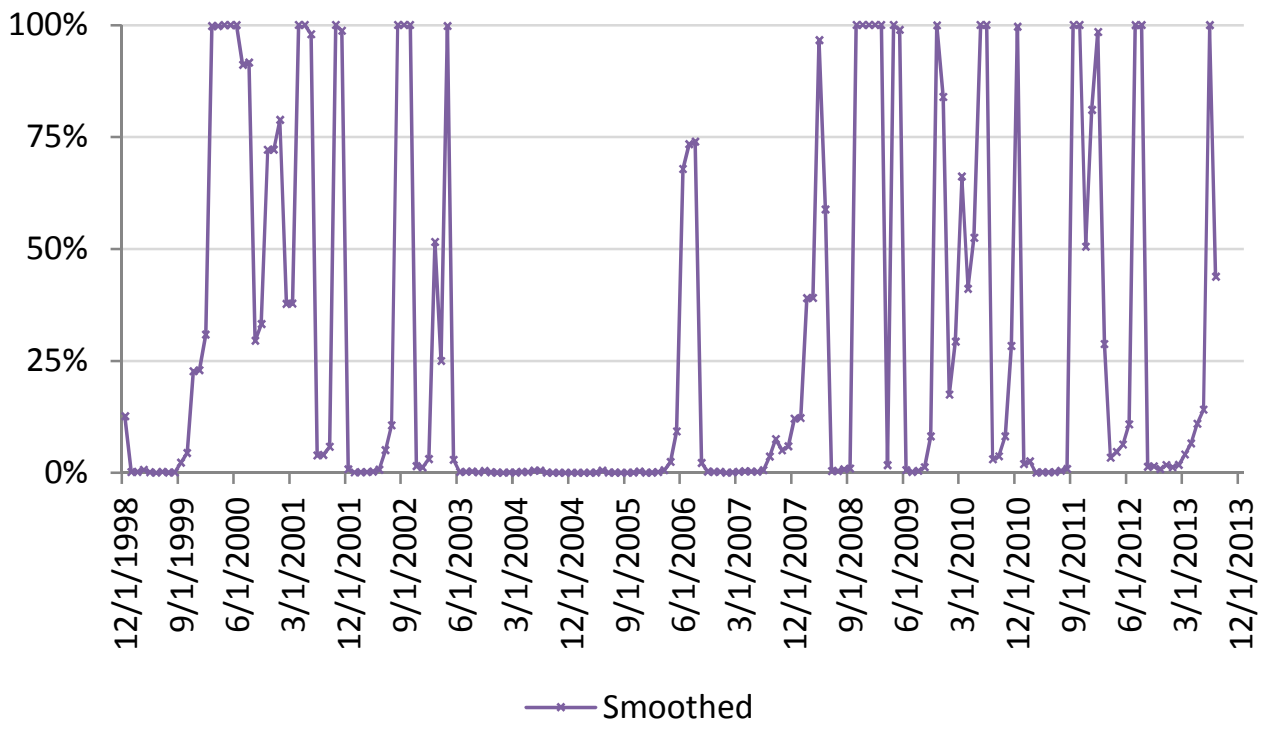

*Source: Bloomberg Finance LLP, MSCI, Russell, S&P, Worldscope, Deutsche Bank Quantitative Strategy*

Figure 62: Comparison of four different probabilities

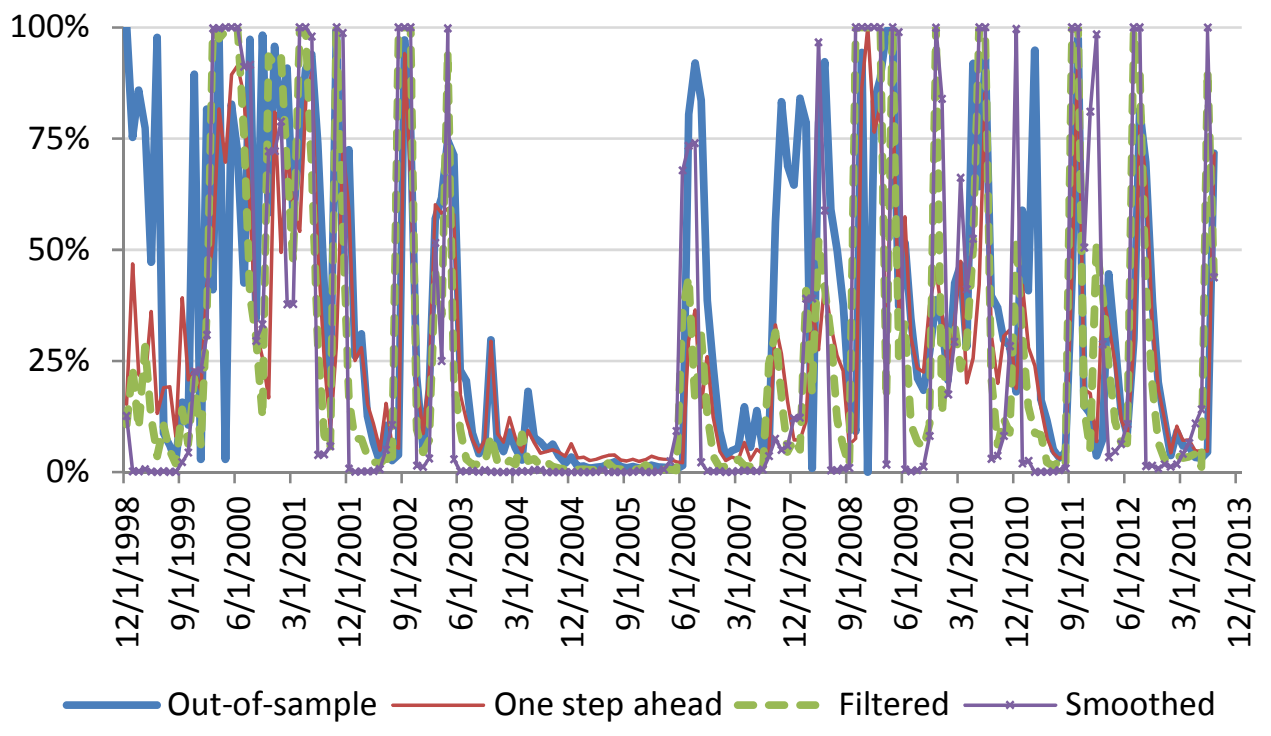

*Source: Bloomberg Finance LLP, MSCI, Russell, S&P, Worldscope, Deutsche Bank Quantitative Strategy*

Figure 63: Correlation matrix

| | Out-of-sample | One step ahead | Filtered | Smoothed |
|---|---|---|---|---|
| Out-of-sample | 100% | | | |
| One step ahead | 68% | 100% | | |
| Filtered | 56% | 76% | 100% | |
| Smoothed | 47% | 67% | 90% | 100% |

*Source: Bloomberg Finance LLP, MSCI, Russell, S&P, Worldscope, Deutsche Bank Quantitative Strategy*

## The impact of regimes on asset allocation decisions

Similar to Ang and Bekaert [2002] and [2004], we also find strong evidence that asset classes behave very differently in different regimes, which has significant implications for asset allocations decisions.

### Risk regimes

In Figure 64, we show the out-of-sample predicted risk regime (highlighted areas are high risk regimes), overlaid with predicted risk of our SAA portfolio (measured by CVaR). It's clear that the portfolio risk is much higher during those high risk regimes than during normal regimes (about 47% higher).

As our SAA portfolio is dominated more by equities[21], in predicted high risk regimes, risky assets like equities and REITs have much lower returns. On the other hand, in high risk regimes, fixed income and alternatives (e.g., commodities) tend to fare much better (see Figure 65).

[21] Remember that our SAA portfolio comprises 50% equities, 40% bonds, and 10% commodities.

Figure 64: Out-of-sample predicted risk regime and SAA portfolio risk (measured by CVaR)

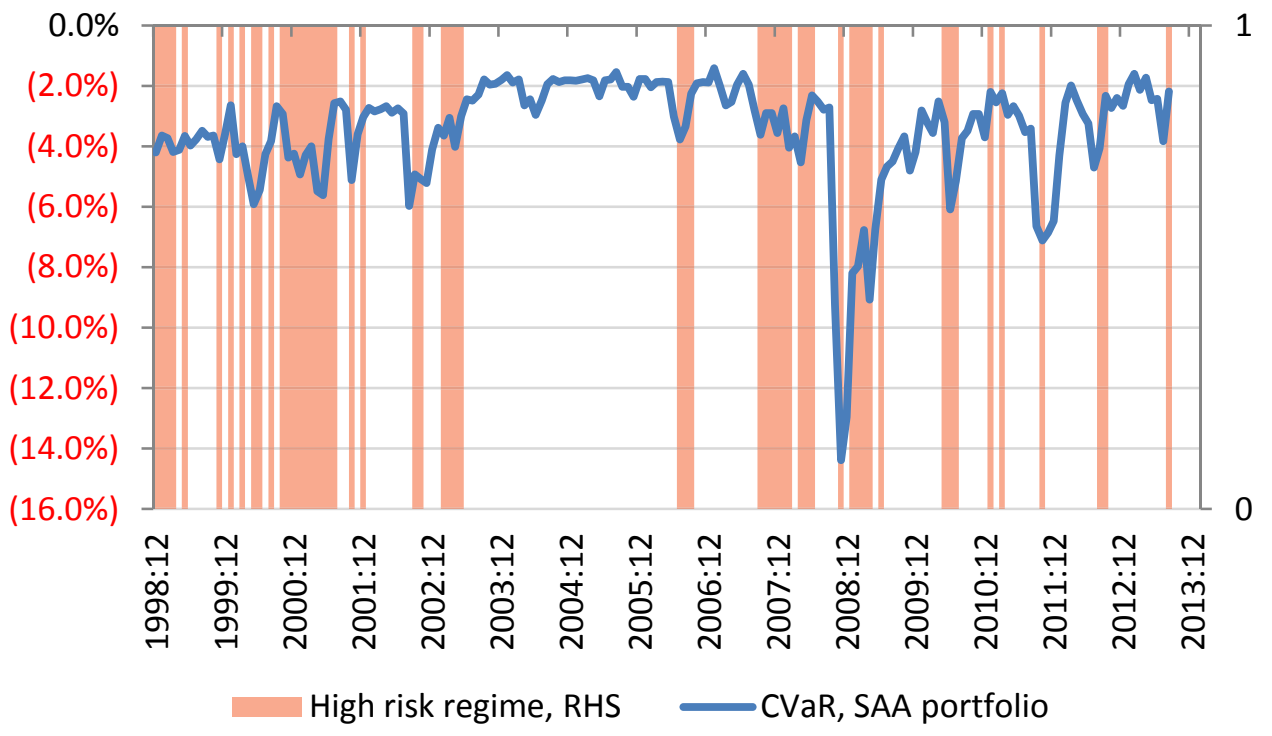

Source: Bloomberg Finance LLP, MSCI, Russell, S&P, Worldscope, Deutsche Bank Quantitative Strategy

Figure 65: Asset returns in high and low risk regimes

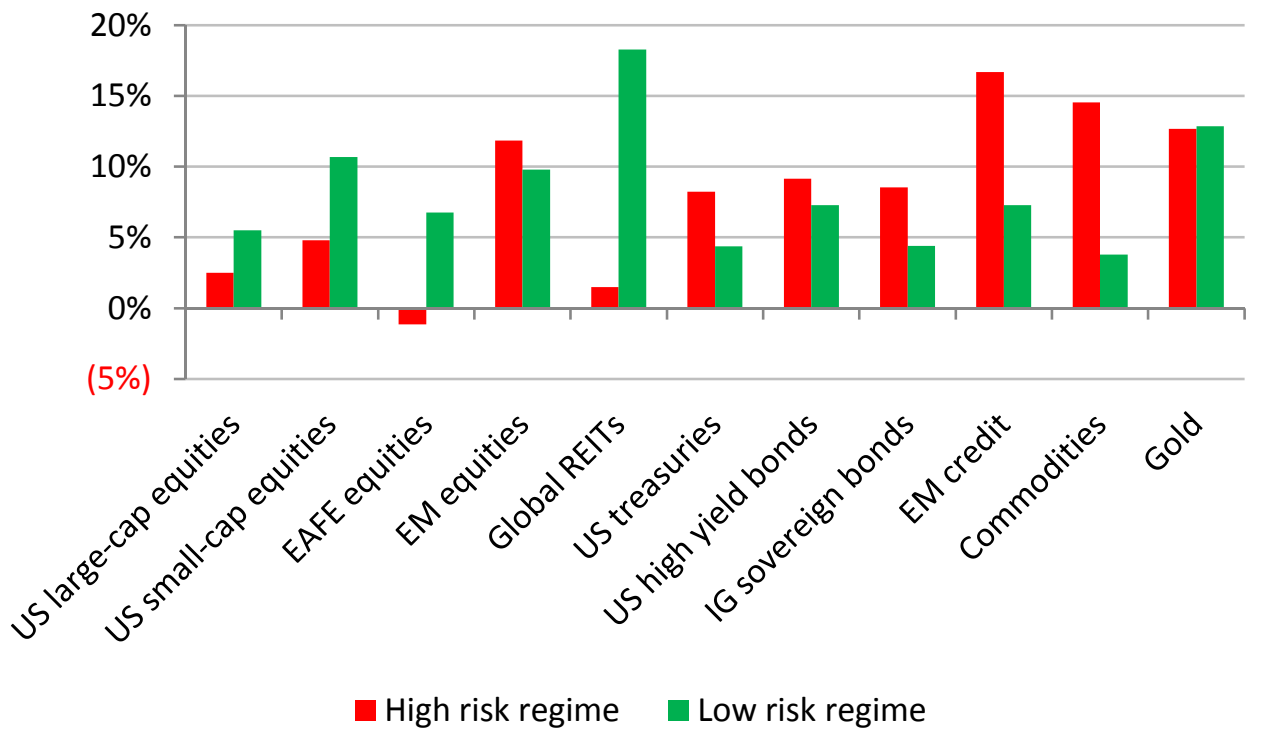

Source: Bloomberg Finance LLP, MSCI, Russell, S&P, Worldscope, Deutsche Bank Quantitative Strategy

## Correlation regimes

The cross-asset class correlation regime also appears to be time varying, but seems to be stickier than risk regimes. In fact, our Markov switching model was only able to identify two high correlation regimes since 1998 – towards the beginning of the sample around 1999 and then in recent years since 2009 until now[22] (see Figure 66).

Interestingly, risky assets (e.g., equities, high yield, and EM credit) perform far better in high cross-asset correlation regimes, while fixed income and alternatives (e.g., commodities) are more likely to benefit from the low correlation environment (see Figure 67).

Figure 66: Out-of-sample predicted correlation regime and SAA portfolio average correlation

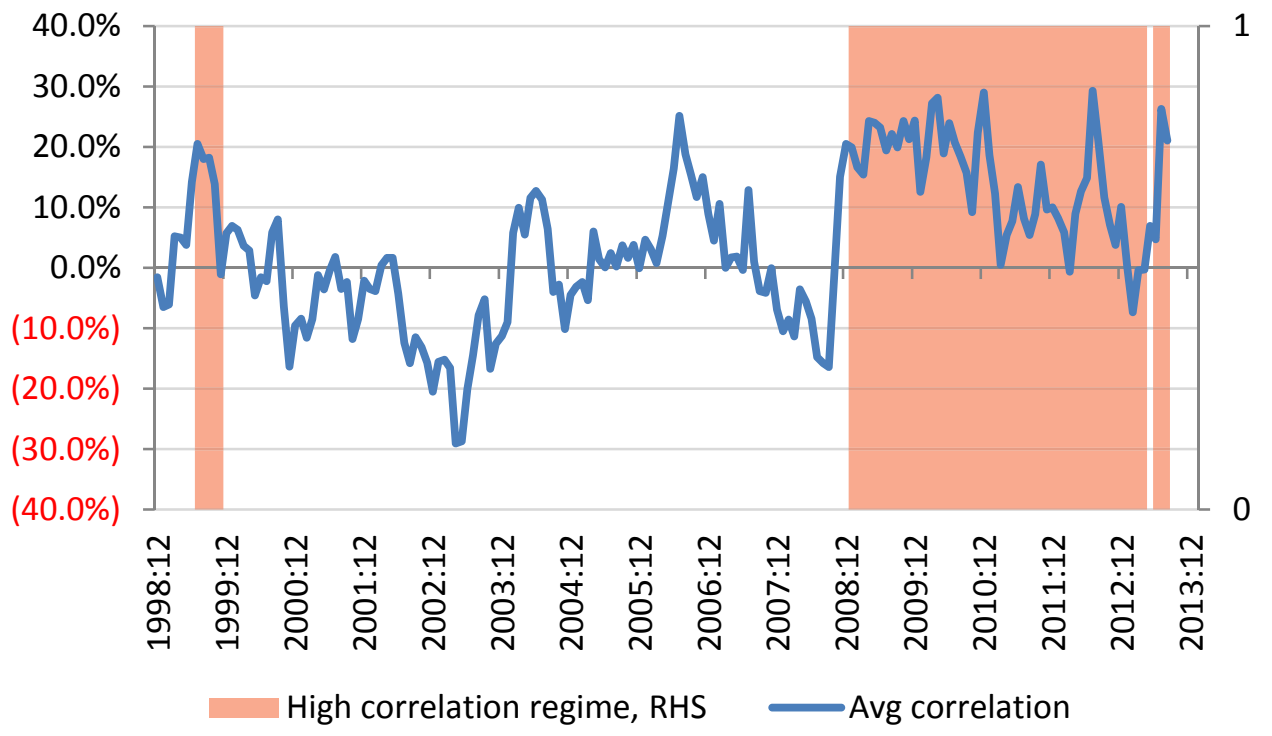

Source: Bloomberg Finance LLP, MSCI, Russell, S&P, Worldscope, Deutsche Bank Quantitative Strategy

Figure 67: Asset returns in high and low correlation regimes

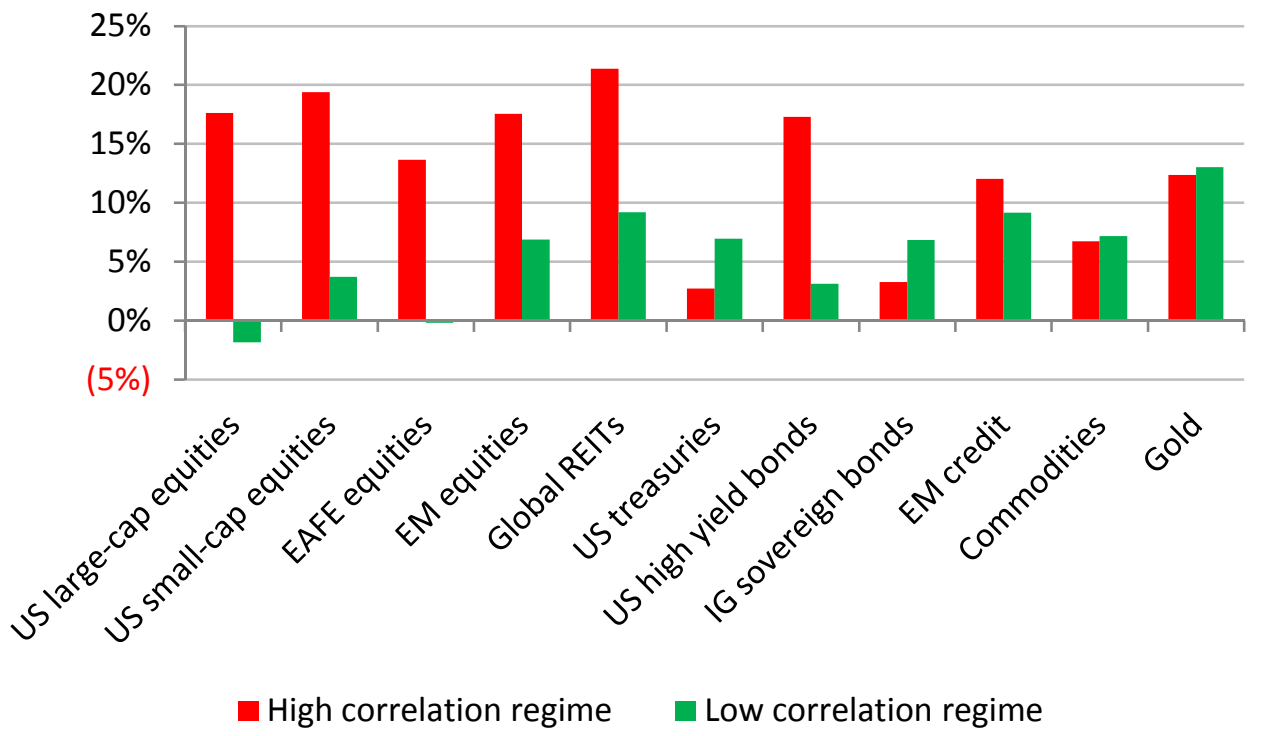

Source: Bloomberg Finance LLP, MSCI, Russell, S&P, Worldscope, Deutsche Bank Quantitative Strategy

[22] The 1999 high correlation was mostly driven by rising correlation between equities and bonds, while the post-2009 high correlation regime seems to be mostly driven by the heightened correlation between equities and commodities.

## Risk and correlation regimes

We can further combine risk regimes with correlation regimes to create a four-state model:

- High risk and high correlation
- High risk and low correlation
- Low risk and high correlation
- Low risk and high correlation

Figure 68 shows our four-state of the world in the past 15 years. The late 1990's and early 2000's Internet bubble and burst were mostly in the "high risk/low correlation" state. The era of 2002-2007 was mostly "low risk/low correlation". The global financial crisis started from 2008, followed by the European debt crisis and periodic of risk-on/risk-off. By our Markov regime switching model, this is a period of shifts between "high risk/high correlation" and "low risk/high correlation", i.e., correlation has been consistently high, while risk was switched on-and-off.

Figure 69 shows the returns of each asset classes in the above four risk/correlation regimes. Figure 70 shows the best and worst performing assets in each of the four regimes. In a "high risk/high correlation" environment, EM equities, high yield, and EM credit tend to outperform, due to their general low correlation with mainstream asset classes like US equities or US treasuries. On the other hand, in a "high risk/low correlation" regime, we may want to overweight commodities, EM credit, and gold" – alternative asset classes again seem to outperform EAFE equities, US large-cap equities, and REITs.

When risk is low but correlation is high (i.e., "low risk/high correlation"), we may want to finance our overweight in REITs, US small-cap, and US large-cap equities with US treasuries and investment grade sovereign bonds – a classic risk-on investment style. Lastly, in a "low risk/low correlation" state, REITs, gold, EM equities are more likely to outpace US large-cap equities and commodities.

Figure 68: Predicted risk and correlation regimes

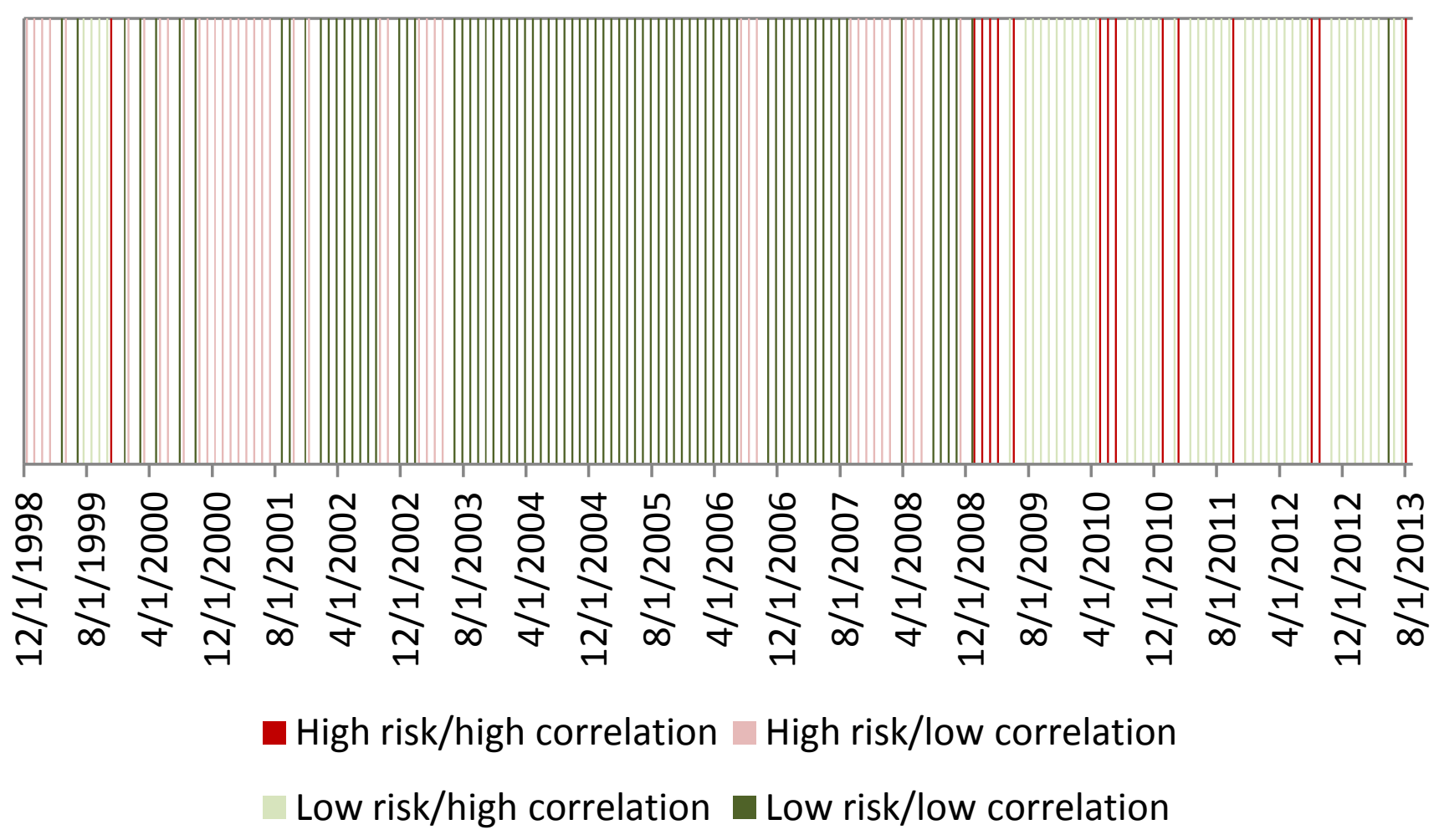

*Source: Bloomberg Finance LLP, MSCI, Russell, S&P, Worldscope, Deutsche Bank Quantitative Strategy*

Figure 69: Asset returns in different risk and correlation regimes

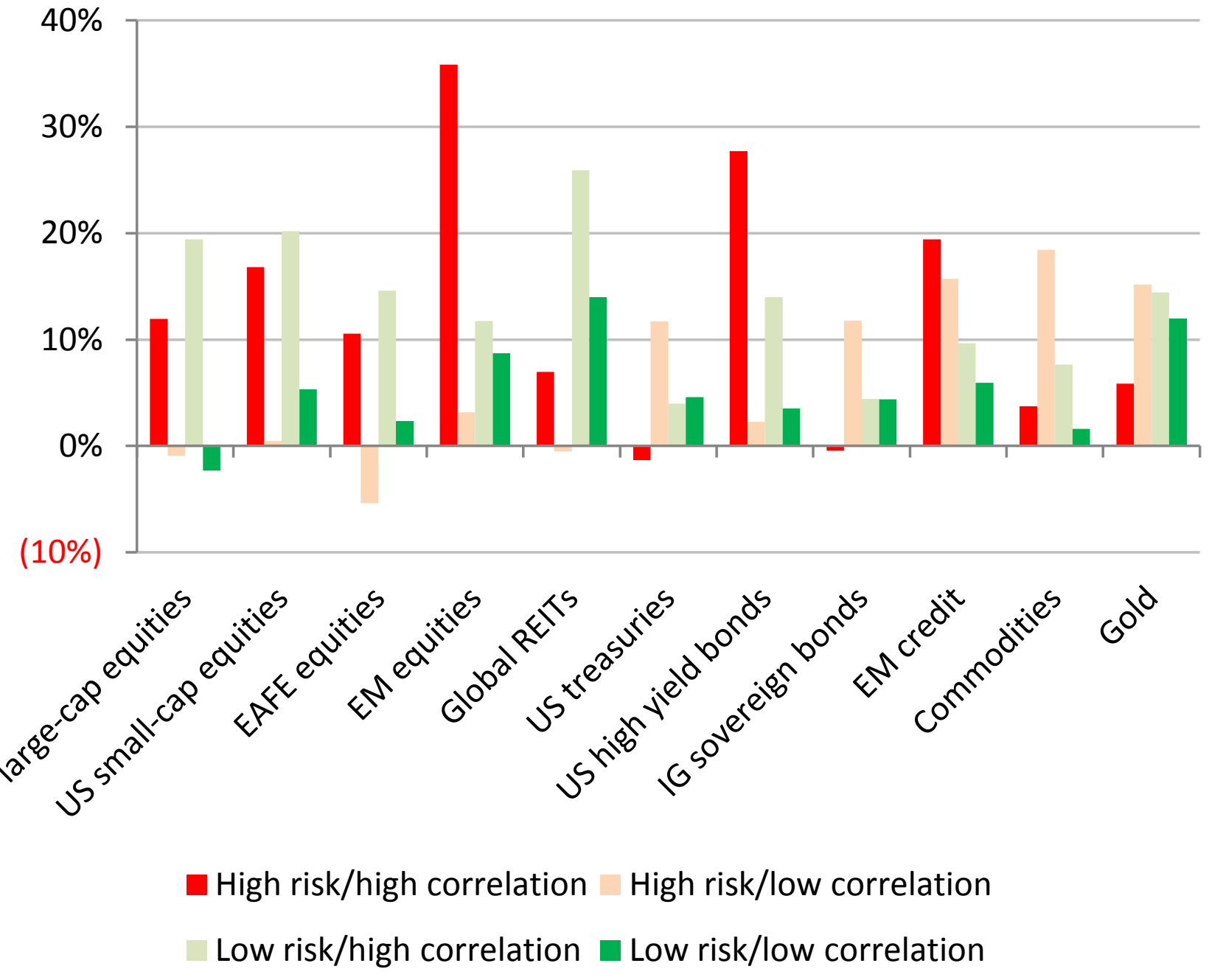

*Source: Bloomberg Finance LLP, MSCI, Russell, S&P, Worldscope, Deutsche Bank Quantitative Strategy*

Figure 70: Best performing assets in each one of the four risk and correlation regimes

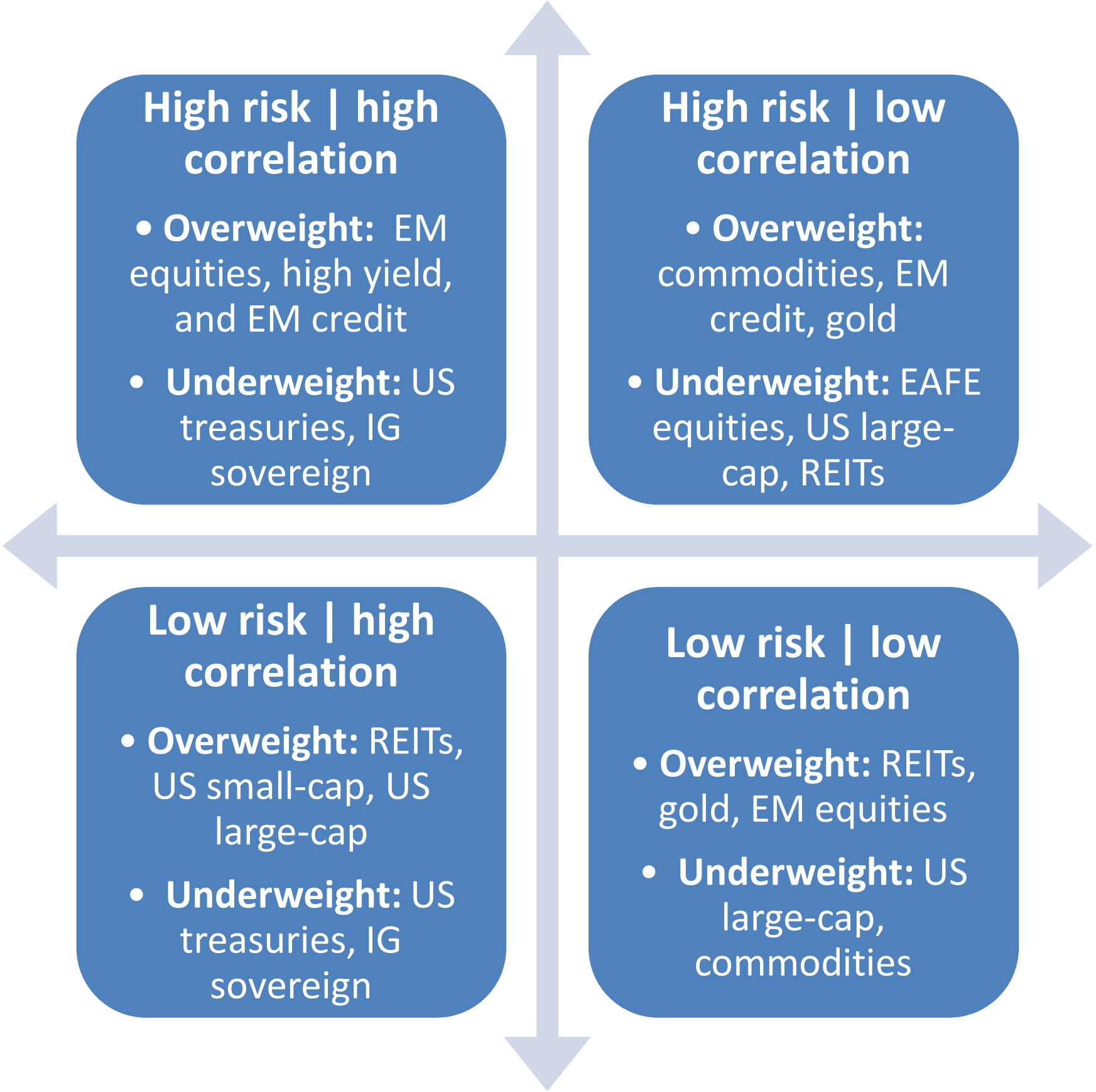

*Source: Bloomberg Finance LLP, MSCI, Russell, S&P, Worldscope, Deutsche Bank Quantitative Strategy*

## An example of how to use risk and correlation regime prediction in asset allocation

Below we show two simple regime-based dynamic asset allocation strategies[23]:

- **Long only:** in each of the four risk/correlation regimes, we long the top performing assets shown in Figure 70, equally weighted, without leverage.
- **Long/short dollar neutral portfolio:** in each of the four risk/correlation regimes, we long the top performing assets, while short the bottom performing assets (see Figure 70) – all equally weighted.

As shown in Figure 71 and Figure 72, both strategies do deliver reasonable risk-adjusted returns.

[23] Please note that both investment strategies suffer from look-ahead bias. This is because we use the same data to identify the best and worst performing assets, and then we use the same data to backtest the performance of the strategy. We simply want to use these two examples to show the potential upside of regime switching models. In the next section, we show a real investable strategy without look-ahead bias.

Figure 71: Cumulative performance of regime-dependent strategies

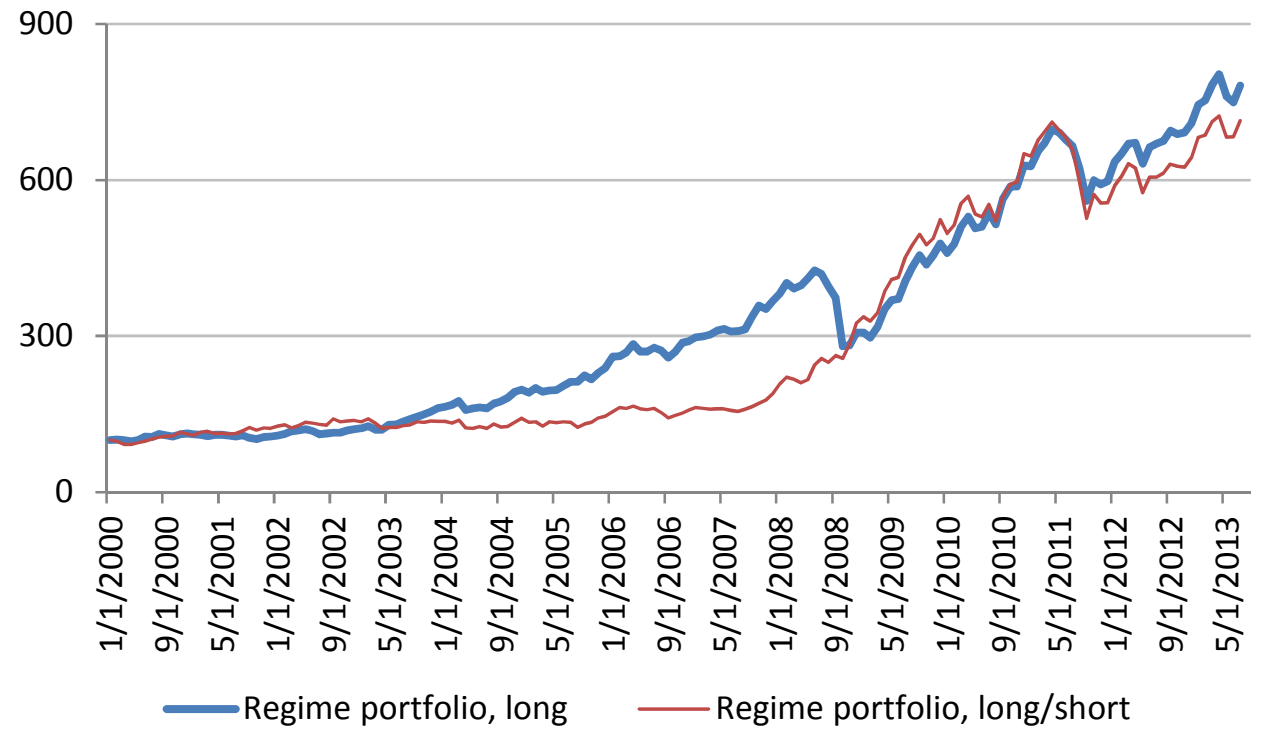

Source: Bloomberg Finance LLP, MSCI, Russell, S&P, Worldscope, Deutsche Bank Quantitative Strategy

Figure 72: Return, risk, and Sharpe ratio

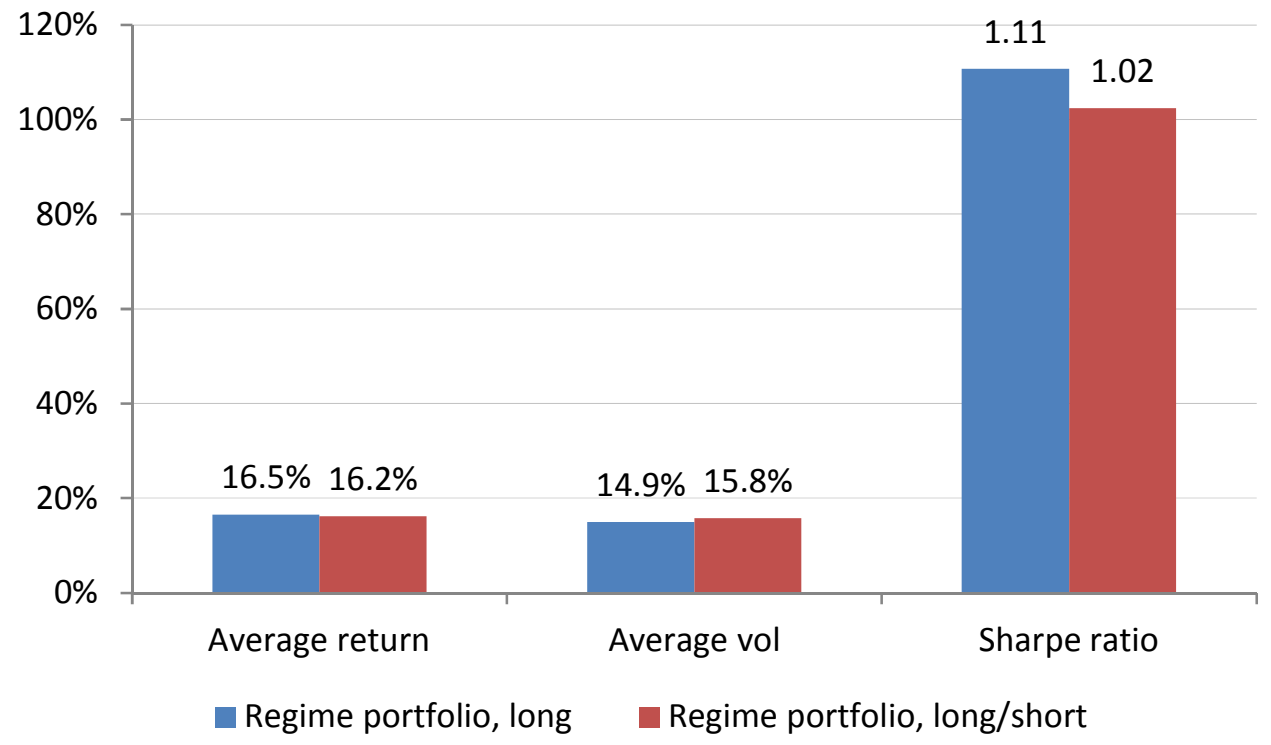

Source: Bloomberg Finance LLP, MSCI, Russell, S&P, Worldscope, Deutsche Bank Quantitative Strategy

## Using regimes to predict asset returns

Our regime switching model does appear to differentiate asset returns intuitively well. The next question is whether we can use this information in systematic investing. The answer turns out to be yes. In this section, we illustrate how to use our predicted real-time risk regimes to enhance our GTAA model presented in the previous section. We first take the simple GTAA model conditional on VRP (as described in the previous sections), then we add our pure out-of-sample predicted high risk/low risk regime indicator as an additional variable:

$$r_t = \hat{\beta}_{0,t} + \hat{\beta}_{1,t} VRP_{t-1} + \hat{\beta}_{2,t} I(P(s_t = 1)) + \varepsilon_t$$

Where,

$P(s_t = 1)$ is the predicted probability in regime $1$, i.e., high risk regime, and

$I(\bullet)$ is an indicator function that takes a value of either $1$ (if $P(s_t = 1) \geq 0.5$) or $0$ (if $P(s_t = 1) < 0.5$).

The predicted return for period $t+1$ is therefore:

$$\hat{r}_{t+1} = \hat{\beta}_{0,t} + \hat{\beta}_{1,t} VRP_t + \hat{\beta}_{2,t} I(P(s_{t+1} = 1))$$

To measure the accuracy of our prediction, we use the cross-sectional information coefficient (IC) metric. At each month end, we calculate the correlation between our predicted returns of the 11 assets (at the previous month end) and the actual realized returns of the 11 assets during the current month.

Figure 73 and Figure 74 show the time series of IC for the GTAA model without and with our regime switching indicator. The GTAA model with our regime switching indicator can improve the average performance (IC) by over 40% and the risk-adjusted performance ($\mu(IC)/\sigma(IC)$) by over 50%.

Figure 73: GTAA model using VRP only

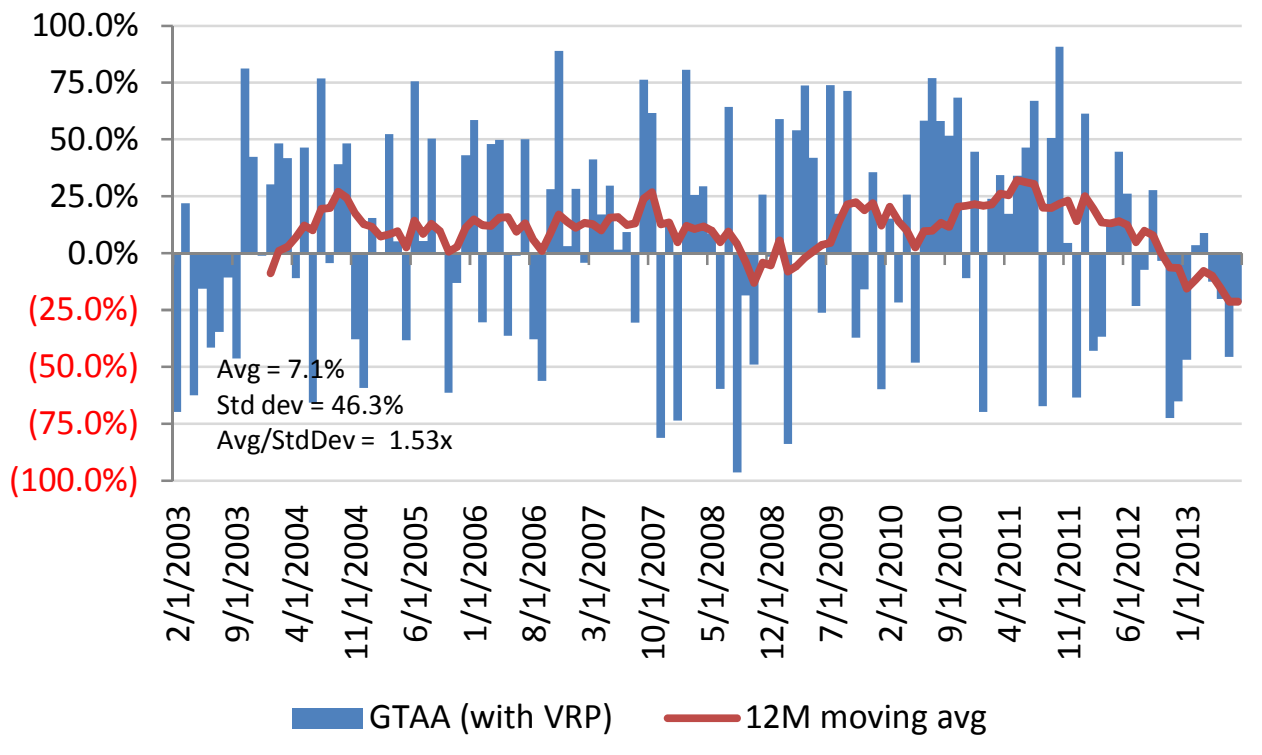

*Source: Bloomberg Finance LLP, MSCI, Russell, S&P, Worldscope, Deutsche Bank Quantitative Strategy*

Figure 74: GTAA model using VRP and regimes

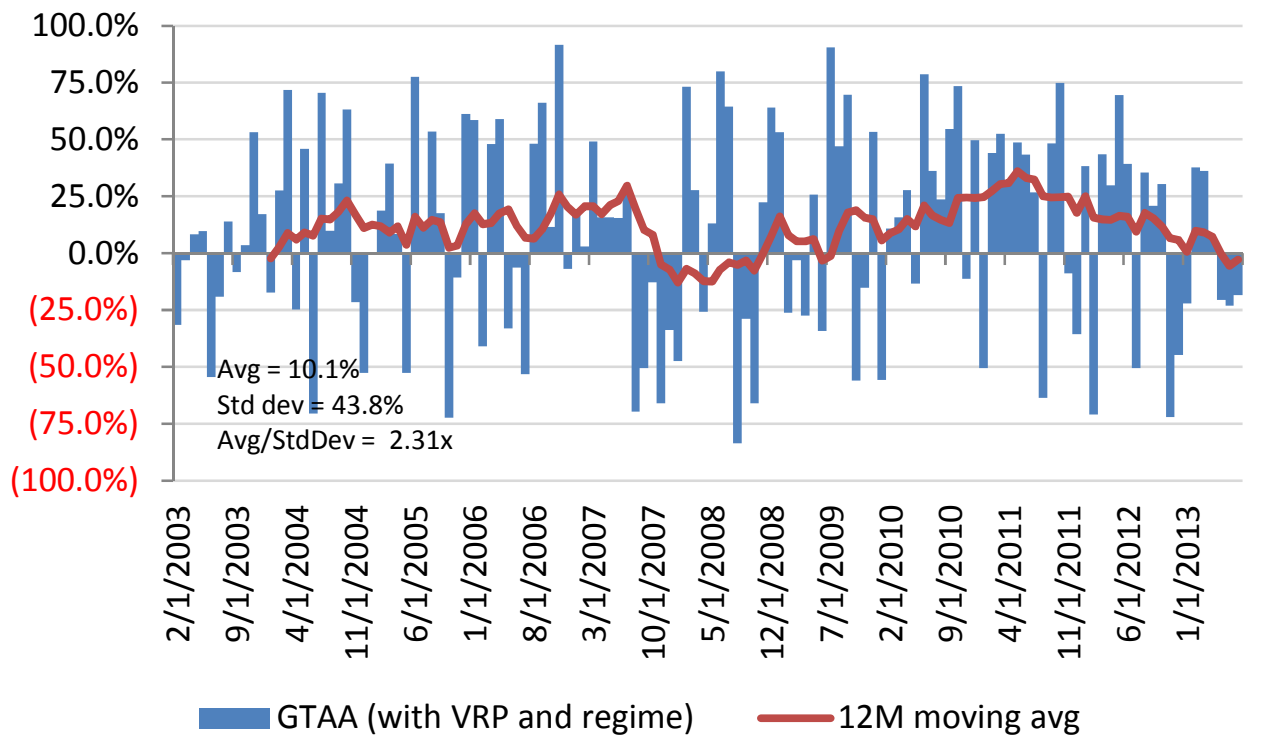

*Source: Bloomberg Finance LLP, MSCI, Russell, S&P, Worldscope, Deutsche Bank Quantitative Strategy*

## Performance summary

In this section, we can combine our alpha prediction and risk models to build active asset allocation strategies (see Figure 75 to Figure 80).

- Our GARCH-DCC-Copula risk model consistently outperforms the sample-based risk models, for our alpha-based strategies, showing higher Sharpe ratios, lower downside risk, better diversification, and lower crowding (tail dependence).
- The performance of MaxSharpe and MaxReturnCVaR portfolios is comparable.
- The three return prediction models (one-year rolling average returns, long-term average returns, and GTAA model) have comparable performance[24].

[24] Please note that the long-term alpha model suffers from look-ahead bias, as we use the entire history to calculate the predicted returns of each asset class. In addition, a good alpha model does not always guarantee an outperforming portfolio. In the next series of our *DB Handbook of Portfolio Construction, Part 3*, we will discuss various ways to incorporate alpha forecasts in portfolio construction.

Figure 75: Realized return comparison

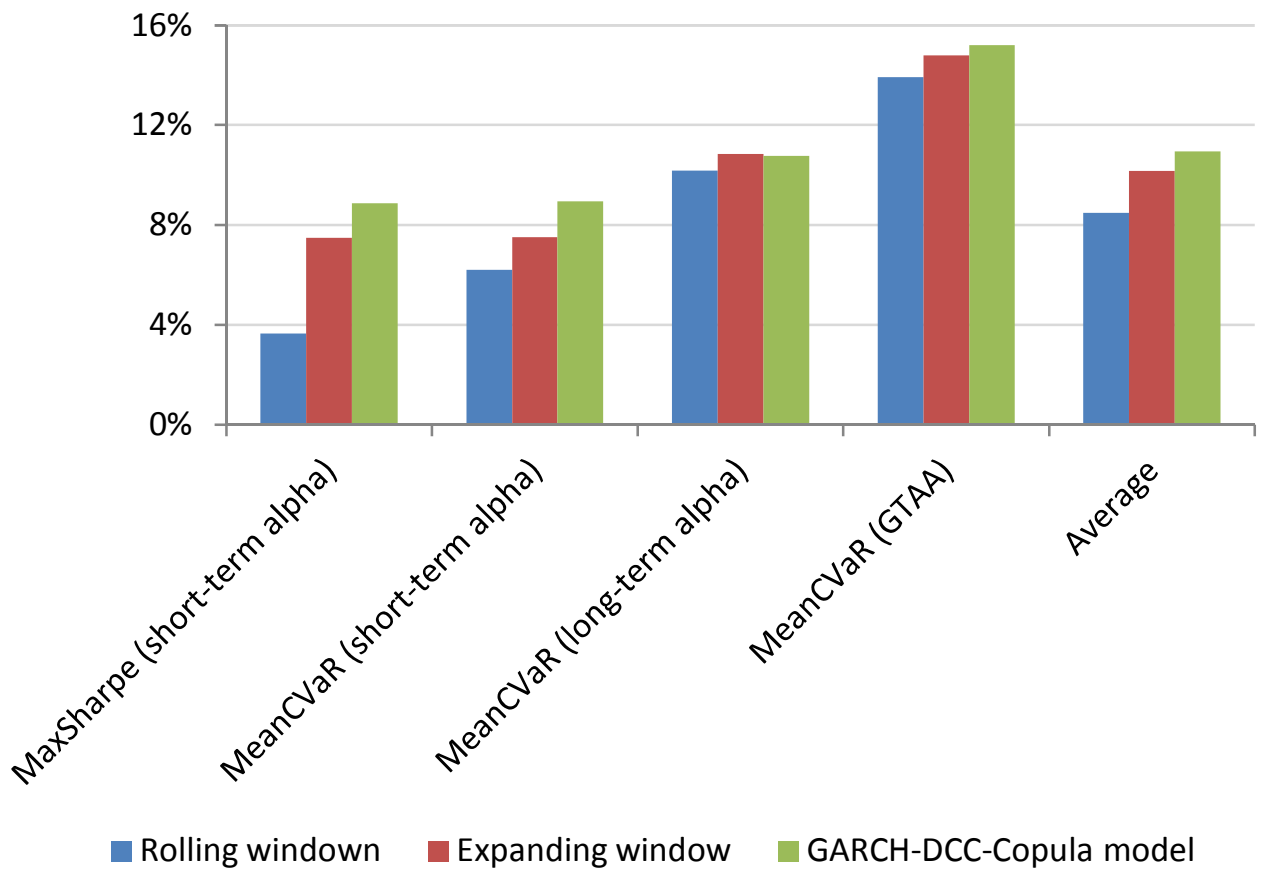

*Source: Bloomberg Finance LLP, MSCI, Russell, S&P, Worldscope, Deutsche Bank Quantitative Strategy*

Figure 76: Realized volatility comparison

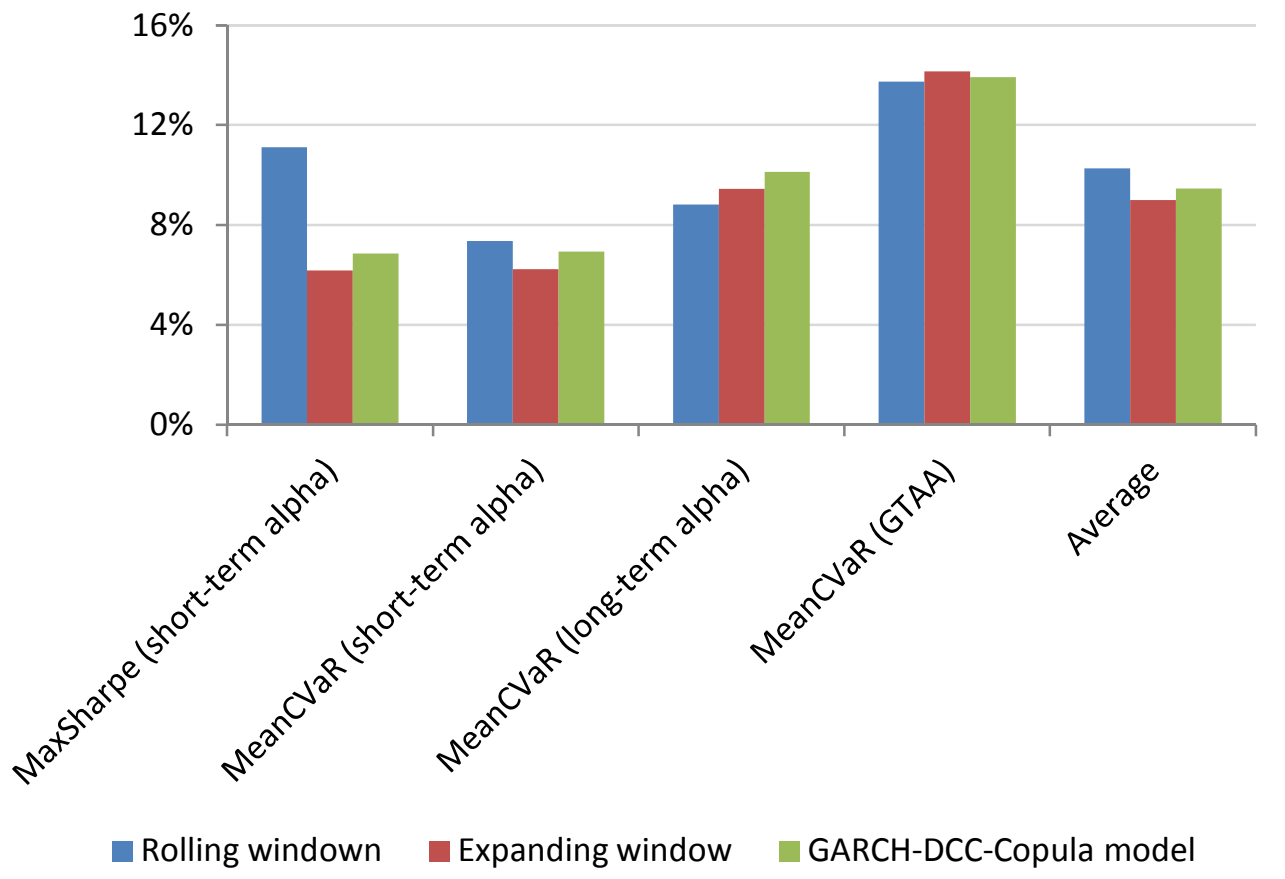

*Source: Bloomberg Finance LLP, MSCI, Russell, S&P, Worldscope, Deutsche Bank Quantitative Strategy*

Figure 77: Realized Sharpe ratio comparison

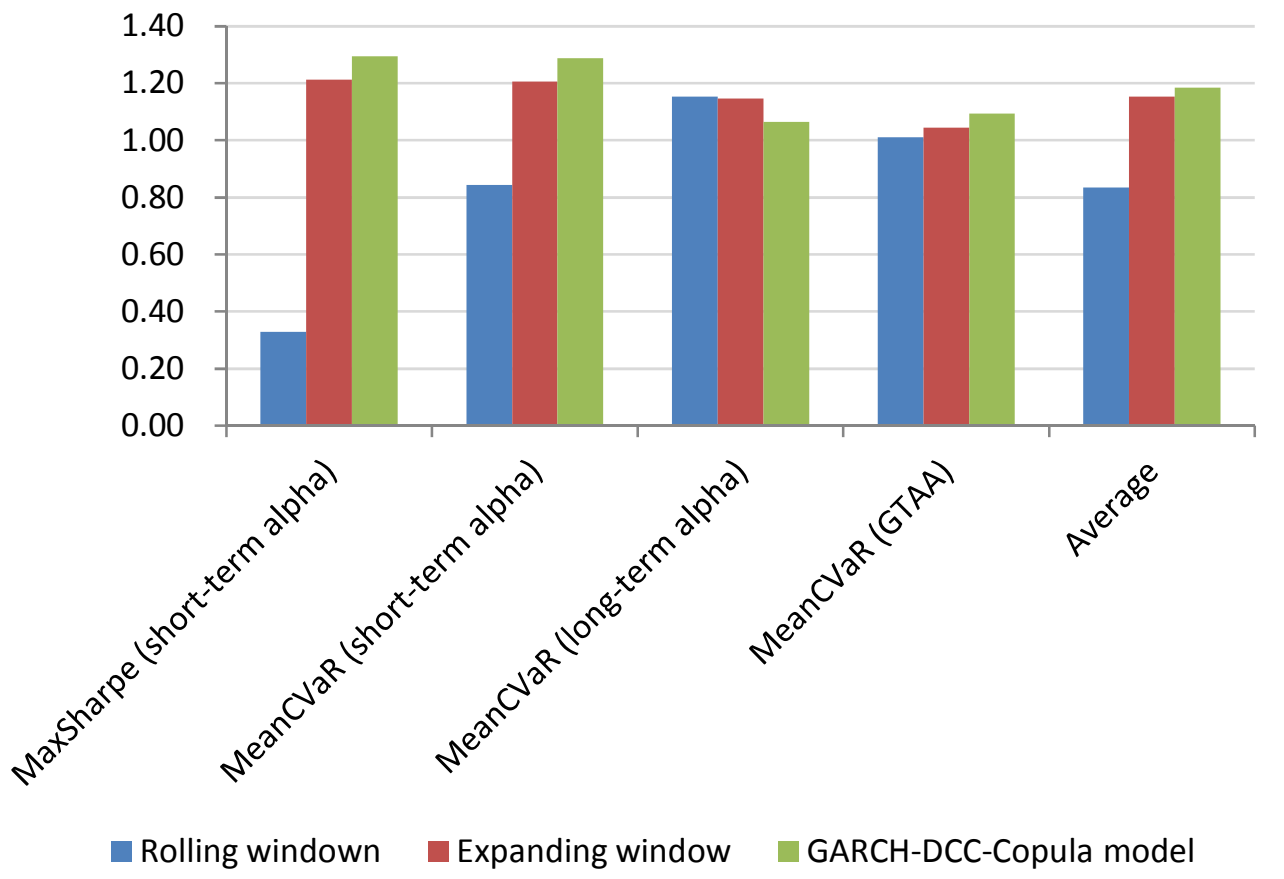

*Source: Bloomberg Finance LLP, MSCI, Russell, S&P, Worldscope, Deutsche Bank Quantitative Strategy*

Figure 78: Realized downside risk, CVaR, comparison

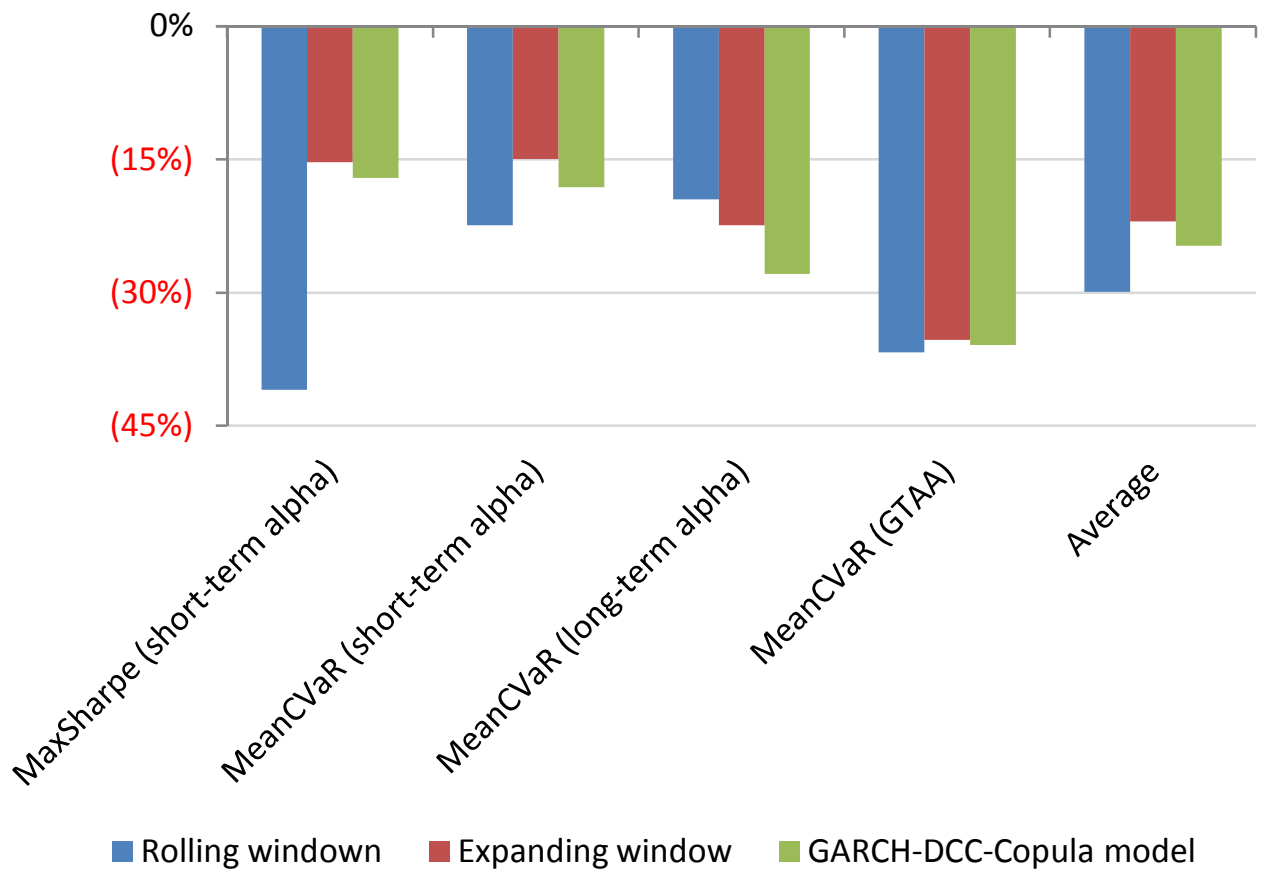

*Source: Bloomberg Finance LLP, MSCI, Russell, S&P, Worldscope, Deutsche Bank Quantitative Strategy*

Figure 79: Portfolio diversification ratio

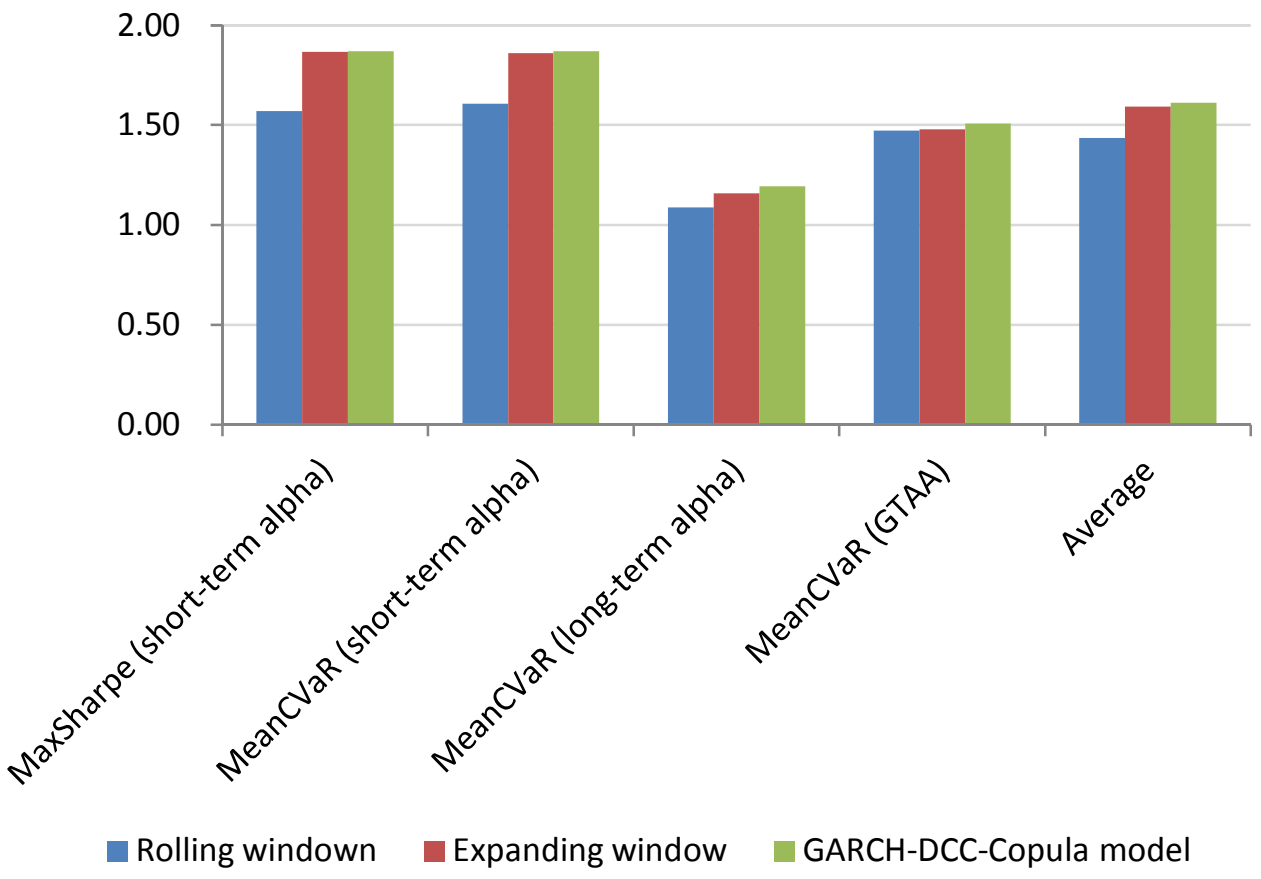

*Source: Bloomberg Finance LLP, MSCI, Russell, S&P, Worldscope, Deutsche Bank Quantitative Strategy*

Figure 80: Weighted portfolio tail dependence

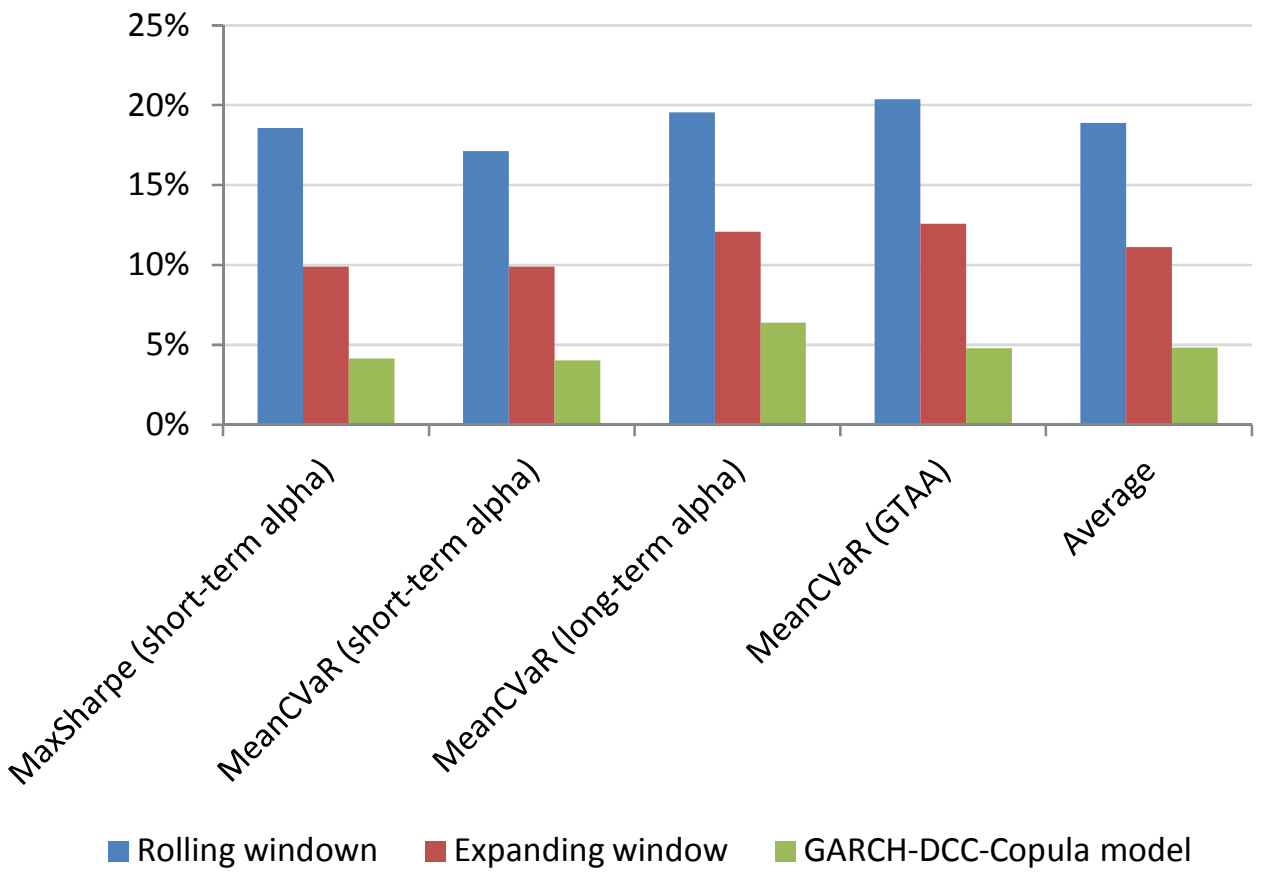

*Source: Bloomberg Finance LLP, MSCI, Russell, S&P, Worldscope, Deutsche Bank Quantitative Strategy*

## How are all these portfolio construction techniques related?

As shown in Figure 81 and Figure 82, cluster analysis reveals that our portfolio construction techniques can be roughly broken down into the following four categories[25]:

- Naïve benchmark: traditional 60-40 benchmark and simple EquallyWgted portfolio
- Diversification based: InvVol, RiskParity, MaxDiversification, and MinTailDependence fall into this bucket. Interestingly, our two short-term alpha strategies using rolling one-year returns (MaxSharpe and MaxMeanCVaR) also seem to belong to this category, possibly due to the low predictive power of naïve alphas.
- Risk reduction based: MinVar, MinCVaR, MinTailVar, and surprisingly, naïve alpha based on long-term average returns also falls into this category
- Alpha based: our VRP and regime based GTAA model using MaxMeanCVaR optimization

[25] Details on how to interpret the cluster analysis graphs can be found in Luo, et al [2013].

Figure 81: Cluster analysis

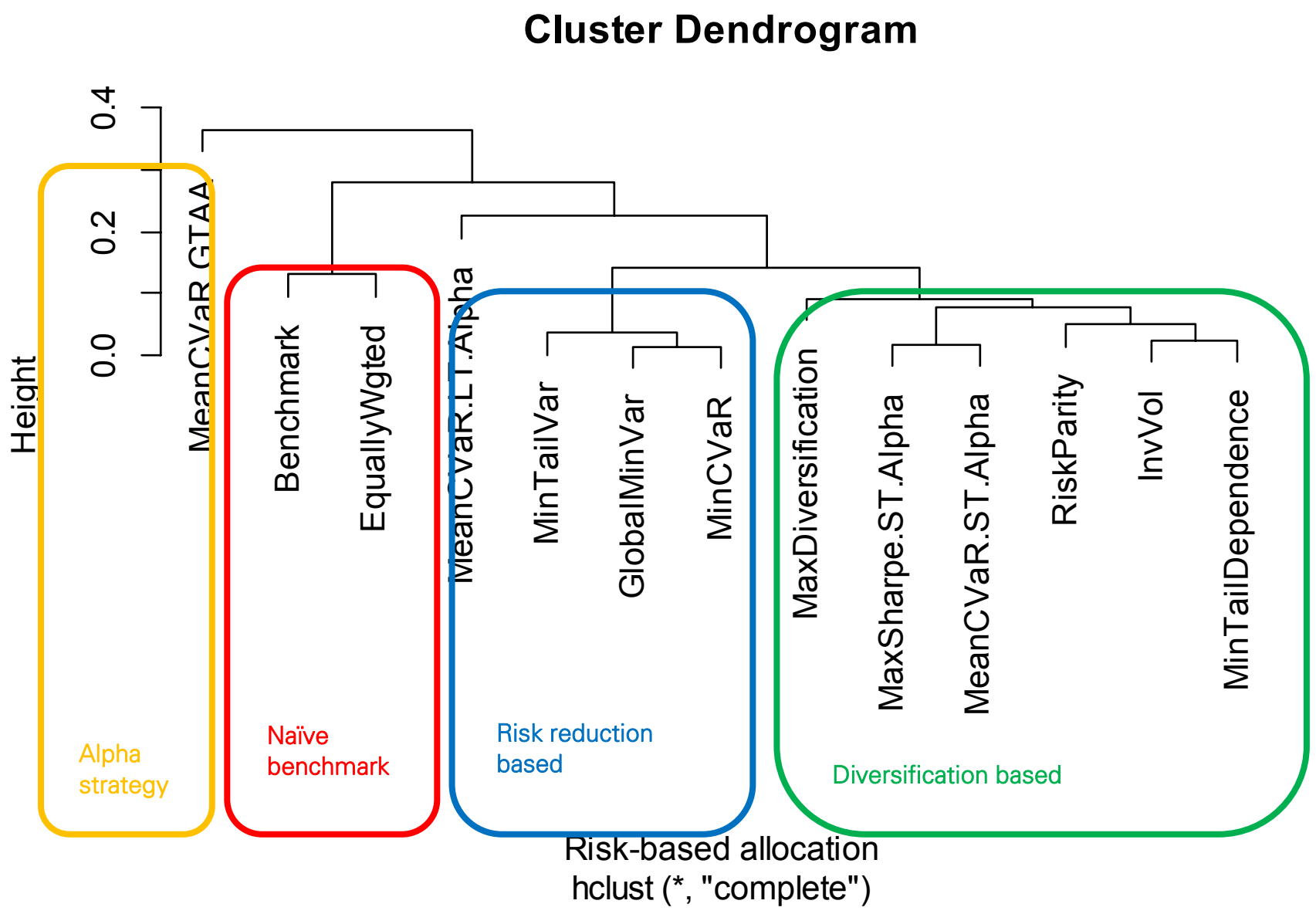

*Source: Bloomberg Finance LLP, MSCI, Russell, S&P, Worldscope, Deutsche Bank Quantitative Strategy*

Figure 82: Eigenvalue ratio chart

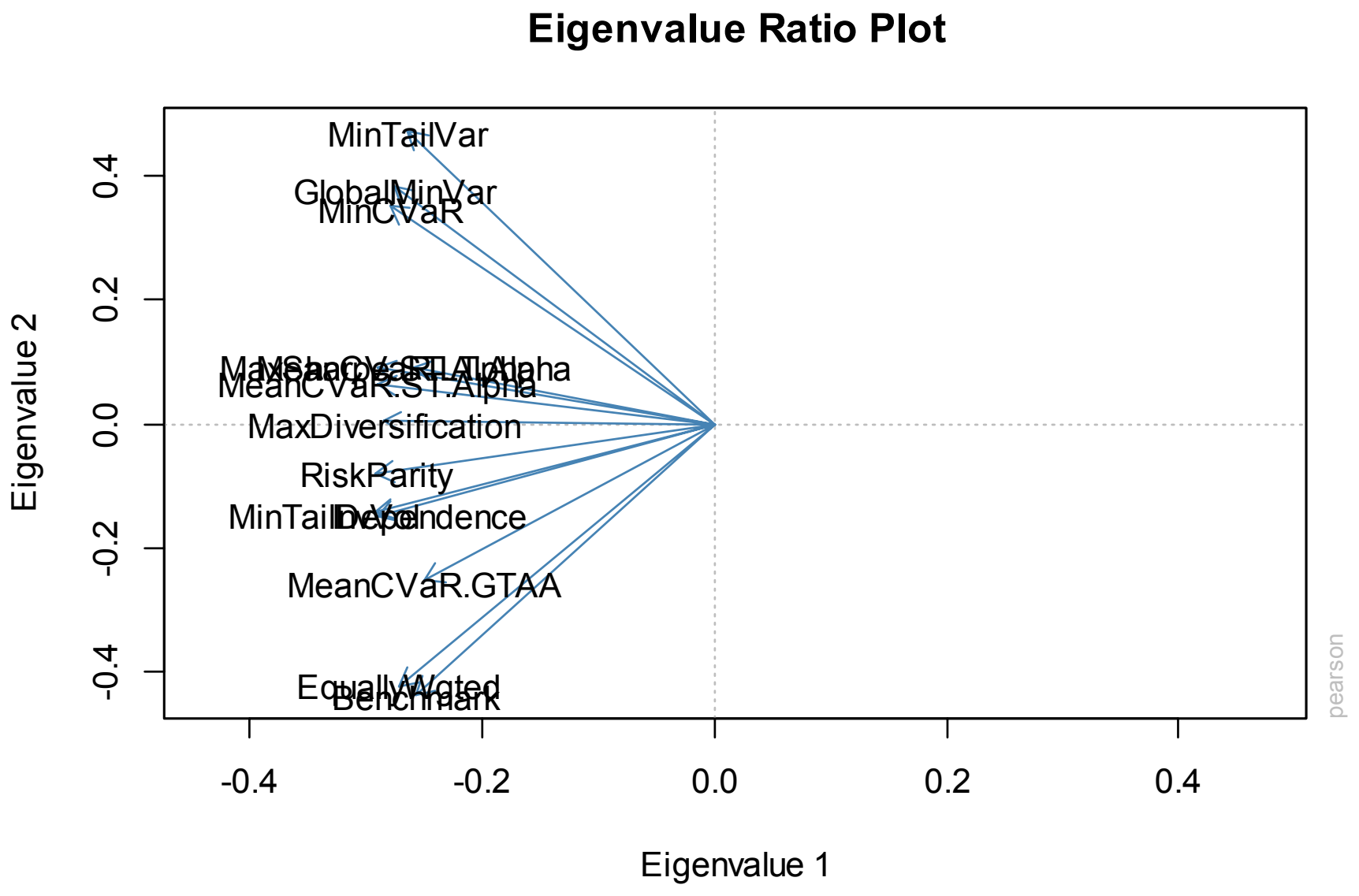

*Source: Bloomberg Finance LLP, MSCI, Russell, S&P, Worldscope, Deutsche Bank Quantitative Strategy*

## Portfolio construction techniques are risk (and correlation) regime dependent

Not only asset classes are regime dependent; interestingly, we also find portfolio construction techniques perform differently in different regimes.

### Risk regime

Given that the benchmark heavily overweights equities (60%), it is not surprising to see that benchmark portfolio delivers higher returns in low risk environment (see Figure 83). On the other hand, it is very interesting to note that all risk-based strategies and almost all alpha-based strategies actually produce higher returns in high risk regimes. More importantly, the outperformance of these risk-based and alpha-based portfolio construction techniques mostly comes from those high-risk periods. This is because almost all optimized strategies have been overweighting bonds compared to the benchmark.

Figure 83: Strategy performance in different risk regimes

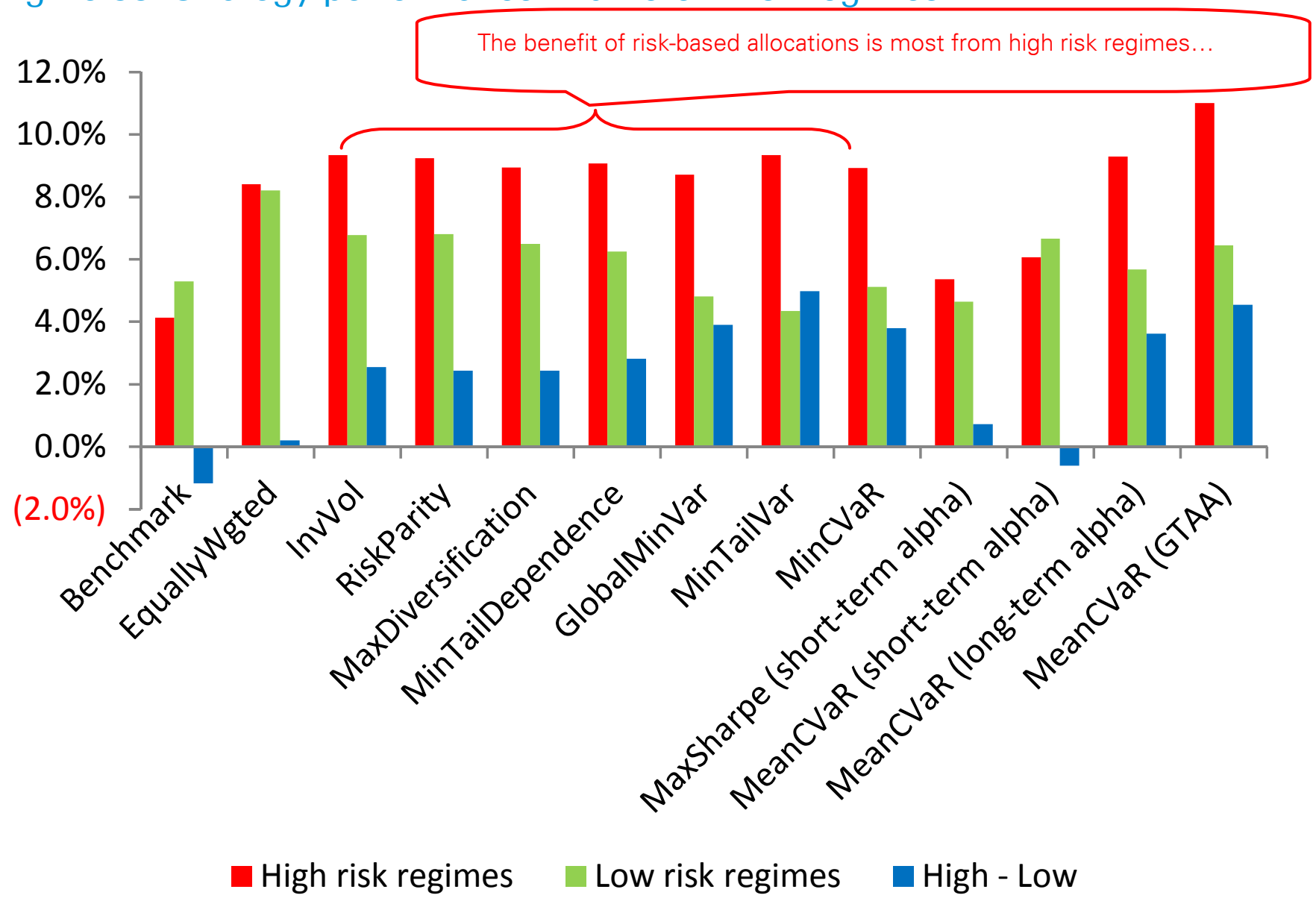

*Source: Bloomberg Finance LLP, MSCI, Russell, S&P, Worldscope, Deutsche Bank Quantitative Strategy*

### Correlation regime

Based on correlation regimes, it is a little surprising to see that the benchmark, all risk-based and all alpha-based strategies yield higher returns in high correlation regimes (see Figure 84).

The benchmark portfolio performs extremely well in the high correlation environment, beating most risk- and alpha-based strategies. On the other hand, in low correlation regimes, all risk- and alpha-based strategies tend to significantly outperform the benchmark.

Figure 84: Strategy performance in different correlation regimes

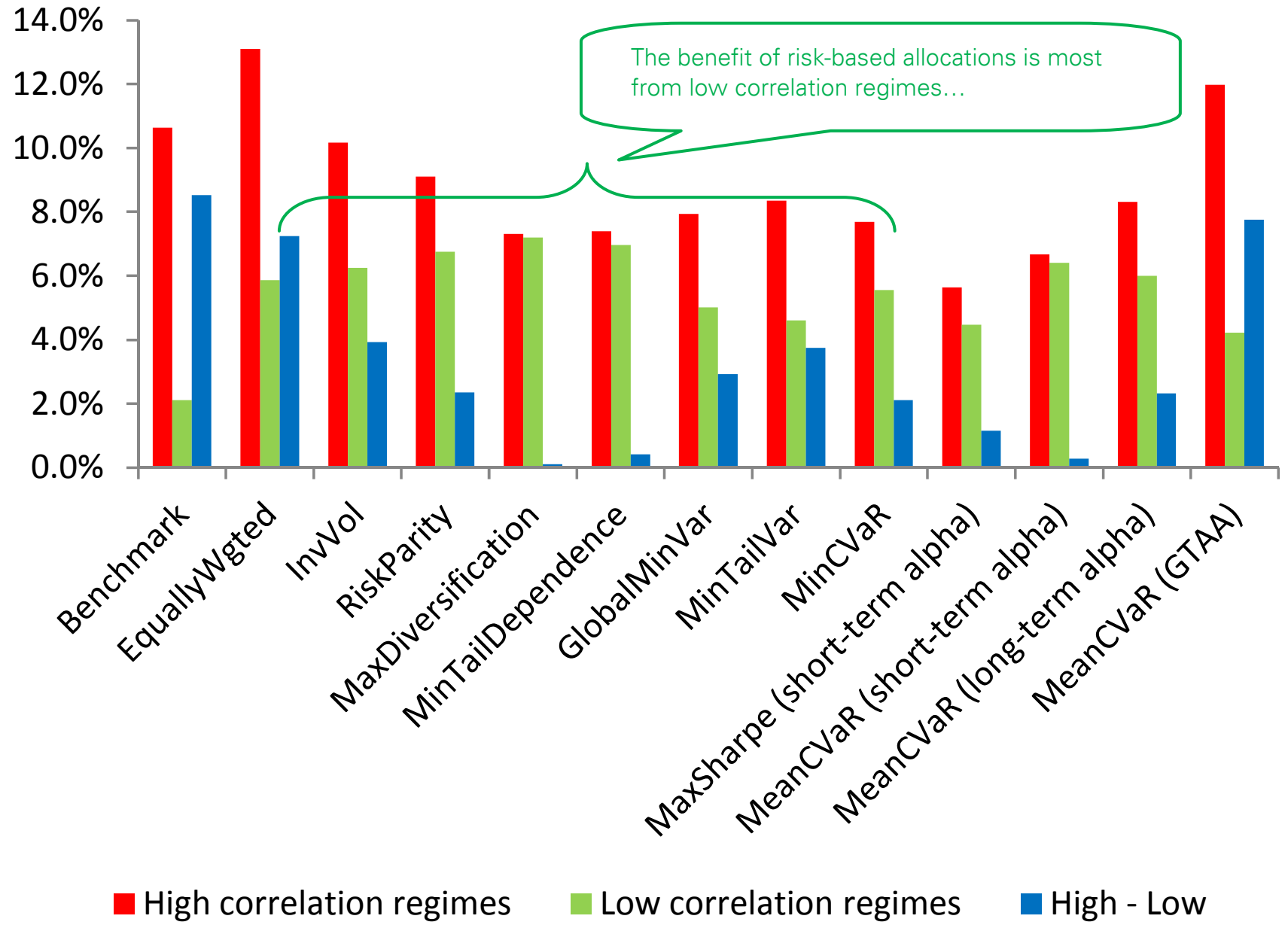

*Source: Bloomberg Finance LLP, MSCI, Russell, S&P, Worldscope, Deutsche Bank Quantitative Strategy*

## Risk and correlation regime

If we divide the world into the four-state risk and correlation grid (see Figure 85), most of the outperformance of risk- and alpha-based strategies comes from high risk/low correlation regimes, which are typically marked by falling equities and rising bonds, i.e., classic risk-off state. Compared to the benchmark, all of our risk- and alpha-based strategies tend to significantly overweight bonds and underweight equities, which explains the outperformance in such high risk/low correlation environment.

Figure 85: Strategy performance in different risk and correlation regimes

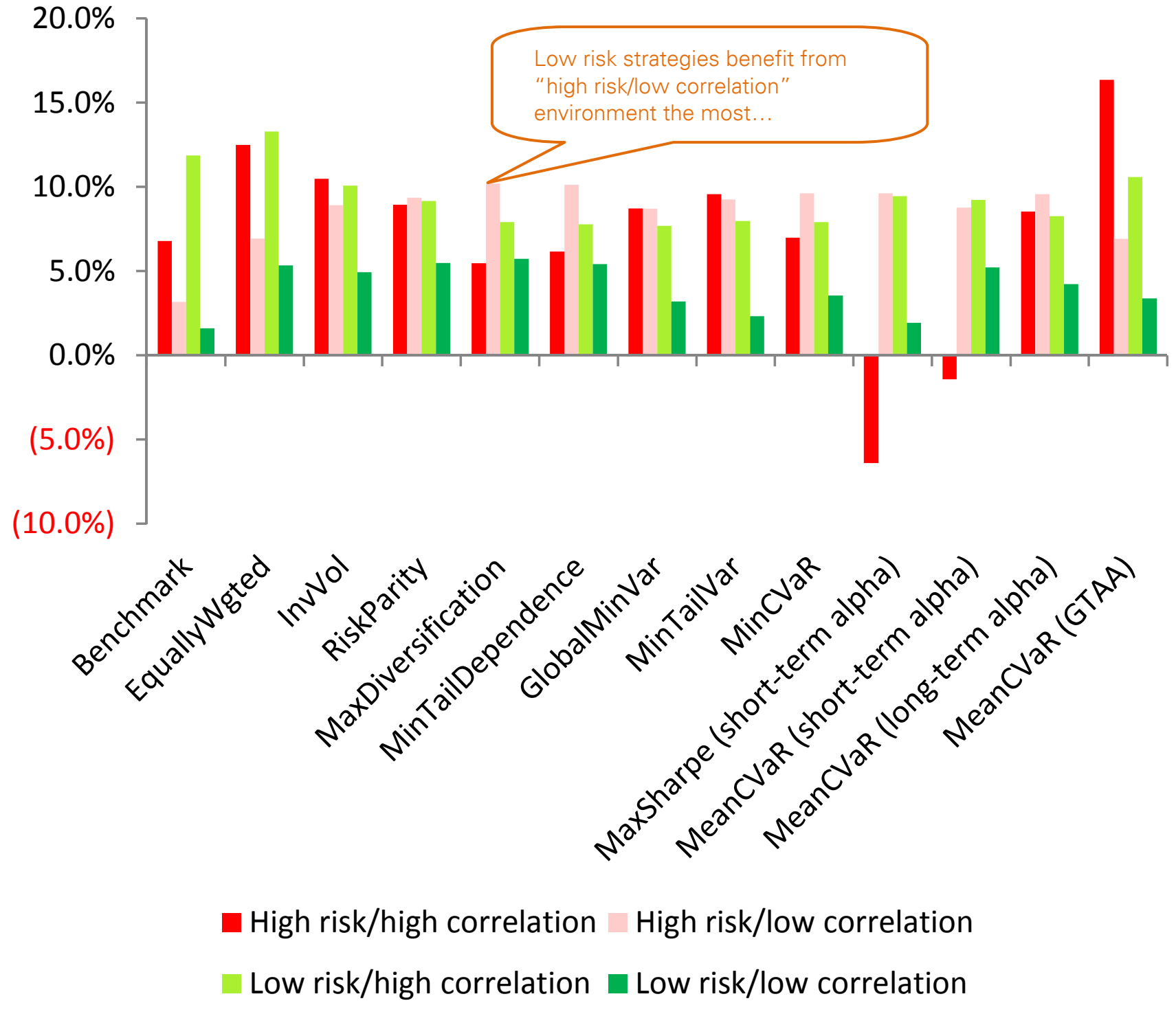

Source: Bloomberg Finance LLP, MSCI, Russell, S&P, Worldscope, Deutsche Bank Quantitative Strategy

## Why can't we use regime switching model on returns directly?

The traditional application of regime switching models in the academic world lies around identifying recessions and business cycles (see seminal work by Hamilton [1989] and more recently, Krolzig [1997]). In the practitioner's world, we have been focusing on how to apply regime switching models to predict asset returns directly (see Luo [2010]).

Returns can be calculated easily *ex post*, but are difficult to predict *ex ante*. On the other hand, risk is not well defined and has to be estimated even *ex post*. However, risk tends to be much easier to predict than returns. Return regimes are much more volatile and difficult to identify. Risk regimes tend to be more stable and are likely to be predictable.

If we try to use similar regime switching models on asset returns directly, we find a few difficulties. For example, for many fixed income types of assets, e.g., US treasuries and US high yield securities, most Markov switching models fail to converge at all, due to numerical issues.

As another example, let's build a regime switching model for EM equities. Figure 86 shows the one-step ahead predicted probabilities in low and high return regimes. The model suggests that the probability in the low return regime is never above 50%; therefore, in the past 14 years, the regime switching model predicts that we have lived in a constant high return regime for EM equities. Such a model is unlikely to be very useful in real-life prediction and investing.

In some of our previous work (see Luo, *et al* [2000]), we also attempted to fit regime switching models to factor returns and found they had inferior predictive power to traditional linear regression approach.

Figure 86: One-step ahead predicted probabilities in low and high return regimes for EM equities

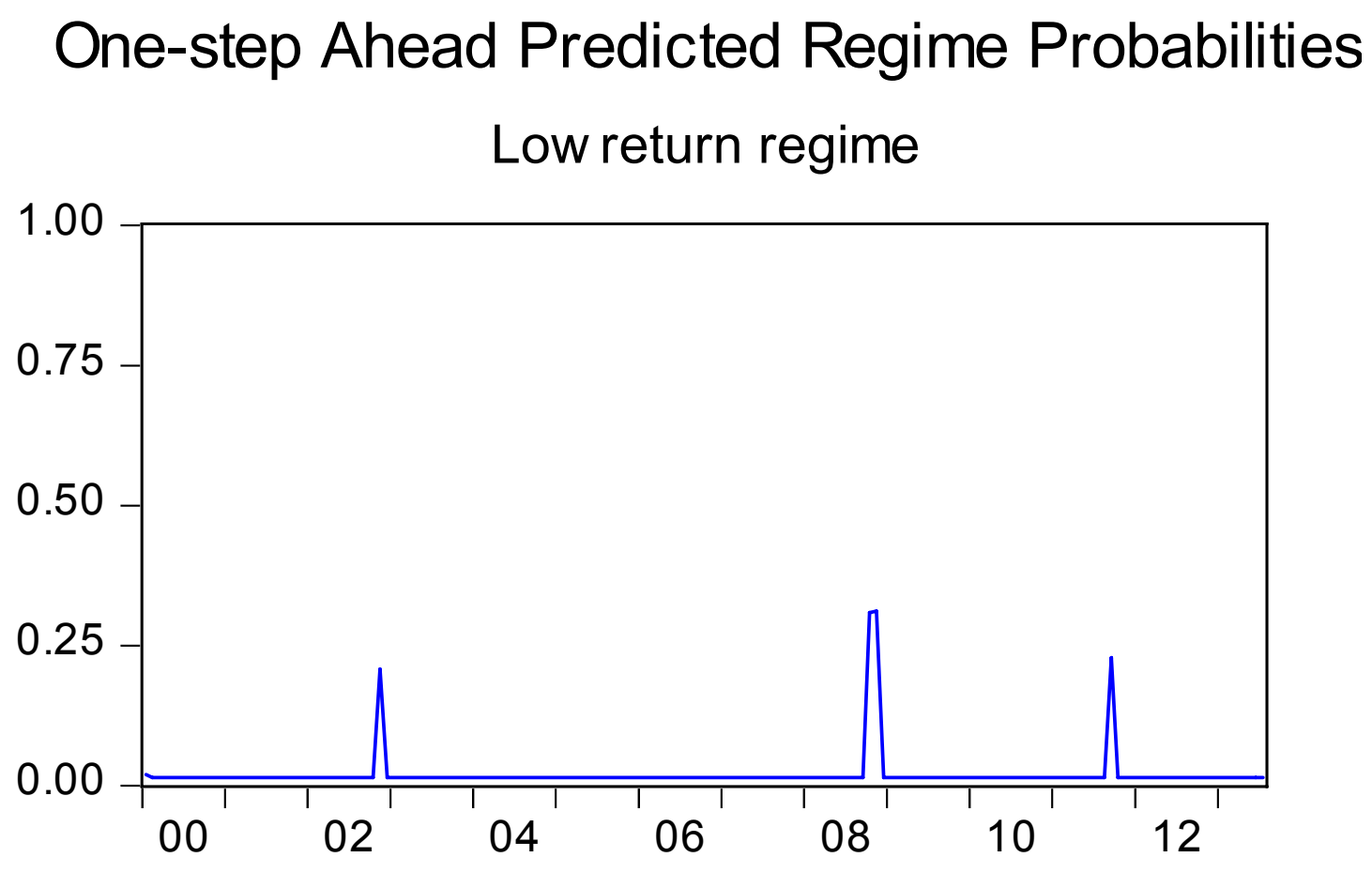

*Source: Bloomberg Finance LLP, MSCI, Russell, S&P, Worldscope, Deutsche Bank Quantitative Strategy*

# VI. Risk premia allocation and factor weighting

In the last section, we use a very different example to demonstrate the same GARCH-Copula-Regime Switching risk model, as well as our global financial risk regime indicator, can also be applied truly out-of-sample. We apply the same suite of algorithms on five simple quantitative stock selection factors. In different contexts, we call these factors differently. In the asset allocation space, they are typically called risk premium factors, alternative betas, or smart betas. In the quantitative equity world, they are just simple stock-selection factors[26]. Therefore, this section can be useful to both asset allocation and quantitative equity investors.

To be comparable with our previous research, we choose the same five simple alternative beta portfolios in global equities as defined in Luo, *et al* [2013]:

- Value, based on trailing earnings yield (long the cheapest quintile/short the most expensive quintile)
- Momentum, based on 12-month total returns excluding the most recent one month[27] (long stocks with the highest momentum quintile/short the worst momentum quintile)
- Quality, based on return on equity (long stocks with the highest ROE quintile/short stocks with the lowest ROE quintile)
- Size. MSCI World SmallCap total return index – MSCI World LargeCap total return index
- Low volatility/low risk, based on trailing one-year daily realized volatility (long stocks with the lowest volatility quintile/short stocks with the highest volatility quintile)

All portfolios (other than size) are constructed from the MSCI World universe, in a regional and sector neutral way, by forming a long/short quintile portfolio, equally weighted within the long and the short portfolios. We divide the MSCI World into the following regions: US, Canada, Europe ex UK, UK, Asia ex Japan Developed, Japan, and Australia/New Zealand. We select equal numbers of stocks from each region in each of the 10 GICS sectors. We exclude those region/sector buckets where we have fewer than five stocks.

In the past 15 years, value has generated the highest cumulative return (see Figure 87), followed by size. Low vol factor actually delivered negative return on average. Consistent with Cahan, *et al* [2012], although minimum variance portfolios tend to outperform capitalization-weighted indices in almost all markets (see Luo, *et al* [2013]), a simple long/short quantile portfolio may not.

[26] In the asset allocation world, investors typically construct factor portfolios first, and then combine these factor portfolios using risk parity type of techniques. Quantitative equity investors usually use more complicated factors to build multi-factor stock-selection models. Then, they use these multi-factor models to construct equity portfolios – either optimized or based on simple heuristic techniques.

[27] The reason to exclude the most recent one month return is to account for the typical one-month reversal effect observed in most markets, but especially US equities and Japanese equities.

In terms of Sharpe ratio (see Figure 88), value and size again dominate. The combined risk premia portfolio is formed using a risk parity approach, rebalanced monthly (see Luo, *et al* [2013] for details). The combined multi-factor portfolio certainly has its diversification benefit and delivers the highest Sharpe ratio.

Finally, we measure downside risk using CVaR and maximum drawdown – both metrics point out that momentum and low vol factors have the greatest downside risk (see Figure 89).

Value and momentum are clearly negatively correlated (see Figure 90). We can also see some strong nonlinear relationship between value and quality, between value and low volatility, between size and low volatility, etc.

Figure 87: Cumulative performance

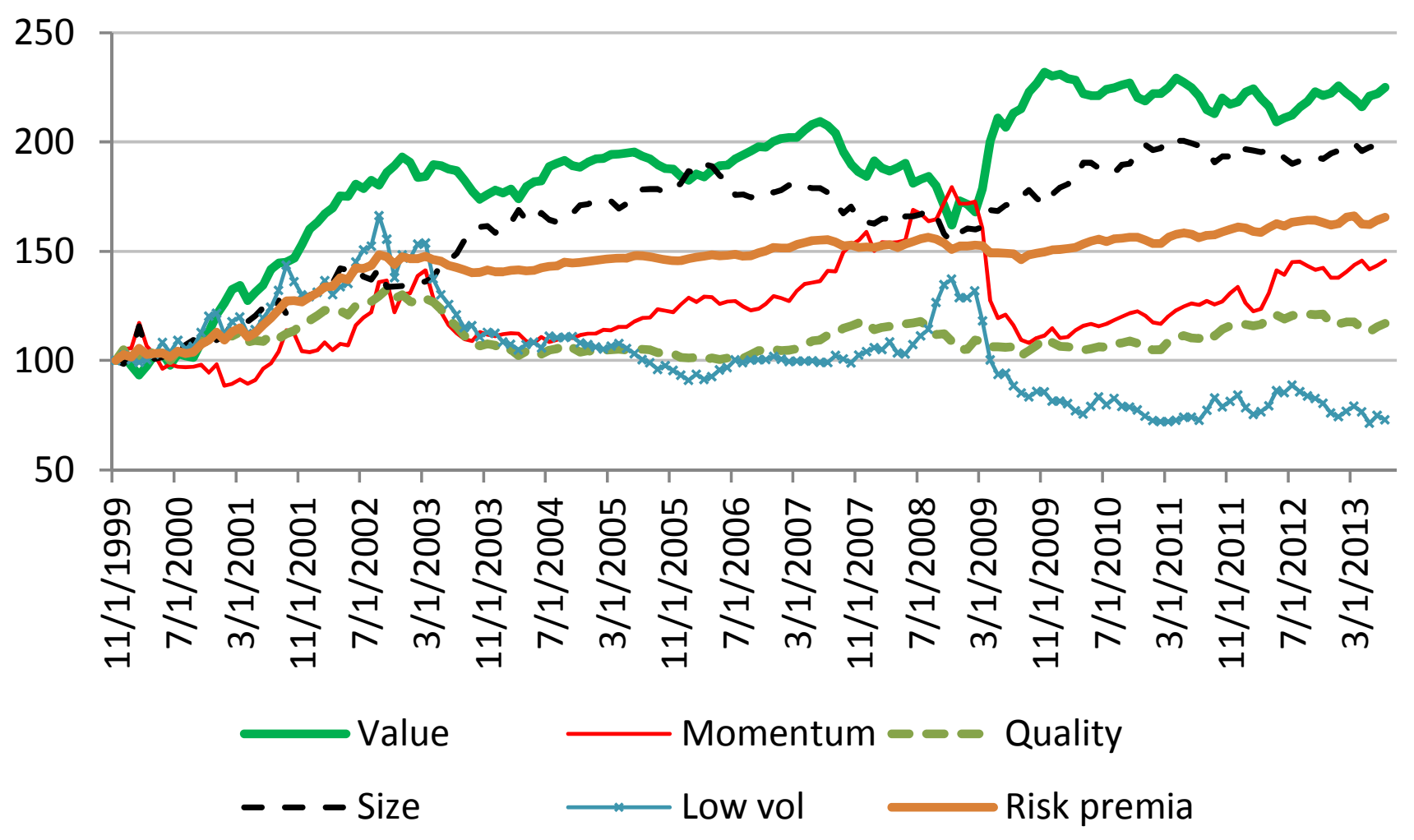

*Source: Bloomberg Finance LLP, MSCI, Russell, S&P, Worldscope, Deutsche Bank Quantitative Strategy*

Figure 88: Risk, return, and Sharpe ratio

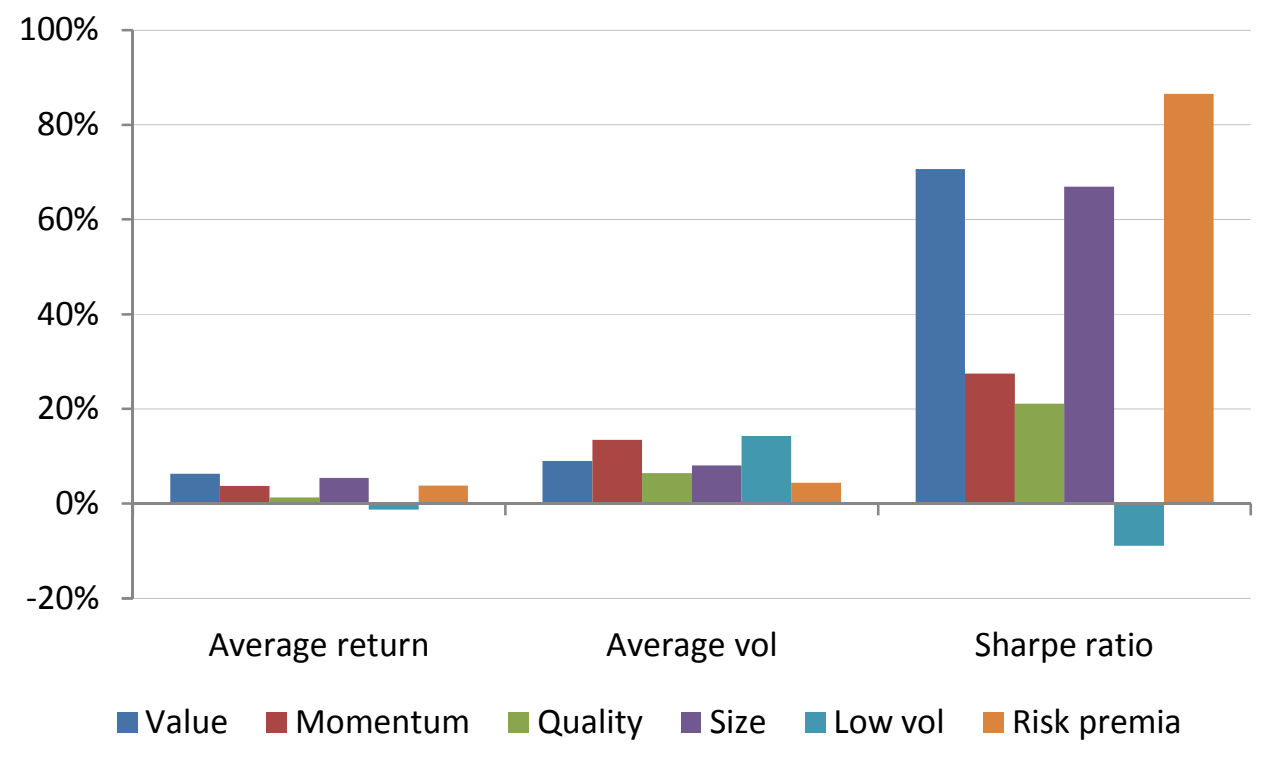

*Source: Bloomberg Finance LLP, MSCI, Russell, S&P, Worldscope, Deutsche Bank Quantitative Strategy*

Figure 89: Downside risk

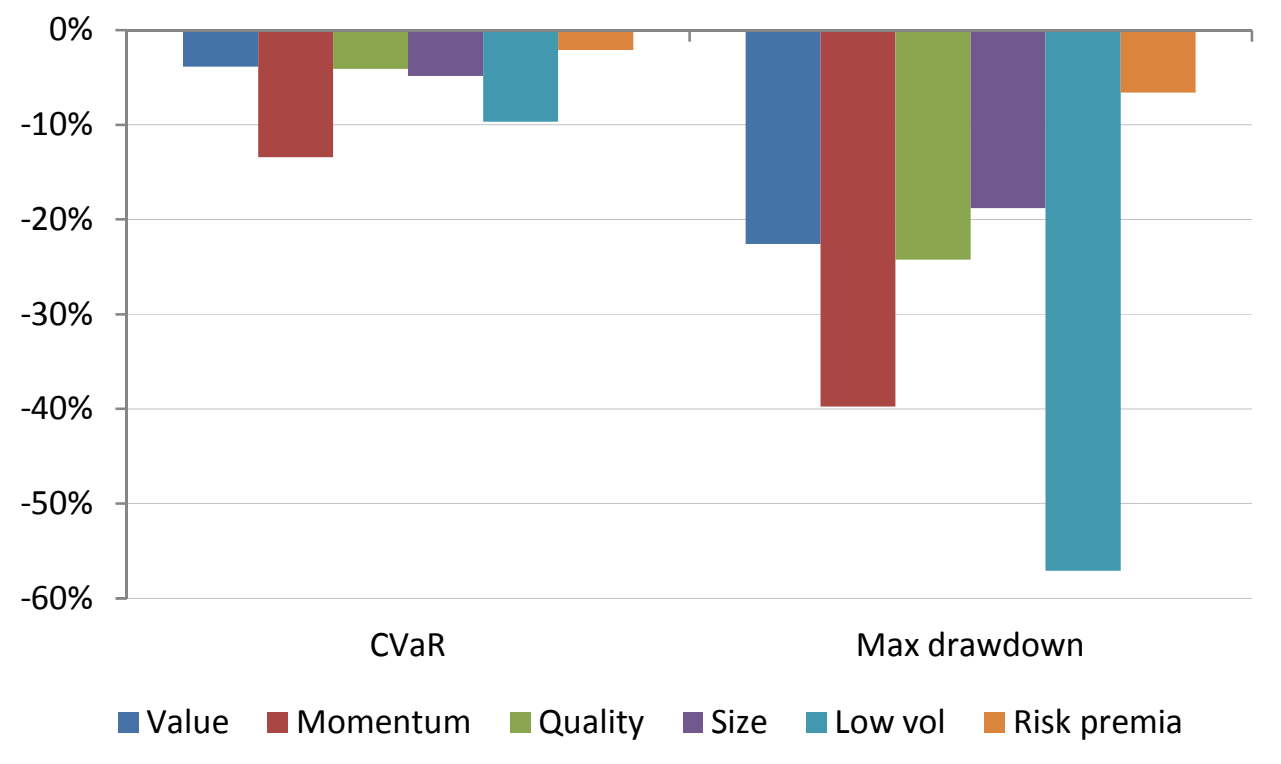

*Source: Bloomberg Finance LLP, MSCI, Russell, S&P, Worldscope, Deutsche Bank Quantitative Strategy*

Figure 90: Scatterplot of factor portfolios

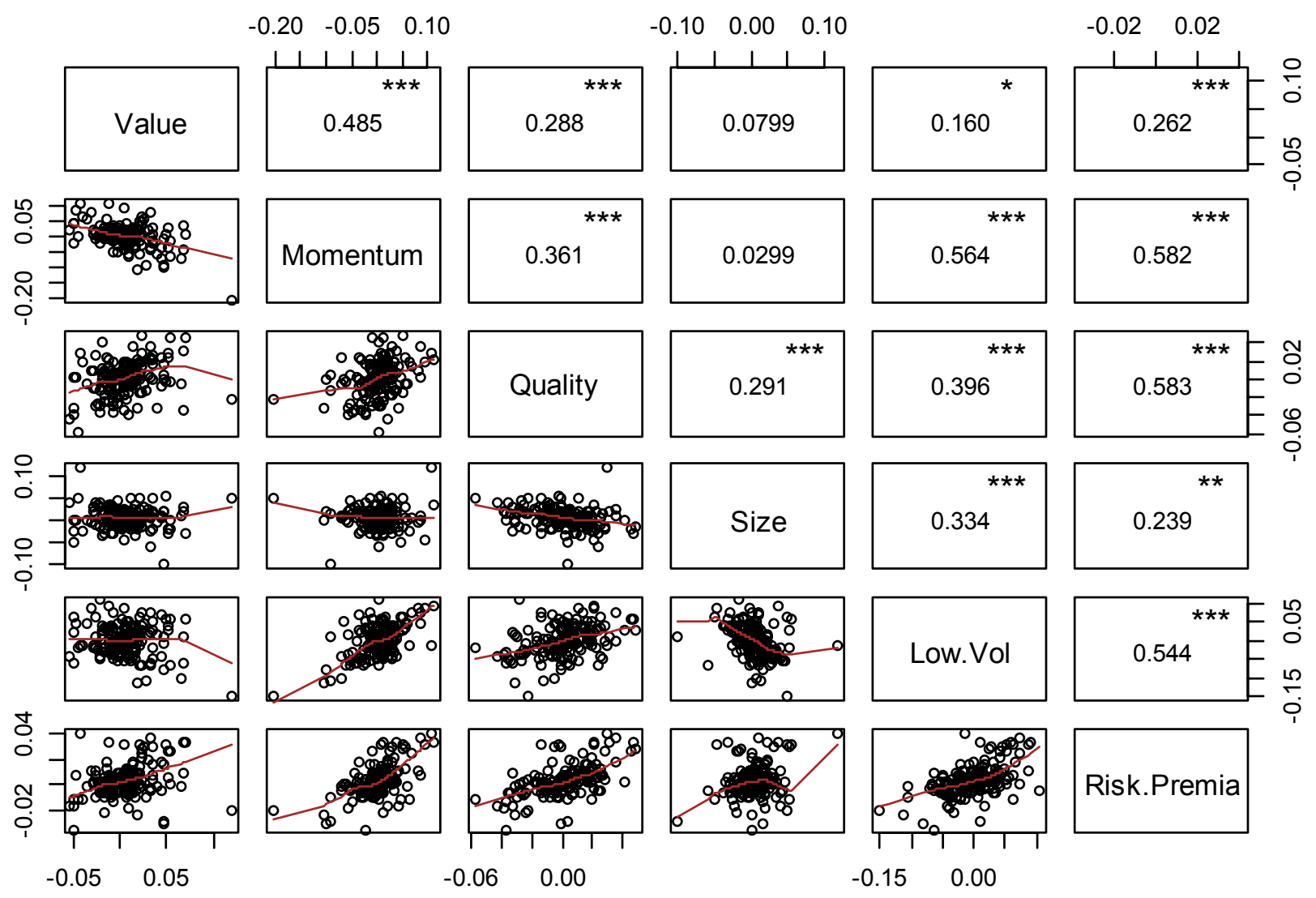

Source: Deutsche Bank

## GARCH-DCC-Copula risk model

We then compare the performance of the same suite of risk-based allocations as in Luo, *et al* [2013], using three sets of risk models:

1. Trading one-year actual returns

2. Expanding window, minimum five years of daily data

3. GARCH-Copula risk model

To test the real out-of-sample predictive power, we decide not to fine tune the "best" statistical model. Rather, we use exactly the same GARCH-DCC-Copula setup as in the previous asset allocation example. We estimate our GARCH-DCC-Copula model using:

1. Each one of the five factors are fitted separately to an ARMA(1,1)-GARCH(1,1) model

2. The residuals from the ARMA(1,1)-GARCH(1,1) model are then used to fit a joint multivariate DCC model (for correlation) and t-Copula model (for tail dependence)

3. The estimated parameters of the ARMA(1,1)-GARCH(1,1)-DCC-Copula model are used to simulate daily returns

4. The simulated daily returns are used as the estimation universe for the risk models used in the various risk-based allocations

We do want to introduce another added feature. The simulated returns from our GARCH-DCC-Copula model can be quite volatile; therefore, it might be subject to dispute whether the outperformance is random. We borrow the philosophy of portfolio resampling (see Michaud [1998]) and our own robust mean-CVaR optimization (see Luo, *et al* [2013]). Now, let's add the following additional steps to our process:

5. We repeat the simulation/optimization process 20 times

6. The final portfolio weights for the month is the simple average of the weights of the 20 portfolios from step 6

Two of the most widely used approaches to construct factor portfolios are the risk parity approach and maximum diversification method. Our GARCH-DCC-Copula risk model seems to be able to meaningfully improve the portfolio performance for both approaches, compared to a more traditional risk model using sample data (see Figure 91 and Figure 92). For risk parity strategy, the improvement is mostly from reduced risk (see Figure 93), while for maximum diversification portfolio, the enhancement comes more from higher returns (see Figure 94). More importantly, as shown in Figure 95 and Figure 96, our GARCH-DCC-Copula risk model also helps to build a less crowded portfolio, if we measure crowding using weighted portfolio tail dependence (see Luo, *et al* [2013] for definition).

Figure 91: Cumulative performance – risk parity portfolios with different risk models

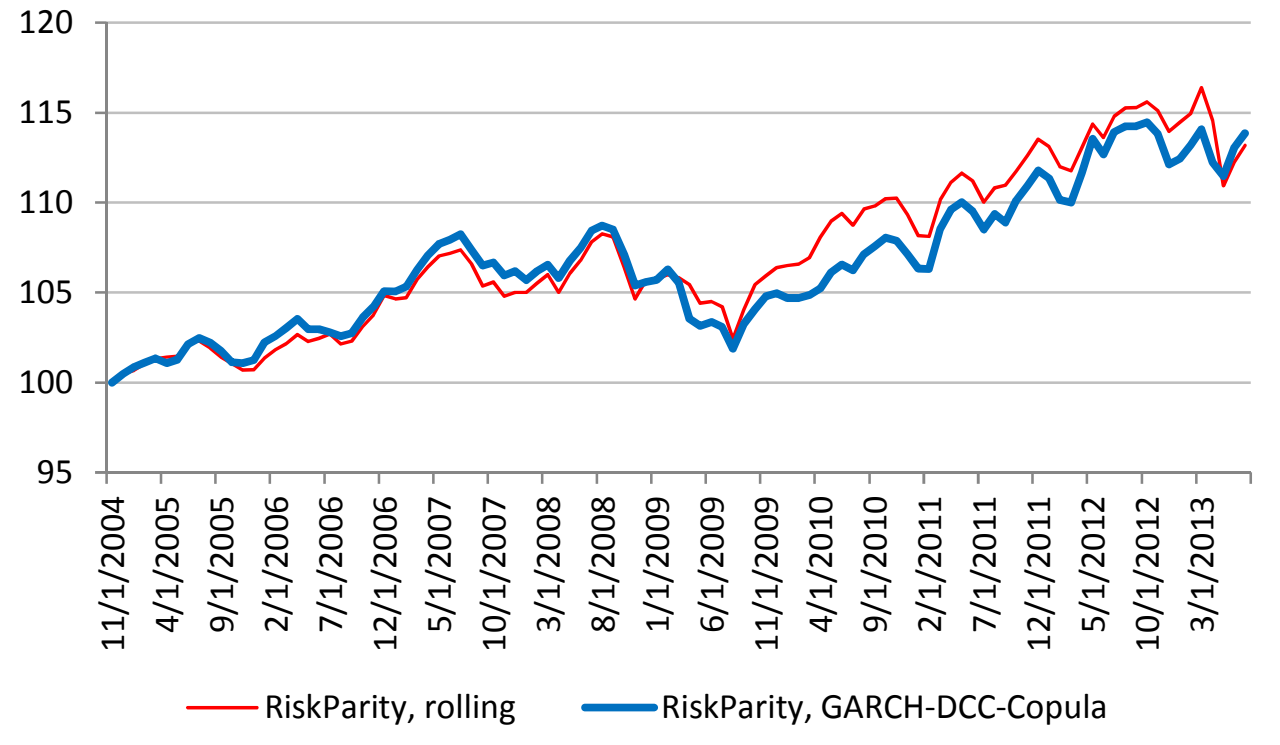

*Source: Bloomberg Finance LLP, MSCI, Russell, S&P, Worldscope, Deutsche Bank Quantitative Strategy*

Figure 92: Cumulative performance – maximum diversification portfolios with different risk models

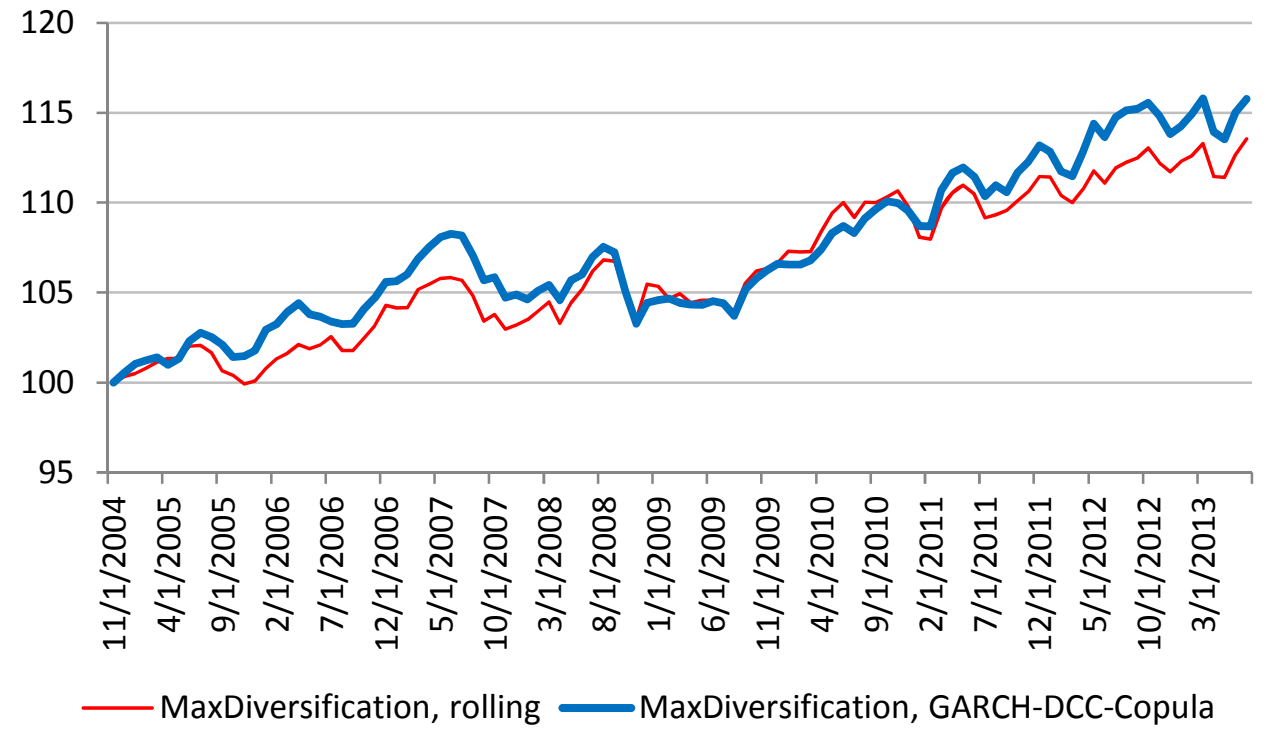

*Source: Bloomberg Finance LLP, MSCI, Russell, S&P, Worldscope, Deutsche Bank Quantitative Strategy*

Figure 93: Return, risk, and Sharpe ratio – risk parity portfolios with different risk models

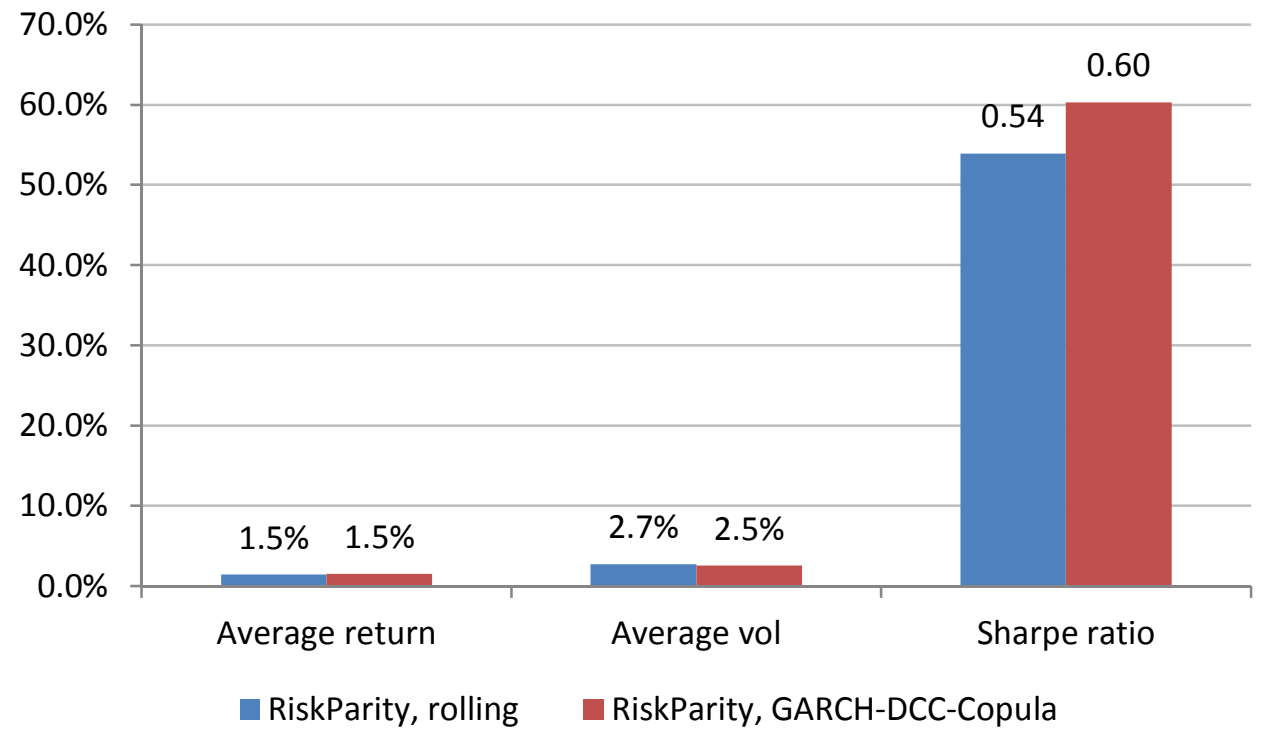

Source: Bloomberg Finance LLP, MSCI, Russell, S&P, Worldscope, Deutsche Bank Quantitative Strategy

Figure 94: Return, risk, and Sharpe ratio – maximum diversification portfolios with different risk models

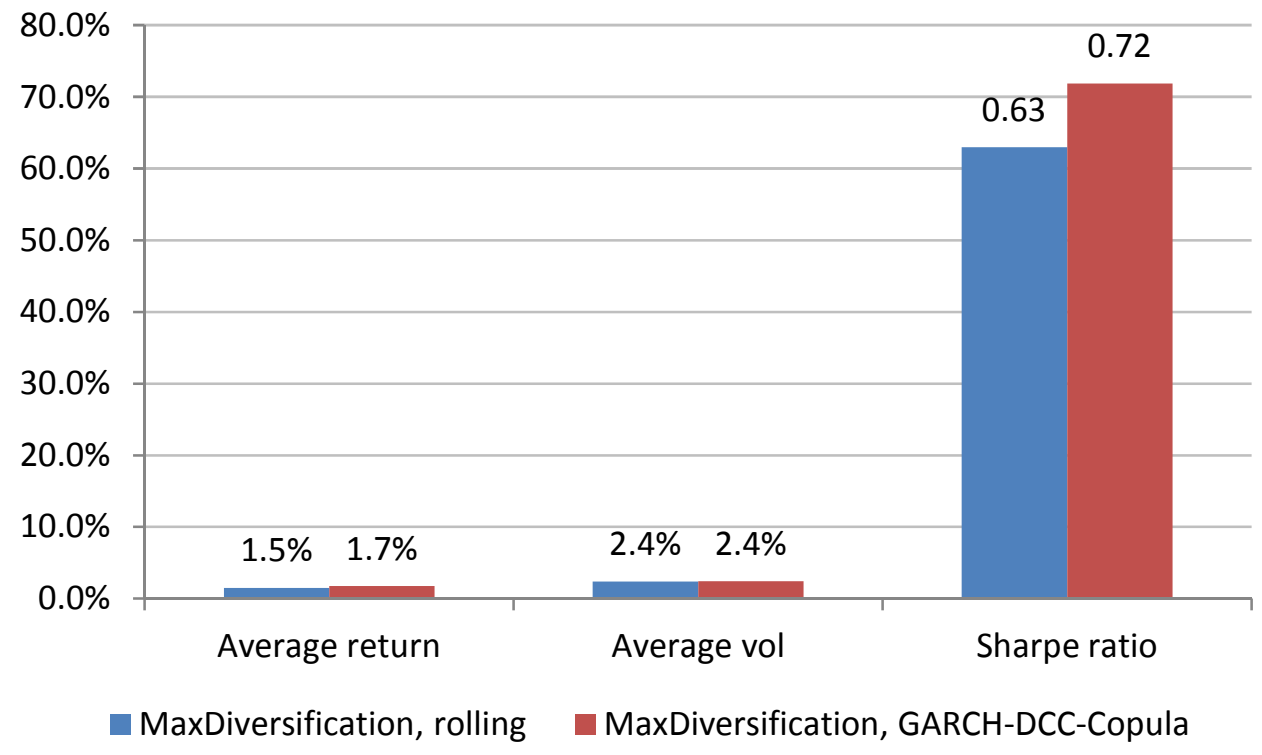

Source: Bloomberg Finance LLP, MSCI, Russell, S&P, Worldscope, Deutsche Bank Quantitative Strategy

Figure 95: Weighted portfolio tail dependence – risk parity portfolios with different risk models

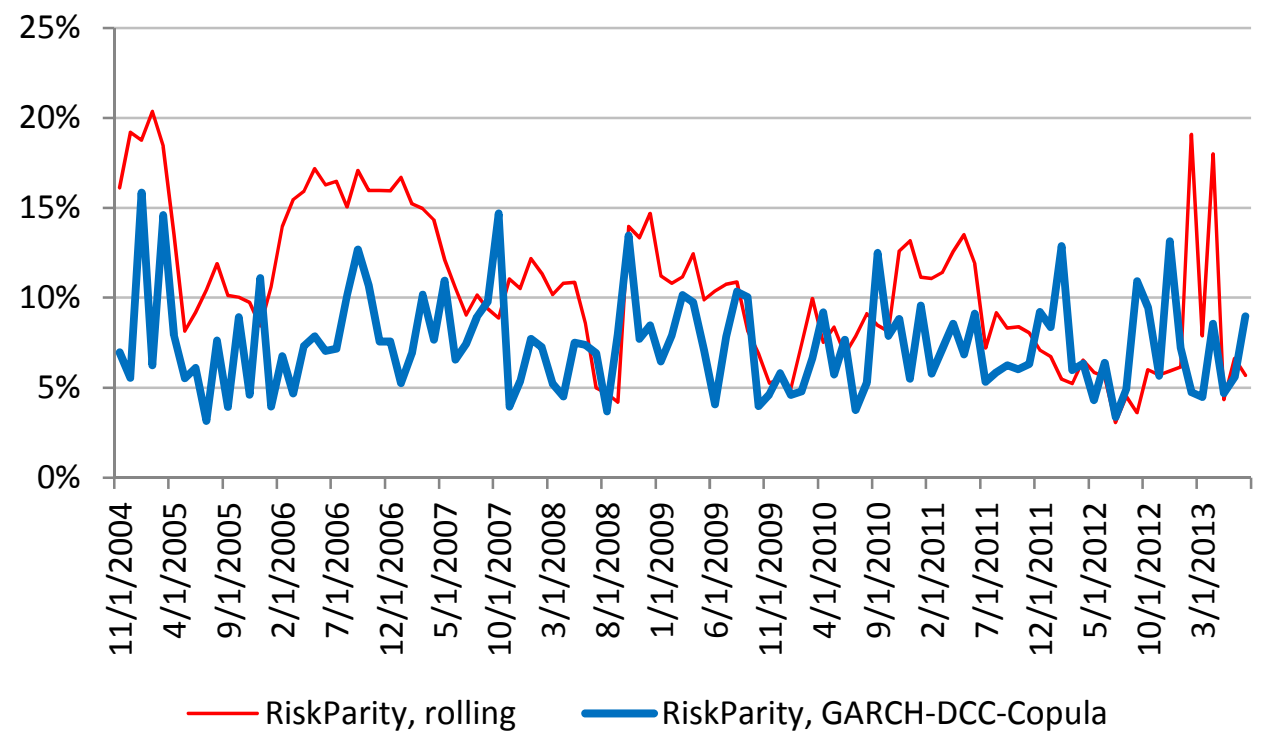

Source: Bloomberg Finance LLP, MSCI, Russell, S&P, Worldscope, Deutsche Bank Quantitative Strategy

Figure 96: Weighted portfolio tail dependence – max diversification portfolios with different risk models

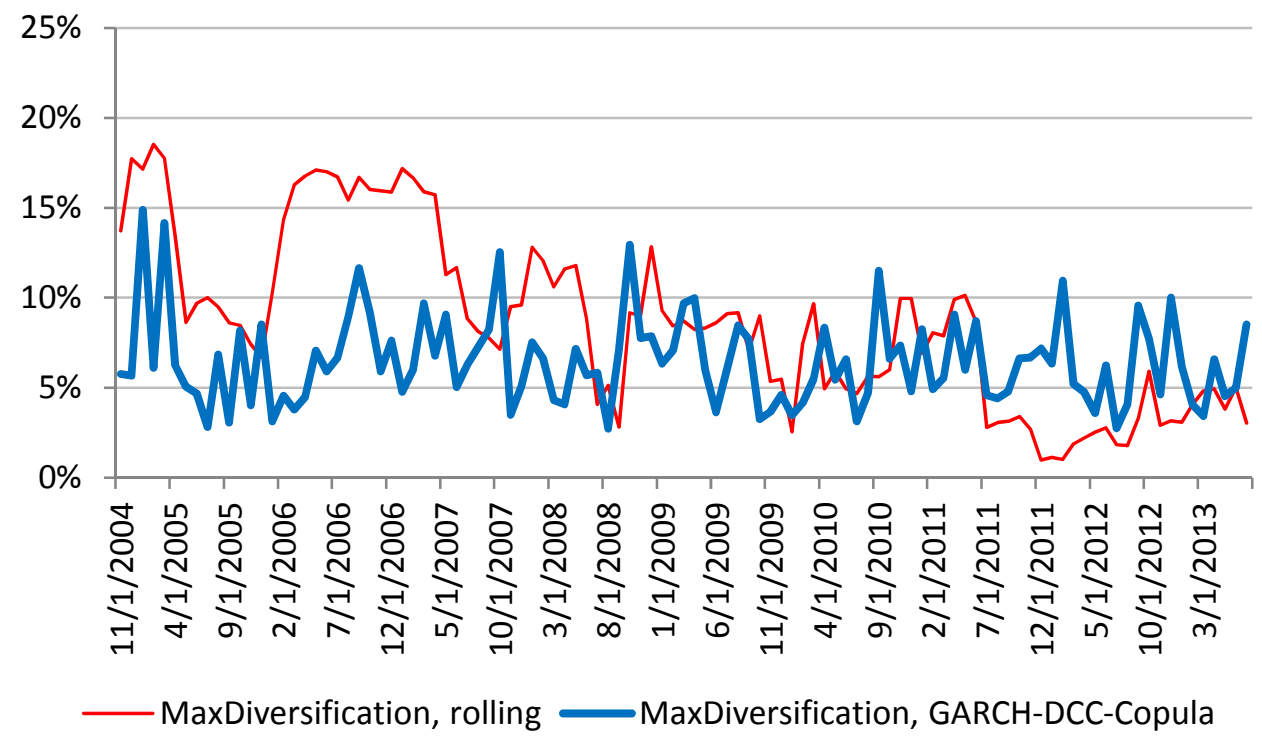

Source: Bloomberg Finance LLP, MSCI, Russell, S&P, Worldscope, Deutsche Bank Quantitative Strategy

## Factor performance under different risk and correlation regimes

In previous sections, we discussed how asset returns and portfolio construction techniques are regime dependent. We are encouraged to see that factor returns also vary considerably by risk (and correlation) regimes.

### Risk regimes

It is quite intuitive that value and quality, as more defensive factors, tend to perform strongly in high risk states, while momentum and size – as more cyclical styles, are exactly the opposite (see Figure 97). Our risk premia factor seems to be able to provide reasonable downside protection in high risk regimes, while delivers decent returns in low risk states.

The biggest surprise is our low volatility factor. Intuition would suggest that low volatility factor should do well in high risk regimes, as an anti-risk factor.

However, as shown in Figure 97, the low volatility factor actually tumbles in a high risk environment. This once again highlights the difference between cross-sectional low risk (of investing in low risk stocks) and time series of low risk (in low risk economic environment).

To dig a little deeper, Figure 98 plots the relationship between low volatility factor performance and the equity market (MSCI World) – it's clear that they are negatively correlated. As expected, low volatility factor does perform better in bear market. Figure 99 displays the scatterplot of low volatility factor returns versus the global financial market CVaR. They also appear to be positively correlated, i.e., low volatility strategy tends to do better when the overall market risk is higher. Furthermore, Figure 100 shows the low volatility factor return distribution in high and low risk regimes. The average return of low volatility factor in high risk regimes is lower than it in low risk regimes, but that is mostly driven by a few outliers. The median return of low volatility factor in high risk regimes is actually significantly higher than in low risk regimes.

Figure 97: Strategy performance in different risk regimes

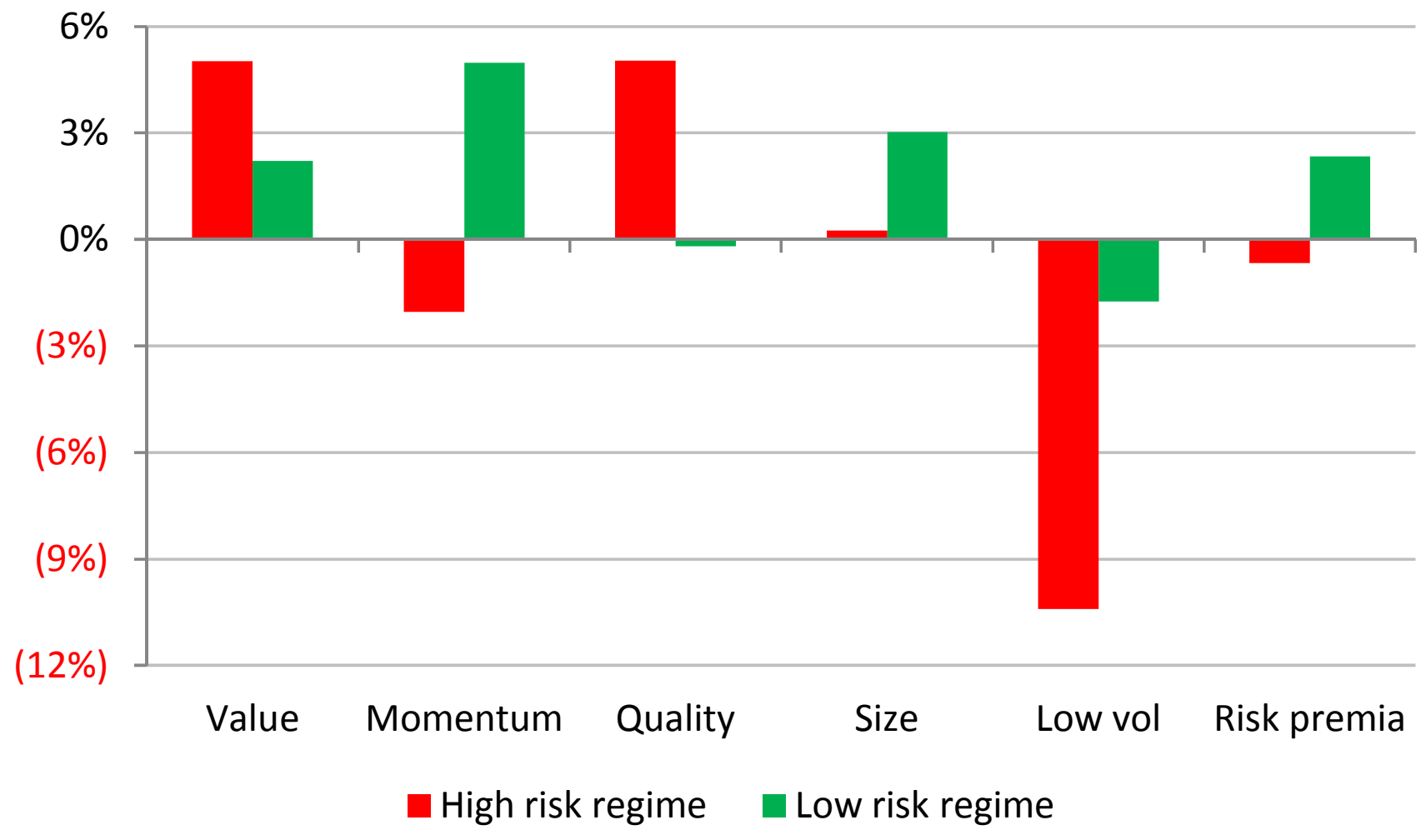

*Source: Bloomberg Finance LLP, MSCI, Russell, S&P, Worldscope, Deutsche Bank Quantitative Strategy*

Figure 98: Low vol vs. equities

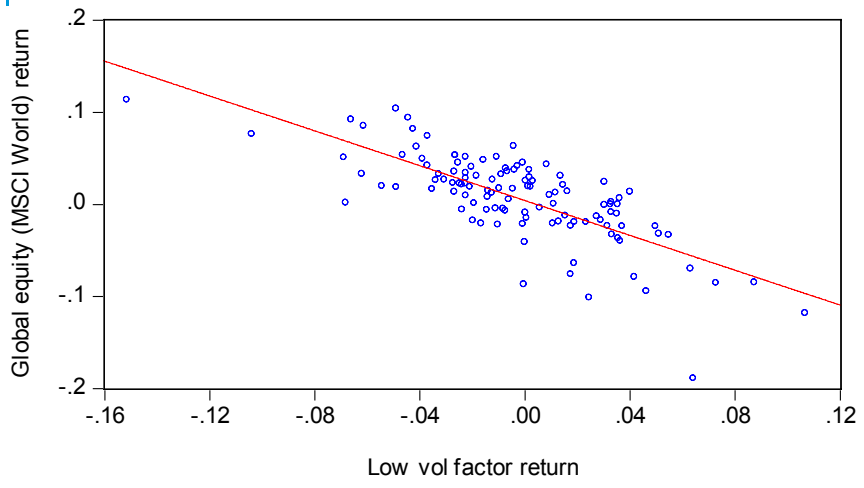

*Source: Bloomberg Finance LLP, MSCI, Russell, S&P, Worldscope, Deutsche Bank Quantitative Strategy*

Figure 99: Low vol vs. market risk

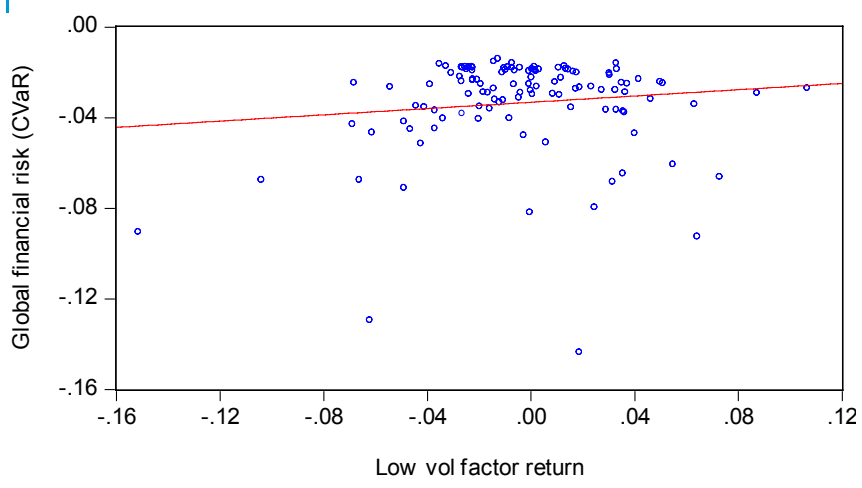

*Source: Bloomberg Finance LLP, MSCI, Russell, S&P, Worldscope, Deutsche Bank Quantitative Strategy*

Figure 100: Low vol factor return in high and low risk regimes

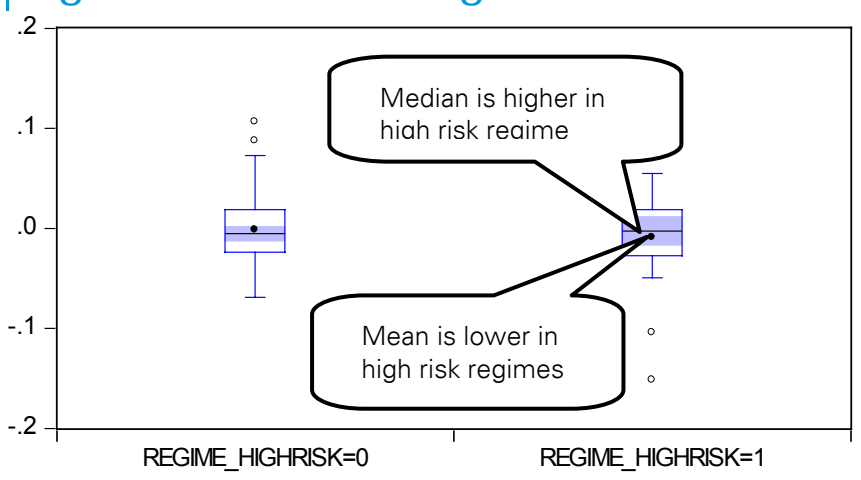

*Source: Bloomberg Finance LLP, MSCI, Russell, S&P, Worldscope, Deutsche Bank Quantitative Strategy*

## Correlation regime

Factor performance in correlation regimes is also quite interesting (see Figure 101). Value, quality, size, and the composite risk premia all deliver higher returns in high correlation regimes. On the other hand, momentum and low volatility factors seem to benefit more from low correlation environment.

Figure 101: Strategy performance in different correlation regimes

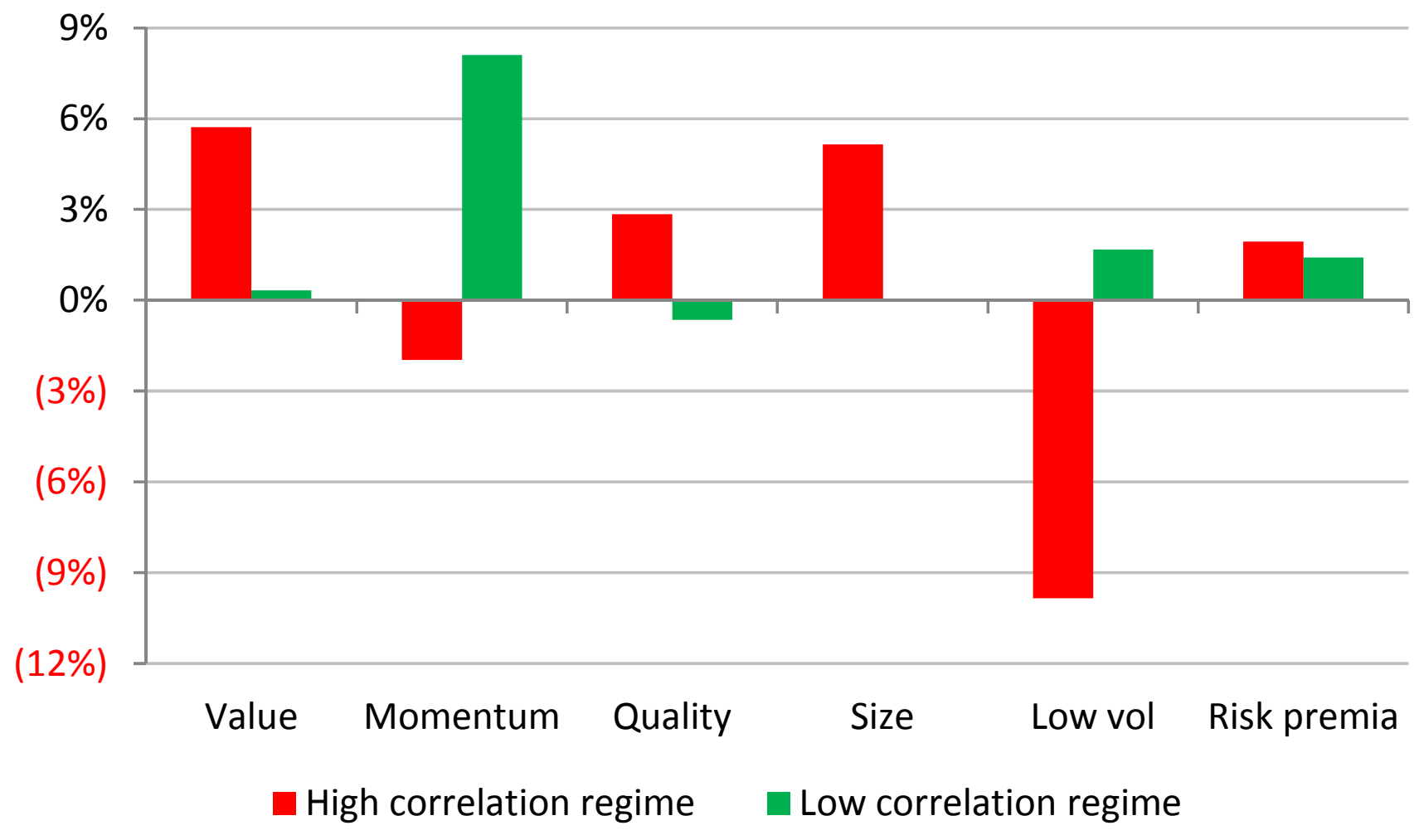

*Source: Bloomberg Finance LLP, MSCI, Russell, S&P, Worldscope, Deutsche Bank Quantitative Strategy*

## Risk and correlation regime

Once we combine risk regimes and correlation regimes, we have four possible states:

- High risk/high correlation
- High risk/low correlation
- Low risk/high correlation
- Low risk/low correlation

Value, momentum, size, and low volatility factors show most variation in "high risk/high correlation" and "high risk/low correlation" regimes (see Figure 102). On the other hand, quality and our composite risk premia seem more immune to changes in risk and correlation environment.

Figure 102: Strategy performance in different risk and correlation regimes

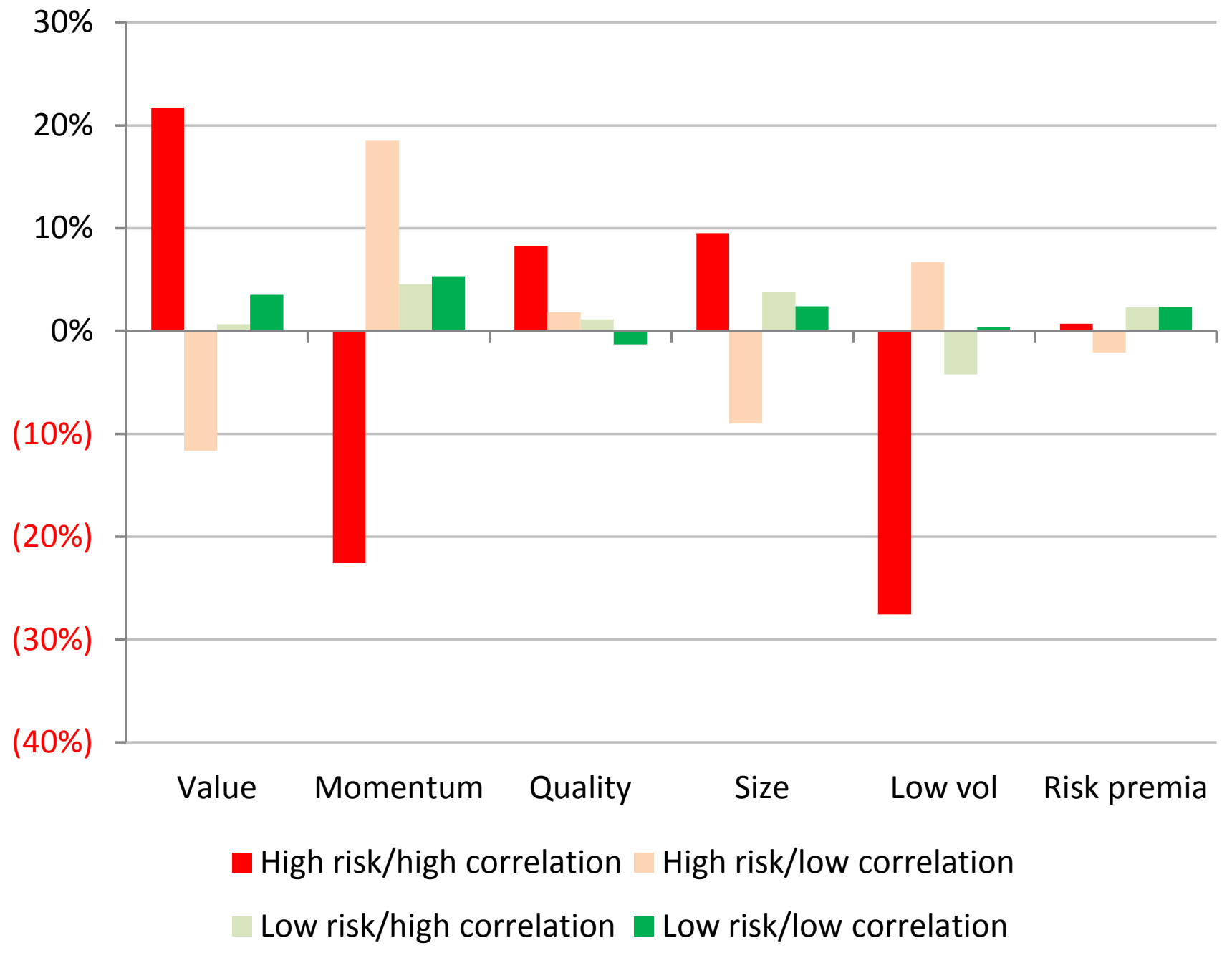

*Source: Bloomberg Finance LLP, MSCI, Russell, S&P, Worldscope, Deutsche Bank Quantitative Strategy*

## Style rotation with SAA risk regimes

As shown in the previous section, our five factors indeed show very different performance in high versus low risk regimes. The next question is whether we can use this information systematically. In this section, we illustrate how we use our regime prediction to improve our factor return prediction. We first take a simple style rotation model conditional on VRP, then we add our pure out-of-sample predicted high risk/low risk regime indicator as an additional variable to assess the incremental value of risk regimes.

- Baseline model:

$$r_t = \hat{\beta}_{0,t} + \hat{\beta}_{1,t} VRP_{t-1} + \varepsilon_t$$

- Style rotation model:

$$r_t = \hat{\beta}_{0,t} + \hat{\beta}_{1,t} VRP_{t-1} + \hat{\beta}_{2,t} I(P(s_t = 1)) + \varepsilon_t$$

Where,

$r_t$ is the return of a factor at time $t$,

$P(s_t = 1)$ is the predicted probability in regime $1$, i.e., high risk regime, and

$I(\bullet)$ is an indicator function that takes a value of either $1$ (if $P(s_t = 1) \geq 0.5$) or $0$ (if $P(s_t = 1) < 0.5$).

The predicted return for period $t+1$ is therefore:

$$\hat{r}_{t+1} = \hat{\beta}_{0,t} + \hat{\beta}_{1,t} VRP_t + \hat{\beta}_{2,t} I(P(s_{t+1} = 1))$$

To measure the accuracy of our prediction, we use the cross-sectional information coefficient (IC) metric. At each month end, we calculate the correlation between our predicted returns of the five factors (at the previous month end) and the actual realized returns of the five factors during the current month.

Figure 103 and Figure 104 show the time series of IC for the style rotation model without and with our regime switching indicator. The GTAA model with our regime switching indicator can improve the average performance (IC) by 21% and the risk-adjusted performance ($\mu(IC)/\sigma(IC)$) by 18%.

Figure 103: Factor rotation model using VRP only

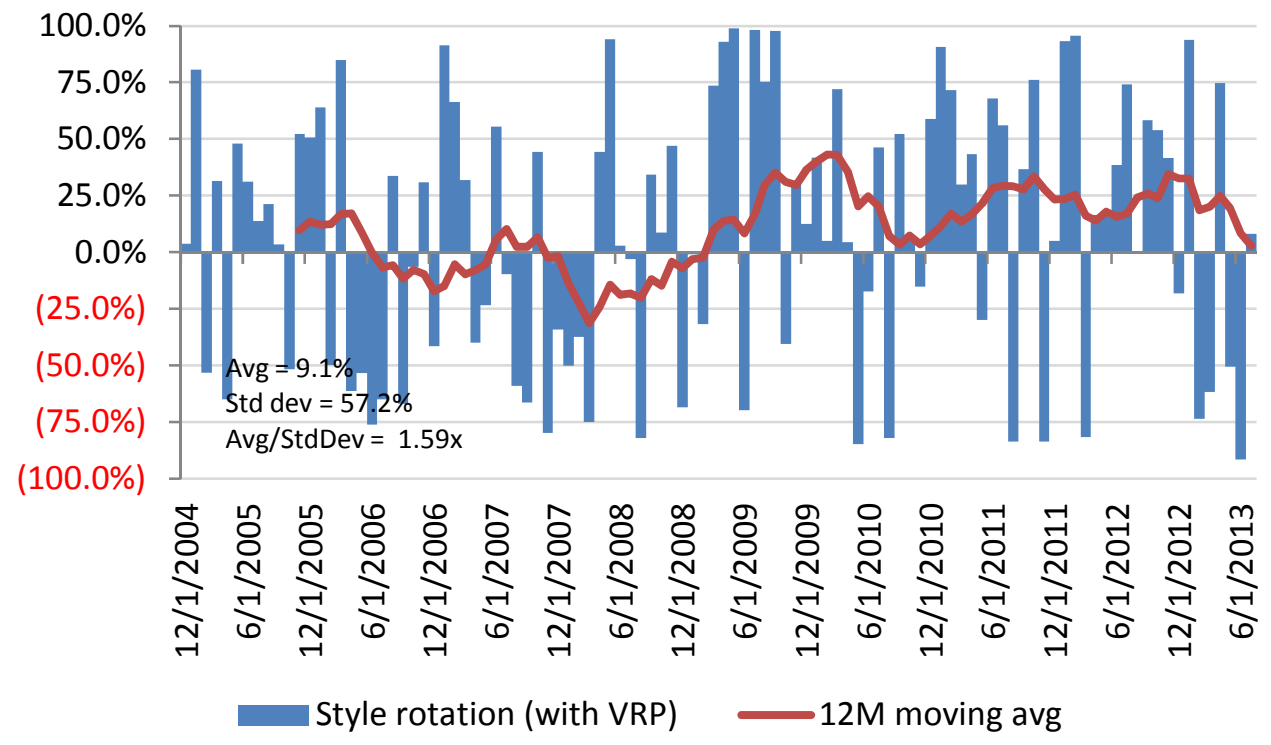

Source: Bloomberg Finance LLP, MSCI, Russell, S&P, Worldscope, Deutsche Bank Quantitative Strategy

Figure 104: Factor rotation model using VRP and SAA risk regimes

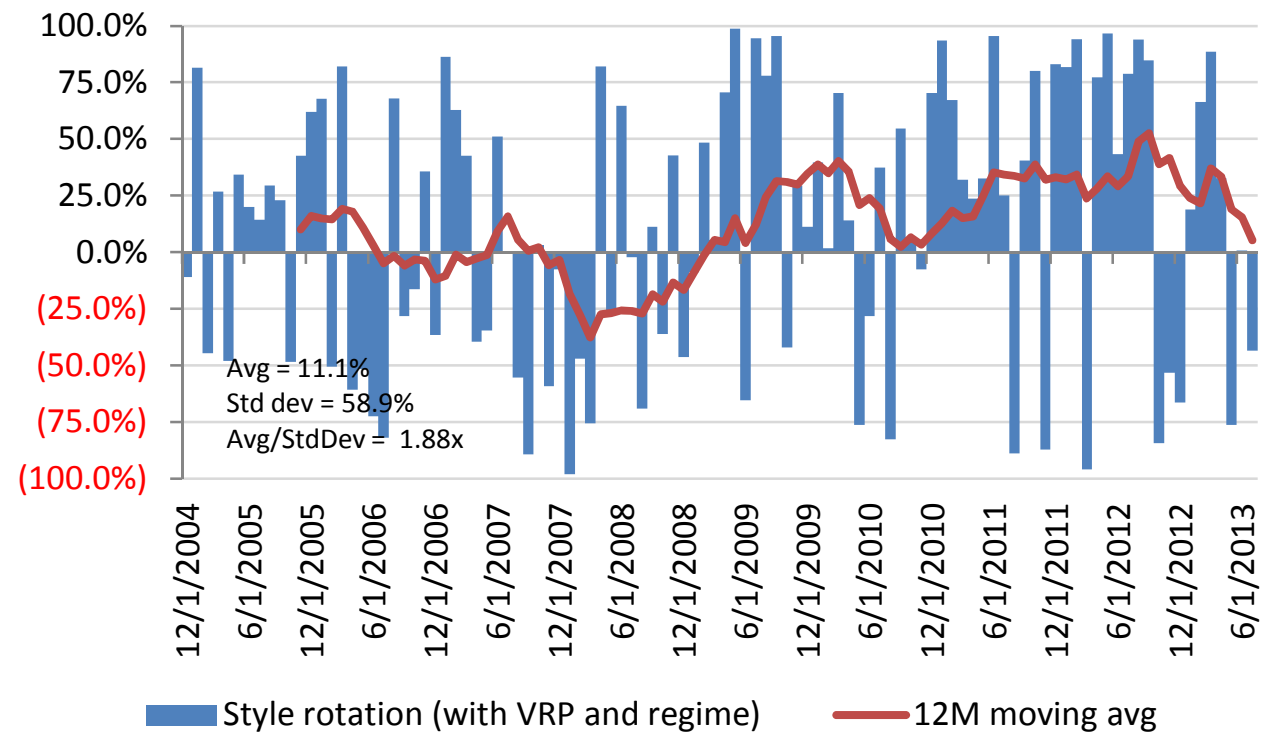

Source: Bloomberg Finance LLP, MSCI, Russell, S&P, Worldscope, Deutsche Bank Quantitative Strategy

## Active and dynamic factor rotation model

In this section, we use the style factor return prediction model built in the previous section as our alpha model, our GARCH-DCC-Copula risk model, and maximum mean-CVaR optimization to construct an active and dynamic factor rotation model.

Starting from November 2004, at the end of each month, we use our style rotation model to predict the return of each factor for the following month as our alpha. We then fit a GARCH-DCC-Copula model using an expanding window. We use the fitted GARCH-DCC-Copula model to simulate return scenarios for our risk models. We use the same robust procedure by repeating the simulation process 20 times, as presented in the previous section. Finally, we choose maximum mean-CVaR as our optimization tool.

As show in Figure 105 and Figure 106, our factor rotation strategy certainly shows higher return and risk when compared to risk-based strategies like RiskParity or MaxDiversification. In terms of Sharpe ratio, our factor rotation model delivers 68% and 44% better performance than RiskParity and MaxDiversification, respectively. Our factor rotation strategy has slightly higher downside risk (see Figure 107), but is less crowded than the risk-based strategies (see Figure 108).

Figure 105: Cumulative performance

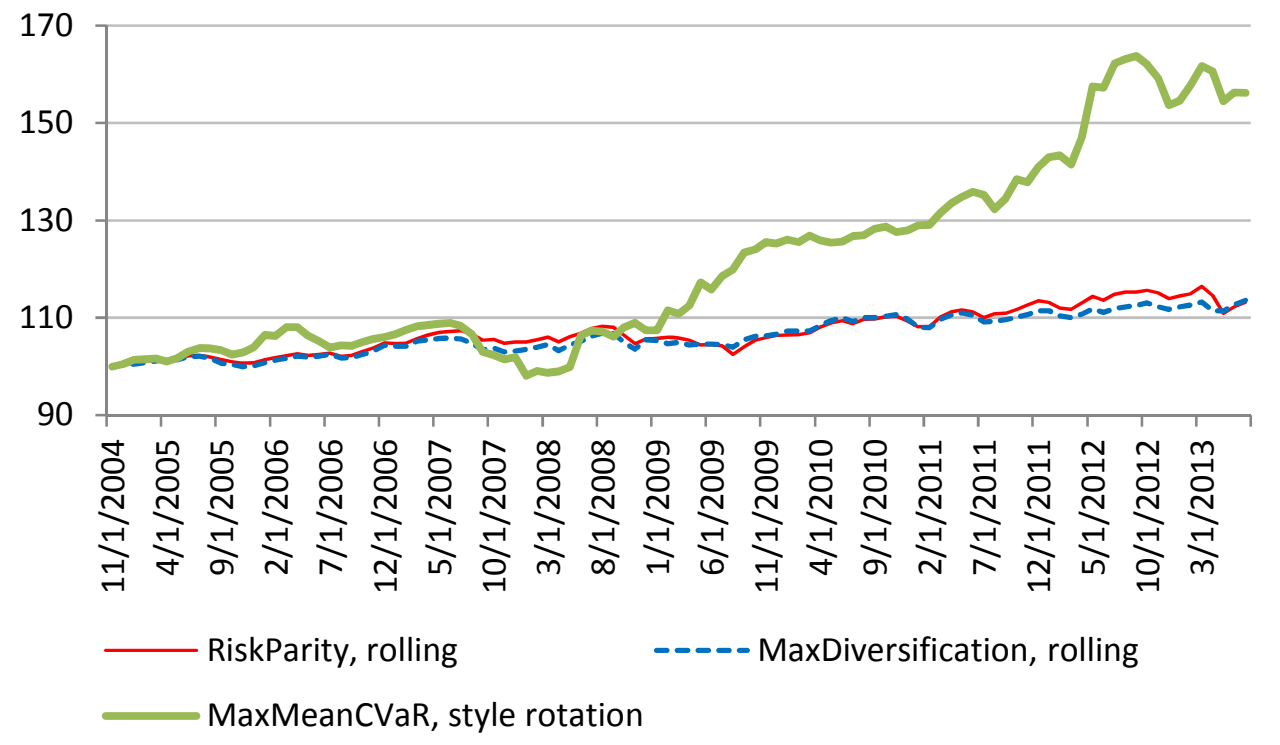

*Source: Bloomberg Finance LLP, MSCI, Russell, S&P, Worldscope, Deutsche Bank Quantitative Strategy*

Figure 106: Return, risk, and Sharpe ratio

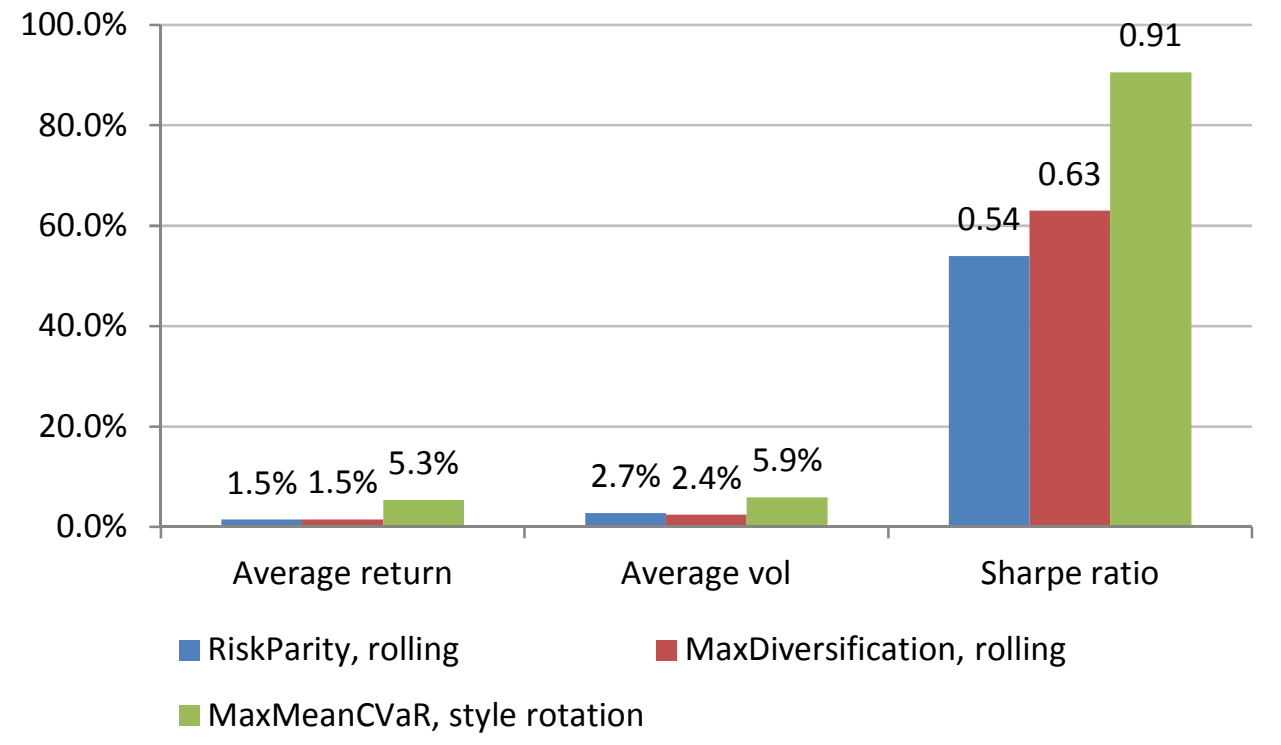

*Source: Bloomberg Finance LLP, MSCI, Russell, S&P, Worldscope, Deutsche Bank Quantitative Strategy*

Figure 107: Downside risk

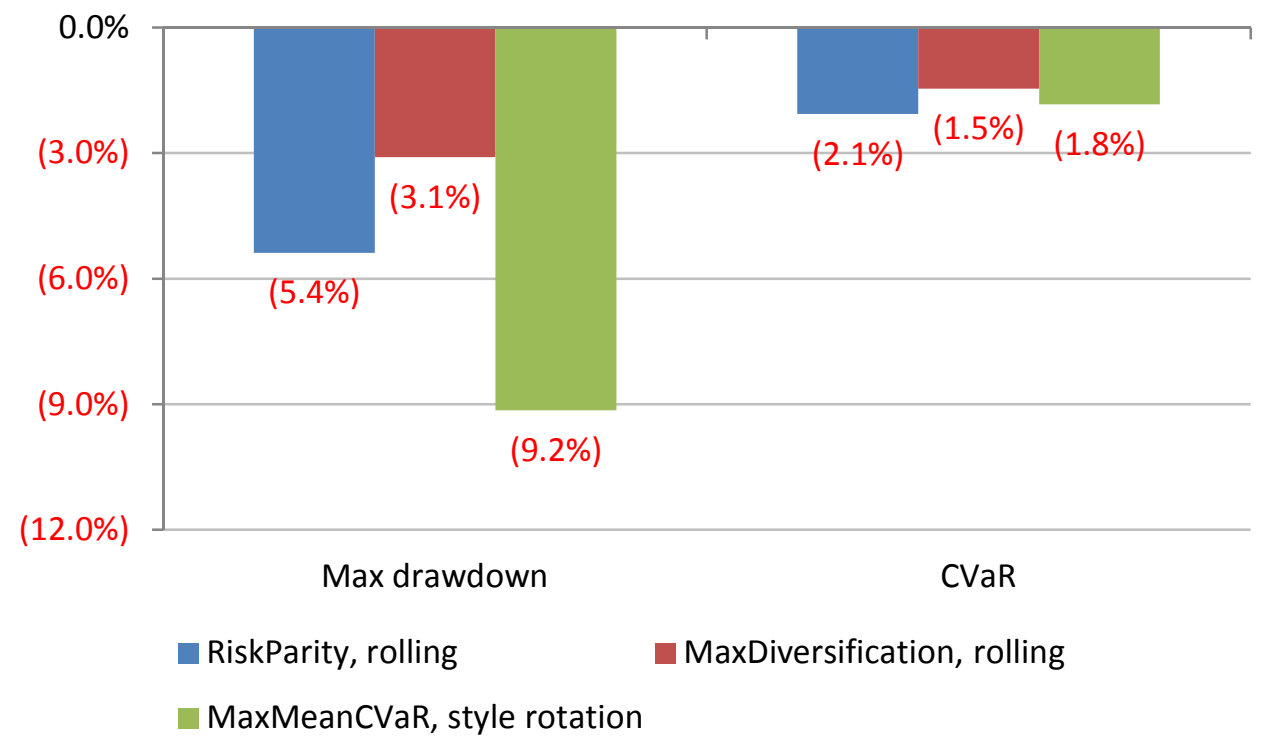

*Source: Bloomberg Finance LLP, MSCI, Russell, S&P, Worldscope, Deutsche Bank Quantitative Strategy*

Figure 108: Weighted portfolio tail dependence

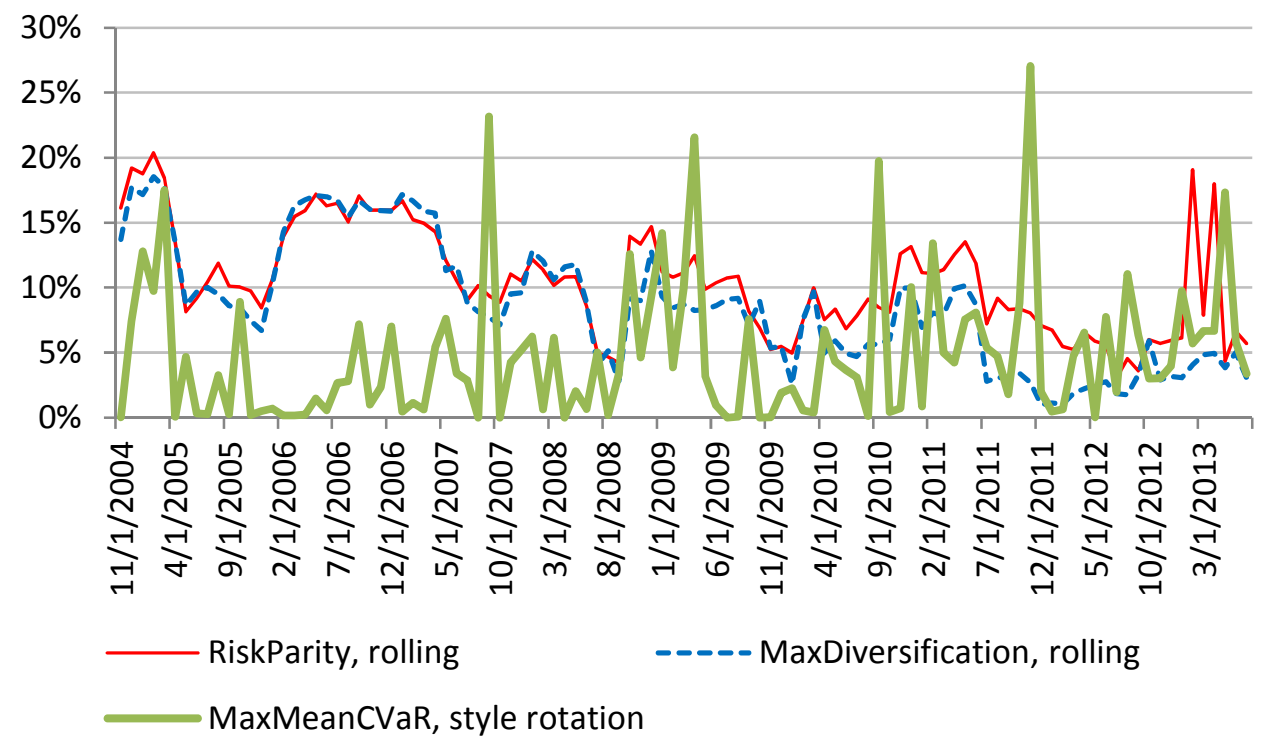

*Source: Bloomberg Finance LLP, MSCI, Russell, S&P, Worldscope, Deutsche Bank Quantitative Strategy*

# VII. Conclusion

In summary, this paper should be useful for both asset allocation managers and quantitative equity investors. We would like to highlight the following key messages:

- Asset returns show serial correlation, extreme outliers, volatility clustering, dynamic correlation, and tail dependence. A GARCH-DCC-Copula model fits asset returns much better than traditional sample-based risk models that only rely on covariance matrices.
- Our GARCH-DCC-Copula risk model significantly improves portfolio performance, for both risk- and alpha-based strategies, for both asset allocation decisions and quantitative equity factor weighting decisions.
- We introduced a representative asset that comprises 50% equities, 40% bonds, and 10% alternatives, to represent the global financial market. We then use our GARCH-DCC-Copula model to simulate and estimate the global financial risk, using conditional value-at-risk (CVaR). Finally, we build a dynamic Markov regime switching model to predict the real-time market risk (and correlation) regimes.
- Our real-time global financial market risk regime indicator can clearly differentiate returns of multi-assets (equities, bonds, and alternative asset classes), risk- and alpha-based portfolio construction techniques, and finally, stock-selection factors (or alternative betas). It can be used in GTAA and style rotation models to improve our return prediction ability.

Lastly, we would like to give our readers a hint of the upcoming Part 3 of our *DB Handbook of Portfolio Construction* series. As shown in this research, a better alpha model is more likely to lead to a better portfolio, but not necessarily all the time. In the upcoming research, we will address the various approaches of incorporating alpha into portfolio construction, discuss their pros and cons, and more importantly, how to choose the best approach for a given problem.

# Appendix 1

## Important Disclosures

### Additional information available upon request

For disclosures pertaining to recommendations or estimates made on securities other than the primary subject of this research, please see the most recently published company report or visit our global disclosure look-up page on our website at http://gm.db.com/ger/disclosure/DisclosureDirectory.eqsr

## Analyst Certification

The views expressed in this report accurately reflect the personal views of the undersigned lead analyst(s). In addition, the undersigned lead analyst(s) has not and will not receive any compensation for providing a specific recommendation or view in this report. Yin Luo/Sheng Wang/Javed Jussa/Zongye Chen/Miguel-A Alvarez

## Hypothetical Disclaimer

Backtested, hypothetical or simulated performance results have inherent limitations. Unlike an actual performance record based on trading actual client portfolios, simulated results are achieved by means of the retroactive application of a backtested model itself designed with the benefit of hindsight. Taking into account historical events the backtesting of performance also differs from actual account performance because an actual investment strategy may be adjusted any time, for any reason, including a response to material, economic or market factors. The backtested performance includes hypothetical results that do not reflect the reinvestment of dividends and other earnings or the deduction of advisory fees, brokerage or other commissions, and any other expenses that a client would have paid or actually paid. No representation is made that any trading strategy or account will or is likely to achieve profits or losses similar to those shown. Alternative modeling techniques or assumptions might produce significantly different results and prove to be more appropriate. Past hypothetical backtest results are neither an indicator nor guarantee of future returns. Actual results will vary, perhaps materially, from the analysis.

# Regulatory Disclosures

## 1.Important Additional Conflict Disclosures

Aside from within this report, important conflict disclosures can also be found at https://gm.db.com/equities under the "Disclosures Lookup" and "Legal" tabs. Investors are strongly encouraged to review this information before investing.

## 2. Short-Term Trade Ideas

Deutsche Bank equity research analysts sometimes have shorter-term trade ideas (known as SOLAR ideas) that are consistent or inconsistent with Deutsche Bank's existing longer term ratings. These trade ideas can be found at the SOLAR link at http://gm.db.com.

## 3. Country-Specific Disclosures

Australia and New Zealand: This research, and any access to it, is intended only for "wholesale clients" within the meaning of the Australian Corporations Act and New Zealand Financial Advisors Act respectively.
Brazil: The views expressed above accurately reflect personal views of the authors about the subject company(ies) and its(their) securities, including in relation to Deutsche Bank. The compensation of the equity research analyst(s) is indirectly affected by revenues deriving from the business and financial transactions of Deutsche Bank. In cases where at least one Brazil based analyst (identified by a phone number starting with +55 country code) has taken part in the preparation of this research report, the Brazil based analyst whose name appears first assumes primary responsibility for its content from a Brazilian regulatory perspective and for its compliance with CVM Instruction # 483.
EU countries: Disclosures relating to our obligations under MiFiD can be found at http://www.globalmarkets.db.com/riskdisclosures.
Japan: Disclosures under the Financial Instruments and Exchange Law: Company name - Deutsche Securities Inc. Registration number - Registered as a financial instruments dealer by the Head of the Kanto Local Finance Bureau (Kinsho) No. 117. Member of associations: JSDA, Type II Financial Instruments Firms Association, The Financial Futures Association of Japan, Japan Investment Advisers Association. Commissions and risks involved in stock transactions - for stock transactions, we charge stock commissions and consumption tax by multiplying the transaction amount by the commission rate agreed with each customer. Stock transactions can lead to losses as a result of share price fluctuations and other factors. Transactions in foreign stocks can lead to additional losses stemming from foreign exchange fluctuations. "Moody's", "Standard & Poor's", and "Fitch" mentioned in this report are not registered credit rating agencies in Japan unless Japan or "Nippon" is specifically designated in the name of the entity. Reports on Japanese listed companies not written by analysts of Deutsche Securities Inc. (DSI) are written by Deutsche Bank Group's analysts with the coverage companies specified by DSI.
Russia: This information, interpretation and opinions submitted herein are not in the context of, and do not constitute, any appraisal or evaluation activity requiring a license in the Russian Federation.

David Folkerts-Landau
Global Head of Research

Marcel Cassard
Global Head
CB&S Research

Ralf Hoffmann & Bernhard Speyer
Co-Heads
DB Research

Guy Ashton
Chief Operating Officer
Research

Richard Smith
Associate Director
Equity Research

Asia-Pacific
Fergus Lynch
Regional Head

Germany
Andreas Neubauer
Regional Head

North America
Steve Pollard
Regional Head

## International Locations

Deutsche Bank AG
Deutsche Bank Place
Level 16
Corner of Hunter & Phillip Streets
Sydney, NSW 2000
Australia
Tel: (61) 2 8258 1234

Deutsche Bank AG
Große Gallusstraße 10-14
60272 Frankfurt am Main
Germany
Tel: (49) 69 910 00

Deutsche Bank AG
Filiale Hongkong
International Commerce Centre,
1 Austin Road West,Kowloon,
Hong Kong
Tel: (852) 2203 8888

Deutsche Securities Inc.
2-11-1 Nagatacho
Sanno Park Tower
Chiyoda-ku, Tokyo 100-6171
Japan
Tel: (81) 3 5156 6770

Deutsche Bank AG London
1 Great Winchester Street
London EC2N 2EQ
United Kingdom
Tel: (44) 20 7545 8000

Deutsche Bank Securities Inc.
60 Wall Street
New York, NY 10005
United States of America
Tel: (1) 212 250 2500

# Global Disclaimer